\documentclass[12pt,preprint]{aastex}

\newcommand\iso[2]{$^{\rm #1}$#2}

\shorttitle{$\omega$ Centauri Abundances}
\shortauthors{Johnson et al.}

\begin{document}

\title{Chemical Abundances for 855 Giants in the Globular Cluster
Omega Centauri (NGC 5139)}

\author{Christian I. Johnson\altaffilmark{1,2,3} and
Catherine A. Pilachowski\altaffilmark{1,3}
}

\altaffiltext{1}{Department of Astronomy, Indiana University,
Swain West 319, 727 East Third Street, Bloomington, IN 47405--7105, USA;
cijohnson@astro.ucla.edu; catyp@astro.indiana.edu}

\altaffiltext{2}{Department of Physics and Astronomy, University of California,
Los Angeles, 430 Portola Plaza, Box 951547, Los Angeles, CA 90095--1547, USA}

\altaffiltext{3}{Visiting astronomer, Cerro Tololo Inter--American 
Observatory, National Optical Astronomy Observatory, which are operated by the 
Association of Universities for Research in Astronomy, under contract with the 
National Science Foundation.}

\begin{abstract}

We present elemental abundances for 855 red giant branch (RGB) stars in the 
globular cluster Omega Centauri ($\omega$ Cen) from spectra obtained with
the Blanco 4m telescope and Hydra multifiber spectrograph.  The sample 
includes nearly all RGB stars brighter than V=13.5, and span's $\omega$ Cen's
full metallicity range.  The heavy $\alpha$ elements (Si, Ca, and Ti) are
generally enhanced by $\sim$+0.3 dex, and exhibit a metallicity dependent
morphology that may be attributed to mass and metallicity dependent Type II
supernova (SN) yields.  The heavy $\alpha$ and Fe--peak abundances suggest
minimal contributions from Type Ia SNe.  The light elements (O, Na, and Al)
exhibit $>$0.5 dex abundance dispersions at all metallicities, and a
majority of stars with [Fe/H]$>$--1.6 have [O/Fe], [Na/Fe], and [Al/Fe]
abundances similar to those in monometallic globular clusters, as well as
O--Na, O--Al anticorrelations and the Na--Al correlation in all but the most
metal--rich stars.  A combination of pollution from intermediate mass
asymptotic giant branch (AGB) stars and \emph{in situ} mixing may explain
the light element abundance patterns.  A large fraction (27$\%$) of $\omega$
Cen stars are O--poor ([O/Fe]$<$0) and are preferentially located within 
5--10$\arcmin$ of the cluster center.  The O--poor giants are spatially 
similar, located in the same metallicity range, and are present in nearly 
equal proportions to blue main sequence stars.  This suggests the O--poor
giants and blue main sequence stars may share a common origin.  [La/Fe] 
increases sharply at [Fe/H]$\ga$--1.6, and the [La/Eu] ratios indicate the 
increase is due to almost pure s--process production.  

\end{abstract}


\section{INTRODUCTION}

For many years, globular clusters were regarded as prototypical simple stellar
populations.  However, recent observations have revealed that several of the
most massive known globular clusters contain multiple main sequence, subgiant,
and/or red giant branch (RGB) populations (Piotto et al. 2007; Marino et al. 
2008; Milone et al. 2008; Anderson et al. 2009; Moretti et al. 2009; Piotto 
2009; Milone et al. 2010).  These data, combined with the well--known and 
pervasive light element abundance correlations and anticorrelations that 
appear to be unique to the globular cluster environment, suggest that many, if 
not all, globular clusters undergo at least some degree of self--enrichment 
(e.g., Carretta et al. 2009a).  While nearly all of these clusters exhibit 
small ($\la$0.1 dex) star--to--star metallicity variations (e.g., see review by 
Gratton et al. 2004), Omega Centauri ($\omega$ Cen) has long been known to 
exhibit both a complex color--magnitude diagram and a metallicity spread of 
more than a factor of ten.

Early color--magnitude diagrams of $\omega$ Cen indicated that it hosts an 
unusually broad RGB (e.g., Woolley 1966; Cannon \& Stobie 1973).  Subsequent
photometric surveys have discovered that this trend continues into both the 
main sequence and subgiant branch regions as well (Anderson et al. 1997; Lee et
al. 1999; Hilker \& Richtler 2000; Hughes \& Wallerstein et al. 2000; Pancino 
et al. 2000; van Leeuwen et al. 2000; Bedin et al. 2004; Ferraro et al. 2004; 
Rey et al. 2004; Sollima et al. 2005; Castellani et al. 2007; Sollima et al. 
2007; Villanova et al. 2007; Bellini et al. 2009a; Calamida et al. 2009).  
Additionally, detailed photometric and spectroscopic analyses have shown
that at least 4--5 discrete populations are present in the cluster (Norris et 
al. 1996; Lee et al. 1999; Hilker \& Richtler 2000; Pancino et al. 2000; Bedin 
et al. 2004; Rey et al. 2004; Sollima et al. 2005; Castellani et al. 2007; 
Villanova et al. 2007; Johnson et al. 2008; Bellini et al. 2009a,b; Calamida et
al. 2009; Johnson et al. 2009; Marino et al. 2010).  These individual
populations span a metallicity range from [Fe/H]\footnote{We make use of the 
standard spectroscopic notations where 
[A/B]$\equiv$log(N$_{\rm A}$/N$_{\rm B}$)$_{\rm star}$--
log(N$_{\rm A}$/N$_{\rm B}$)$_{\sun}$ and
log $\epsilon$(A)$\equiv$log(N$_{\rm A}$/N$_{\rm H}$)+12.0 for elements A
and B.}$\approx$--2.2 to --0.5.  However, few stars are found with 
[Fe/H]$<$--2, and more than half of $\omega$ Cen's stars reside in a population
peaked near [Fe/H]$\approx$--1.7 (Norris \& Da Costa 1995; Suntzeff \& Kraft 
1996; Hilker \& Richtler 2000; Smith et al. 2000; Cunha et al. 2002; Sollima et
al. 2005; Kayser et al. 2006; Stanford et al. 2006; Villanova et al. 2007; 
Johnson et al. 2008; Calamida et al. 2009; Johnson et al. 2009; Marino et al. 
2010).  The rest of the stars reside in the intermediate metallicity 
populations, and a minority ($\la$5$\%$) of stars are found to lie along the 
``anomalous", metal--rich sequence (Lee et al. 1999; Pancino et al. 
2000; Ferraro et al. 2004; Sollima et al. 2005; Villanova et al. 2007).

The large metallicity spread in $\omega$ Cen is commonly believed to be due to
significant self--enrichment induced by multiple star formation episodes 
(e.g., Ikuta \& Arimoto 2000; Tsujimoto \& Shigeyama 2003; Marcolini et al. 
2007; Romano et al. 2007, 2010).  Despite being the most massive known cluster
in the Galaxy, with an estimated mass of $\sim$2--7$\times$10$^{\rm 6}$ 
M$_{\sun}$ (Mandushev et al. 1991; Richer et al. 1991; Meylan et al. 1995; van 
de Ven et al. 2006), Gnedin et al. (2002) showed that $\omega$ Cen does not
currently possess an abnormally deep gravitational potential well.  
Additionally, the cluster's Galactic orbit indicates that it should pass 
through the disk at least every 1--2$\times$10$^{\rm 8}$ years (Dinescu et al.
1999).  This makes it hard to believe that $\omega$ Cen could have experienced 
the 2--4 Gyr period of star formation that seems required
to fit observations of the main sequence turnoff (e.g., Stanford et al. 2006).
However, the cluster's retrograde motion through the Galaxy (Dinescu et al. 
1999) suggests that it may be a captured system and therefore could have been 
more massive in the past.  In fact, the most popular scenario is that $\omega$
Cen, and perhaps several other globular clusters containing multiple 
populations, are the remnant cores of tidally disrupted dwarf galaxies (e.g.,
Dinescu et al. 1999; Majewski et al. 2000; Smith et al. 2000; Gnedin et al.
2002; McWilliam \& Smecker--Hane 2005; Bekki \& Norris 2006).  This is now
favored over an accretion or merger scenario because the individual stellar 
populations within $\omega$ Cen all exhibit the same proper motion, rotation, 
and average radial velocity (e.g., Pancino et al. 2007; Bellini et al. 2009a).

Although the observed evolutionary sequences have now been mostly matched to
the different populations derived from spectroscopy, one of the remaining 
puzzles is how these populations relate to $\omega$ Cen's bifurcated main 
sequence.  The discovery that $\omega$ Cen's main sequence splits into a red 
and blue sequence over a span of at least two magnitudes (e.g., Anderson 1997; 
Bedin et al. 2004) is difficult to explain because the blue main sequence
is more metal--rich than the red main sequence (Piotto et al. 2005).
A possible explanation for this is that the blue main sequence stars 
are selectively enhanced in helium at Y$\sim$0.38 (e.g., Norris 2004; Piotto 
et al. 2005).  While the source of the proposed helium enrichment is not 
clear, the leading candidate appears to be intermediate mass 
($\sim$3--8 M$_{\sun}$) asymptotic giant branch (AGB) stars with perhaps some 
contribution from massive, rapidly rotating main sequence stars (e.g., Renzini 
2008; Romano et al. 2010).  Interestingly, the blue main sequence stars appear 
to be preferentially located near the cluster core (Sollima et al. 2007; 
Bellini et al. 2009b), which is an indication that He--rich material may have 
collected there at some point in the cluster's evolution.  A similar radial 
segregation near the core has been found for stars with [Fe/H]$\ga$--1.2, but 
the more metal--poor stars appear to be rather uniformly distributed across 
the cluster (Suntzeff \& Kraft 1996; Norris et al. 1997; Hilker \& Richtler 
2000; Pancino 2000; Rey 2004; Johnson et al. 2008; Bellini et al. 2009b; 
Johnson et al. 2009).  It is worth noting that helium enrichment may also play 
a role in determining the chemical composition of stars in monometallic 
globular clusters (e.g., Bragaglia et al. 2010).

$\omega$ Cen shows clear signs of extended star formation and chemical
self--enrichment, and the large abundance dispersion is not limited to just the
Fe--peak elements.  Instead, the [X/H] ratios for all elements analyzed so far
are found to vary by at least a factor of ten as well (e.g., Cohen 1981; 
Paltoglou \& Norris 1989; Norris \& Da Costa 1995; Smith et al. 2000; Cunha 
et al. 2002; Johnson et al. 2009; Villanova et al. 2009; Cunha et al. 2010; 
Stanford et al. 2010).  Despite the current interpretation that $\omega$ Cen 
may be the surviving core of a disrupted dwarf galaxy, the [X/Fe] abundance 
ratios for the light elements (O, Na, Mg, and Al) and heavy $\alpha$ elements 
(Si, Ca, and Ti) more closely resemble the patterns found in individual 
globular clusters.  These patterns include the O--Na, O--Al, and Mg--Al 
anticorrelations concurrent with the Na--Al correlation and consistently 
supersolar [$\alpha$/Fe] ratios (e.g., Norris \& Da Costa 1995; Smith et al. 
2000; Johnson et al. 2009).  This suggests that both Type II supernovae (SNe) 
and the products of proton--capture nucleosynthesis have played a significant 
role in shaping $\omega$ Cen's chemical enrichment.  However, the abundance 
patterns of neutron--capture elements in $\omega$ Cen stars with 
[Fe/H]$\ga$--1.5 indicate that the slow neutron--capture process (s--process) 
was also a dominant production mechanism.  This strongly contradicts the 
trends found in other globular clusters, and is instead more similar to 
observations of dwarf galaxies (e.g., see reviews by Venn et al. 2004; Geisler 
et al. 2007).  While dwarf galaxies also contain many s--process enhanced 
stars, the low [X/Fe] ratios for the light and especially $\alpha$ elements 
suggests a significant contribution from Type Ia SNe.  In contrast, the 
enhanced [$\alpha$/Fe] ratios, high [Na/Fe] and [Al/Fe] abundances, and low 
[Cu/Fe] ratios seem to indicate that Type Ia SNe have played only a minimal 
role in $\omega$ Cen.  However, Type Ia SNe may have contributed in the most 
metal--rich stars, as is evidenced by a potential downturn in [$\alpha$/Fe] 
and rise in [Cu/Fe] at [Fe/H]$>$--1 (Pancino et al. 2002; Origlia et al. 2003; 
but see also Cunha et al. 2002).

In this paper we have obtained a nearly complete sample that includes 855 RGB 
stars and covers $\omega$ Cen's full metallicity range down to V=13.5.  We 
present new chemical abundance measurements of several light odd--Z,
$\alpha$, Fe--peak, and neutron--capture elements, and compare these results 
with abundance trends found in the Galactic disk, halo, bulge, globular 
cluster, and nearby dwarf galaxy populations.  We also compare the abundance 
patterns found for the different $\omega$ Cen populations in an effort to 
understand the cluster's formation and chemical enrichment history.

\section{OBSERVATIONS AND REDUCTIONS}

The observations for this project were taken at the Cerro Tololo 
Inter--American Observatory (CTIO) using the Blanco 4m telescope equipped with
the Hydra multifiber positioner and bench spectrograph.  We obtained all of 
the spectra in two separate runs spanning 24--28 March 2008 and 6--8 March 
2009.  We employed two different wavelength setups encompassing 
$\sim$6135--6365 and $\sim$6500--6800 \AA\ with wavelength centers near
6250 and 6670 \AA, respectively.  This required the use of two separate 
order blocking filters for each echelle spectrograph setup.  The red setup
centered near 6670 \AA\ used the echelle filter \#6 (E6757); however, neither
the echelle filter \#7 nor \#8 provides sufficient transmissivity over the 
bluer region spanning 6135--6365 \AA.  Since a primary goal of this project 
is to obtain both oxygen and sodium abundances from the 6300 \AA\ [O I] line
and 6154/6160 \AA\ Na I lines, we purchased a new, single--piece echelle filter
(E6257\footnote{This filter is on long--term loan at CTIO and available for
public use.}) that provides $>$75$\%$ transmissivity from $\sim$6135--6365 \AA\ 
and allowed for the simultaneous observation of both the oxygen and sodium 
lines.  For both setups, the ``large" 300$\micron$ (2$\arcsec$) fibers combined
with the 400 mm Bench Schmidt Camera and 316 line mm$^{\rm -1}$ echelle grating
to yield a resolving power of R($\lambda$/$\Delta$$\lambda$)$\approx$18,000
(0.35 \AA\ FWHM).  A summary of the Hydra observations is provided in Table 1.

\begin{figure}
\plotone{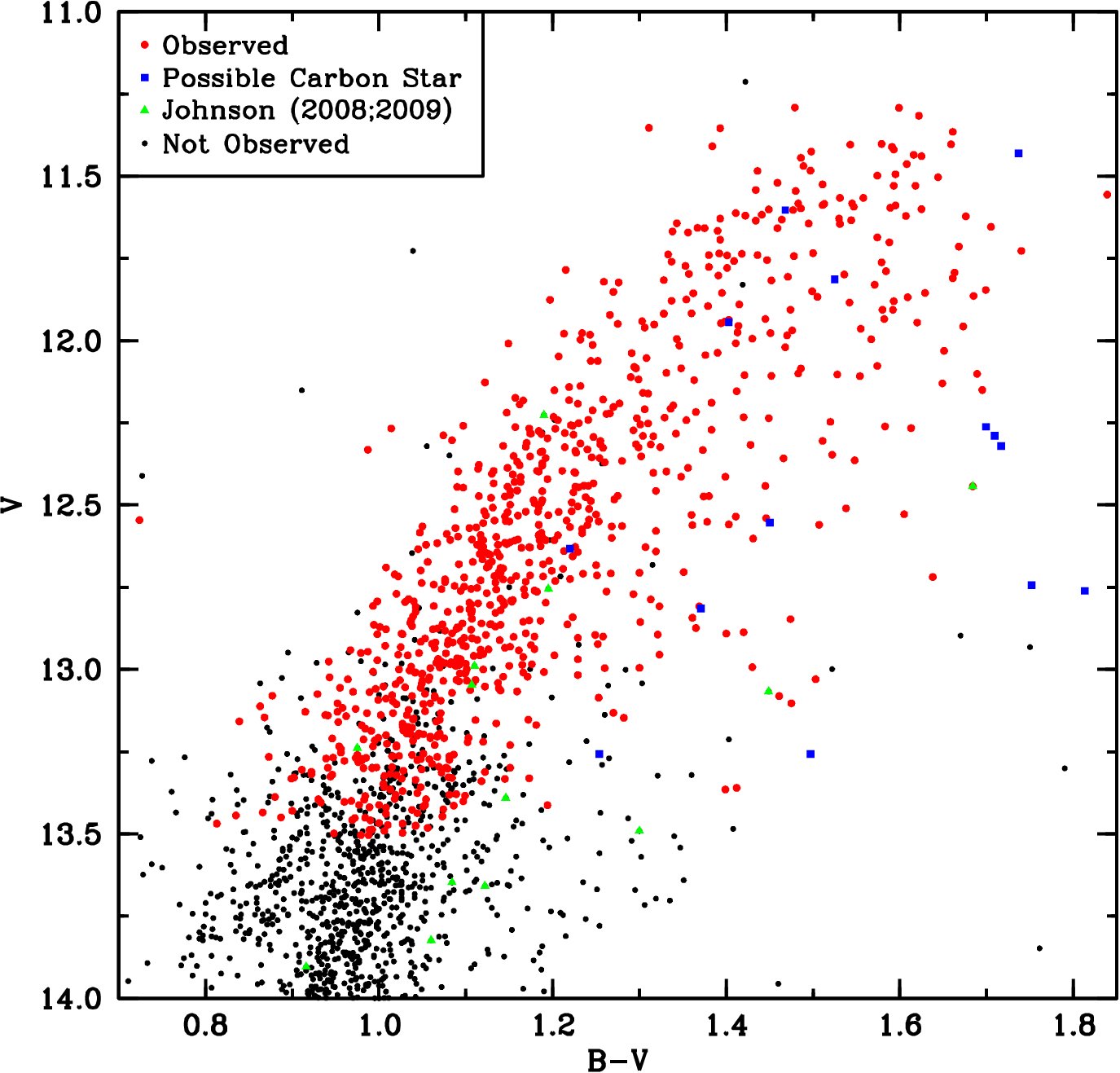}
\caption{A color--magnitude diagram of $\omega$ Cen's RGB with photometry
taken from van Leeuwen et al. (2000).  The filled red circles represent the
stars observed for this study.  The filled blue squares indicate possible 
carbon stars.  Note that the identifiers for the carbon stars are provided
in Table 2.  The filled green triangles show stars that were observed for
Johnson et al. (2008; 2009), but do not overlap with the current sample.  The
complete sample from van Leeuwen et al. is represented by the small black
circles.}
\label{f1}
\end{figure}

Photometry, coordinates, and membership probabilities for all stars were
taken from the proper motion study by van Leeuwen et al. (2000).  We targeted
stars with V$\leq$13.5 and 0.70$\leq$B--V$\leq$1.85 while excluding those 
with membership probabilities below 70$\%$.  Field stars located along 
$\omega$ Cen's line--of--sight are easily removed due to the cluster's 
comparatively large radial velocity and small velocity dispersion 
($\langle$V$_{\rm R}$$\rangle$$\sim$232 km s$^{\rm -1}$; 
$\sigma$$\sim$10 km s$^{\rm -1}$; e.g., Reijns et al. 2006; Sollima et al. 
2009).  The magnitude and color restrictions provide a balance between 
maximizing the signal--to--noise ratio (S/N) of observations and limiting the 
number of required Hydra configurations.  At V=13.5, one can obtain a 
S/N$\approx$100 after three hours of integration.  This luminosity cutoff also 
allows for the observation of all giant branches in $\omega$ Cen, and is at 
least 1 mag. below the RGB tip of the most metal--rich stellar population (see 
Figure \ref{f1}).  

In order to limit the number of repeat observations, stars were given a low 
priority in the Hydra assignment code following their inclusion into a Hydra
configuration, and stars were completely removed from the fiber 
assignment process if incorporated into two Hydra configurations.  The
total number of fibers assigned to objects ranged from 50--110, and the 
co--added S/N ratio for almost all stars extended from about 100 to more than 
350.  The full sample obtained for this project is shown in Figure \ref{f1} 
along with the non--repeat stars from our previous papers on the cluster.  

\begin{figure}
\plotone{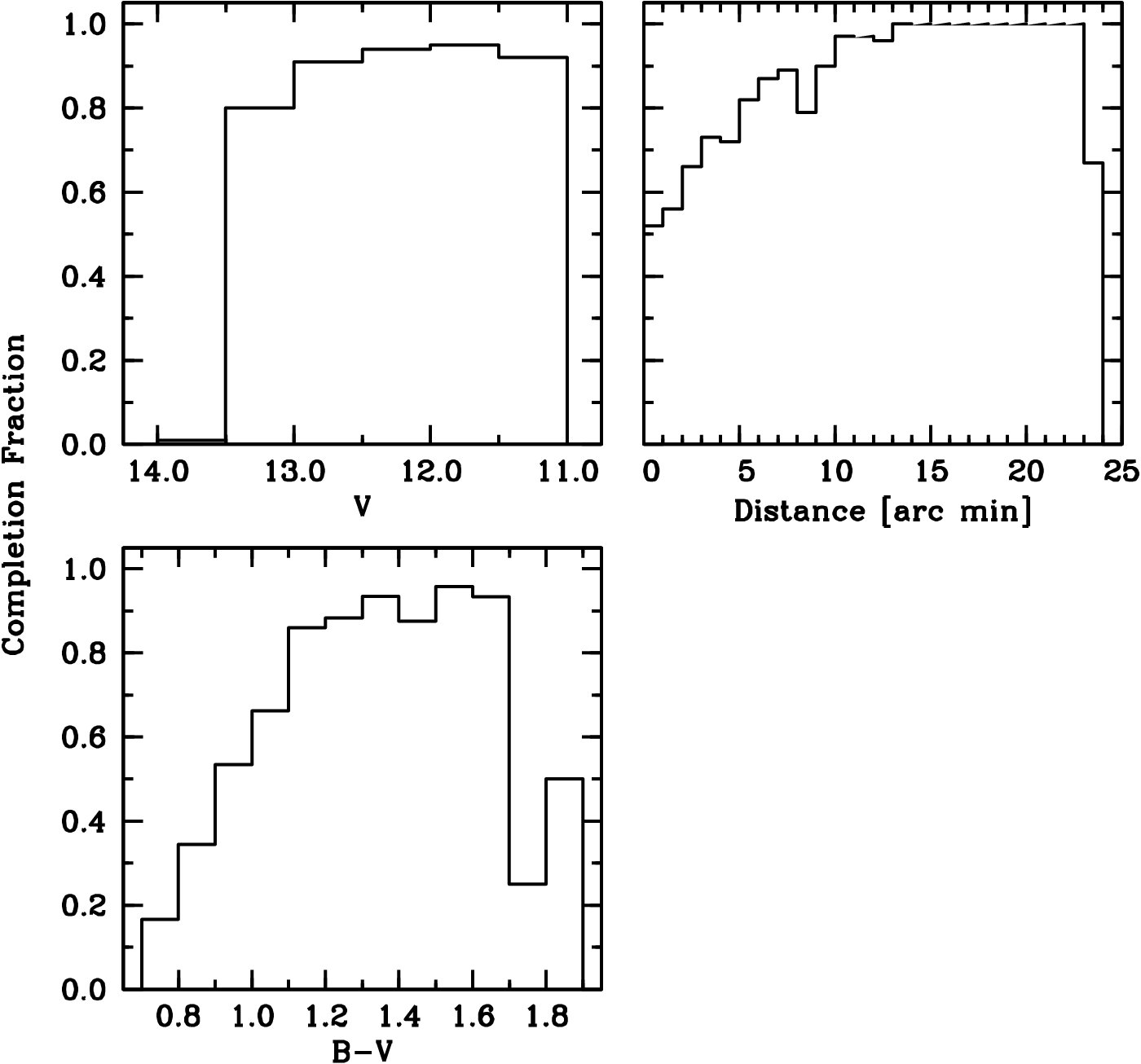}
\caption{The three panels illustrate the observed completion fraction of our
sample in terms of V magnitude, B--V color, and distance from the cluster
center relative to the van Leeuwen et al. (2000) observations.  For the bottom 
left and top right panels, the completion fraction only includes stars 
with V$\leq$13.5, as discussed in $\S$2.}
\label{f2}
\end{figure}

\begin{figure}
\plotone{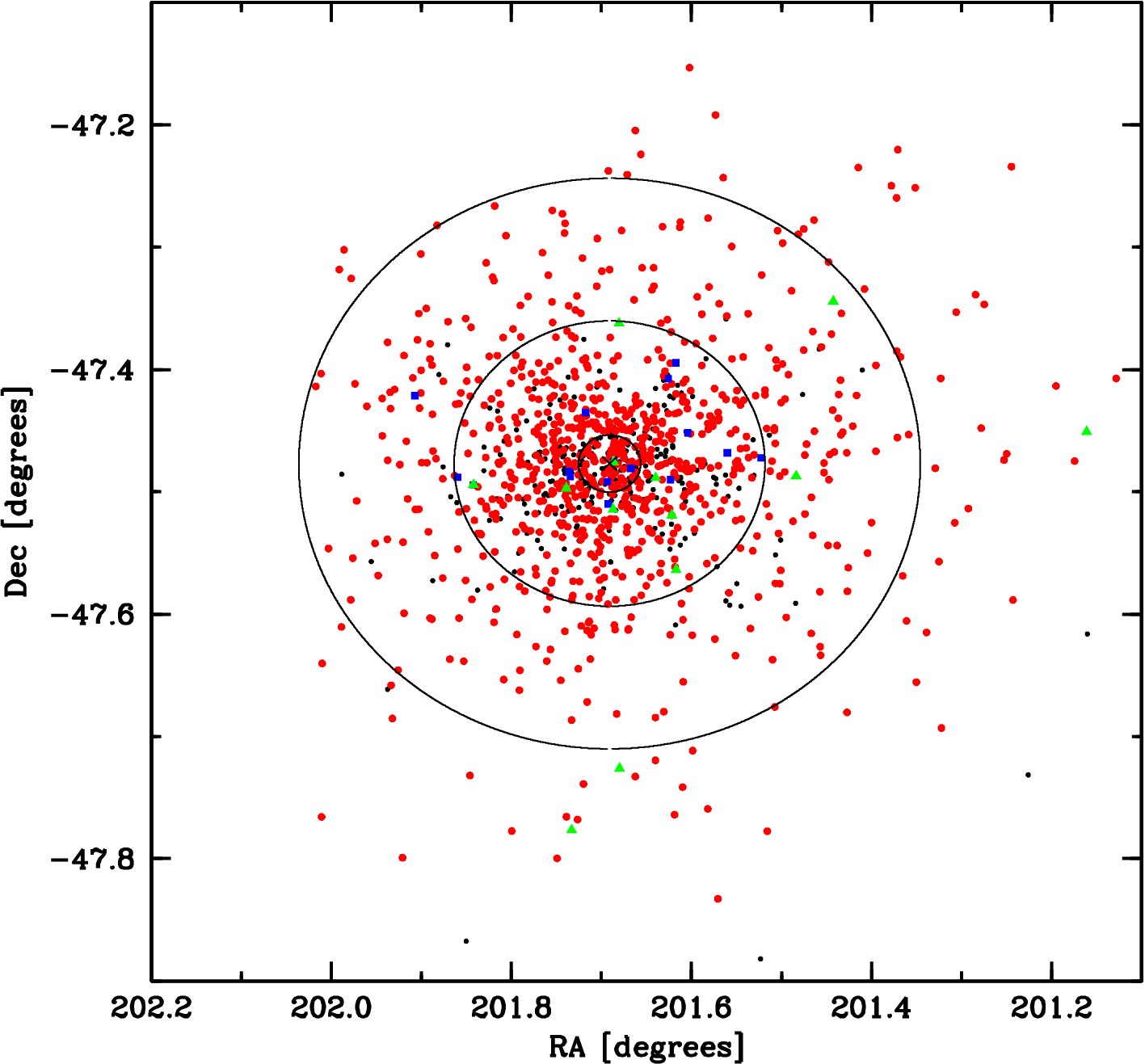}
\caption{This figure shows the coordinate positions of our sample.  The cross
indicates the cluster center defined by van Leeuwen et al. (2000) as 
201.691$\degr$, --47.4769$\degr$ (J2000; 
13$^{\rm h}$26$^{\rm m}$45.9$^{\rm s}$, --47$\degr$28$\arcmin$37.0$\arcsec$).
The ellipses represent 1, 5, and 10 times the core radius of 1.40$\arcmin$
(Harris et al. 1996).  The symbols are the same as those in Figure \ref{f1},
and the van Leeuwen et al. data only represent stars with V$\leq$13.5 and a
membership probability $\geq$70$\%$.}
\label{f3}
\end{figure}

The complexity of $\omega$ Cen's color--magnitude diagram requires a large 
sample of stars to fully interpret its chemical history.  Therefore, we have 
obtained a nearly 100$\%$ complete sample of RGB members with V$\leq$13.0 and 
achieved more than 75$\%$ completion for V$\leq$13.5.  Since $\omega$ Cen 
exhibits a moderate radial metallicity gradient (Norris et al. 1996; Suntzeff 
\& Kraft 1996; Norris et al. 1997; Hilker \& Richtler 2000; Pancino et al. 
2000; Rey et al. 2004; Johnson et al. 2008; Bellini et al. 2009b; Johnson et 
al. 2009), we targeted stars spanning a wide range of cluster radii.  Figure 
\ref{f2} shows the observed completion fraction in terms of V magnitude, B--V 
color, and distance from the cluster center, and Figure \ref{f3} illustrates 
the spatial location of our sample relative to the cluster center.  For 
V$\leq$13.0, B--V$>$1.1, and 10$\arcmin$$<$D$<$24$\arcmin$, the completion 
fraction exceeds 0.90.  However, the completion fraction for the inner 
10$\arcmin$ of the cluster ranges from 0.52--0.90.  The decrease is due to both
stellar crowding near the cluster core and physical limitations on fiber 
placement.  Despite the large sample size, a modest evolutionary selection 
effect is present because the most metal--rich stars have both lower V 
magnitudes and tend to be located closer to the cluster center.  Therefore, we 
have only observed stars along the most metal--rich giant branch that are 
within $\sim$1 mag. of the RGB tip.

Data reduction was handled using the necessary tasks provided in standard
IRAF\footnote{IRAF is distributed by the National Optical Astronomy 
Observatories, which are operated by the Association of Universities for 
Research in Astronomy, Inc., under cooperative agreement with the National 
Science Foundation.} packages.  We used \emph{ccdproc} to trim the overscan
region and apply the bias level correction.  However, the majority of the data 
reduction process was carried out with the \emph{dohydra} package, which was 
used to trace the fiber locations on the detector, remove scattered light, 
apply the flat--field correction, identify lines in the ThAr comparison 
spectrum, apply the wavelength calibration, remove cosmic rays, subtract the 
background sky spectra, and extract the one--dimensional spectra.  The 
reduction processes were identical for both the 6250 and 6750 \AA\ data with 
the exception of the wavelength calibration.  A problem with the calibration 
lamp during the 6750 \AA\ observations meant that we had to use a high S/N, 
daylight solar spectrum for wavelength calibration instead of the ThAr 
comparison source.  

Following completion of the \emph{dohydra} task, the data were continuum fit 
and normalized before being corrected for telluric contamination.  We obtained 
high S/N spectra of multiple bright, rapidly rotating B--stars spanning
air masses ranging from 1.05 to 1.75.  These spectra were used as the 
templates for removing the telluric features in the 6270--6350 \AA\ window.  
Fortunately, the cluster's radial velocity corresponds to a wavelength shift of
roughly $+$4.8 \AA.  This moves the 6300 \AA\ [O I] stellar absorption line 
away from the telluric emission feature at 6300 \AA, and places it cleanly 
between the 6302 and 6306 \AA\ telluric absorption doublets.  After applying 
the telluric correction, the spectra were then co--added to remove any 
remaining cosmic rays and increase the S/N.

\section{Analysis}

\subsection{Model Stellar Atmospheres}

Effective temperatures (T$_{\rm eff}$) were determined via the empirical 
V--K color--temperature relation from Alonso et al. (1999, 2001; their 
equations 8 \& 9), which is based on the infrared flux method (Blackwell \& 
Shallis 1977).  The V magnitudes were taken from van Leeuwen et al. (2000) and 
the K magnitudes were taken from the Two Micron All Sky Survey (2MASS; 
Skrutskie et al. 2006) database\footnote{The 2MASS catalog can be accessed 
online at: http://irsa.ipac.caltech.edu/applications/Gator/.}.  All photometry 
was corrected for interstellar reddening and extinction using the recommended 
values of E(B--V)=0.12 (Harris 1996) and E(V--K)/E(B--V)=2.70 (McCall 2004).
While there is some evidence for minor differential reddening near the 
cluster's core (Cannon \& Strobie 1973; Calamida et al. 2005; van Loon et al. 
2007; McDonald et al. 2009), the well--defined evolutionary sequences observed 
in the photometry by Villanova et al. (2007) seem to suggest that differential 
reddening is not a major issue.  Therefore, we have applied a uniform reddening
correction that is independent of a star's location in the cluster.  Although 
our data set did not contain enough Fe I lines of varying excitation potential 
to strongly constrain T$_{\rm eff}$ via excitation equilibrium, we did not find
any strong, systematic trends in plots of Fe I abundance versus excitation 
potential.  It is likely that our photometric temperatures are accurate to 
within the roughly 25--50 K uncertainty range given by the Alonso et al. (1999)
empirical fits.

Surface gravity (log g) estimates were obtained using the photometric 
temperatures and absolute bolometric magnitudes (M$_{\rm bol}$).  
The bolometric corrections were taken from Alonso et al. (1999; their 
equations 17 and 18) and applied to the absolute visual magnitudes 
(M$_{\rm V}$), which assumed a distance modulus of (m--M)$_{\rm V}$=13.7 (van 
de Ven et al. 2006).  Final surface gravity values were calculated with the 
standard relation,
\begin{equation}
log(g_{*})=0.40(M_{bol.}-M_{bol.\sun})+log(g_{\sun})+4[log(T/T_{\sun})]+
log(M/M_{\sun}),
\end{equation}
and assumed stellar mass of 0.80 M$_{\rm \sun}$.  However, the likely age 
spread of $\sim$2--4 Gyr (e.g., Stanford et al. 2006) among the stars in 
different populations means that a mass spread among RGB stars undoubtedly 
exists as well.  This is further complicated by the inferred existence of a 
helium--rich population (Bedin et al. 2004; Norris 2004; Piotto et al. 2005) in
which stars will evolve more rapidly.  When one also includes ``contamination" 
of first ascent RGB stars with AGB stars, which could account for as much as 
20--40$\%$ of the RGB above the horizontal branch (e.g., Norris et al. 1996), 
it is not unreasonable to assume $\omega$ Cen giants will have a mass range
spanning $\sim$0.60--0.80 M$_{\rm \sun}$.  Fortunately, the surface gravity 
estimates scale with log(M) and are thus relatively insensitive to small 
changes in the assumed stellar mass.  We estimate that the uncertainty 
introduced into our surface gravity values due to the inherent mass range
on $\omega$ Cen's RGB does not exceed $\Delta$log g=0.15.  Comparison between
the abundances of elements in different ionization states seems to 
substantiate this with $\langle$[FeI/H]-[FeII/H]$\rangle$=--0.09 
($\sigma$=0.10) and $\langle$[ScI/Fe]-[ScII/Fe]$\rangle$=--0.18
($\sigma$=0.21).  We provide a more detailed analysis regarding how
surface gravity uncertainties affect abundance ratio determinations in $\S$3.3.

In addition to effective temperature and surface gravity, metallicity and 
microturbulence (v$_{\rm t}$) information are required to generate a suitable
1--D model atmosphere.  For an initial metallicity estimate, we used the
empirical [Ca/H] calibration provided by van Leeuwen et al. (2000; their 
equation 15) with the assumptions that stars with [Fe/H]$<$--1 have 
[Ca/Fe]=$+$0.30 and those with [Fe/H]$>$--1 decline to [Ca/Fe]=0 at solar 
metallicity.  This assumption is verified in our new [Ca/Fe] data 
(see $\S$4.6).  An initial microturbulence value was determined from the 
empirical v$_{\rm t}$--T$_{\rm eff}$ relation given in Johnson et al. (2008;
their equation 2).  The initial T$_{\rm eff}$, log g, [Fe/H], and v$_{\rm t}$ 
values were used to generate the necessary model atmospheres via 
interpolation within the available ATLAS9\footnote{The model atmosphere grids 
can be downloaded from http://cfaku5.cfa.harvard.edu/grids.html.} grid.  The 
final determination of all microturbulence values followed the prescription
outlined by Magain (1984) in which the microturbulence was adjusted until 
trends of Fe I abundance versus line strength were removed.  The overall 
model atmosphere metallicity was then adjusted to match the derived [Fe/H] 
abundance for each star.  This value was also used to further refine the 
calculated effective temperature, which has a slight metallicity dependence.  A
full listing of star identifiers, photometry, model atmosphere parameters, and
S/N ratios is provided in Table 2.

\subsection{Derivation of Abundances}

\subsubsection{Equivalent Width Analysis}

Chemical abundances for Na, Si, Ca, Sc, Ti, Fe, and Ni were determined through
standard equivalent width (EW) analyses using the \emph{abfind} driver
in the LTE line analysis code MOOG\footnote{The MOOG code can be downloaded at:
http://www.as.utexas.edu/~chris/moog.html.} (Sneden 1973).  Individual EWs were 
measured by fitting single or multiple Gaussian profiles to isolated and 
blended stellar absorption lines using the interactive EW fitting code 
developed for Johnson et al. (2008).  The high resolution, high S/N solar and 
Arcturus atlases\footnote{These are available online from the NOAO Data 
Archives at: http://www.noao.edu/archives.html.} (Hinkle et al. 2000) were used
to aid in line identification and continuum placement.  The Arcturus atlas was 
also used as a reference for selecting suitable spectral lines.  However, the 
atomic log gf values were determined by an inverse solar analysis in which the 
EWs measured in the Sun were forced to match the photospheric abundances given
in Anders \& Grevesse (1989)\footnote{The solar log $\epsilon$(Fe) abundance
was assumed to be 7.52 (Sneden et al. 1991a).}.  When comparing our derived 
log gf values to those given in the NIST\footnote{The NIST Atomic Line Database
can be accessed at: http://www.nist.gov/physlab/data/asd.cfm.} (Ralchenko et 
al. 2008), Thevenin (1990), and VALD\footnote{The VALD linelist can be accessed
at: http://www.astro.uu.se/~vald/php/vald.php.} (Kupka et al. 2000) 
compilations, we find very good agreement such that 
$\langle$log gf$_{\rm ours}$--log gf$_{\rm lit.}$$\rangle$=--0.02 
($\sigma$=0.08).  

While most abundances were determined through a straight--forward EW analysis,
the odd--Z Fe--peak and neutron--capture elements have line profiles that may 
be affected by hyperfine structure.  For the purposes of this study, this 
includes the elements Sc, La, and Eu.  Prochaska \& McWilliam (2000) give 
hyperfine log gf values for the 6245 \AA\ Sc II line, but unfortunately their 
work does not include the 6309 \AA\ Sc II nor the 6210 and 6305 Sc I lines
used here.  Similarly, the Zhang et al. (2008) analysis of solar Sc abundances
only includes the 6245 \AA\ line as well.  However, the error introduced by
ignoring hyperfine structure increases as a function of EW, and the small
Sc EWs in our sample ($\langle$EW$\rangle$=42 m\AA, $\sigma$=26 m\AA) lead us 
to believe a standard EW abundance analysis is a reasonable approach for this
element.

For abundance determinations of the neutron--capture elements La and Eu, we 
have employed the hyperfine structure linelists available in Lawler et al.
(2001a) for the 6262 \AA\ La II line and Lawler et al. (2001b) for the 
6645 \AA\ Eu II line.  However, the La abundances were determined by spectrum 
synthesis and are described in $\S$3.2.2.  Eu abundance determinations are
more complicated than those for most elements because the line profiles are
both affected by hyperfine splitting and Eu has two stable, naturally occurring
isotopes (\iso{151}{Eu} and \iso{153}{Eu}) with solar system isotopic fractions
of 47.8$\%$ and 52.2$\%$, respectively (Lawler et al. 2001b).  The Eu EWs were 
measured using the same interactive fitting code mentioned previously, and the 
EWs were combined with the Eu linelist and isotope fractions as inputs into the
MOOG \emph{blends} driver to obtain the final abundances.

All EWs measured for this project and the atomic linelists are provided in 
Tables 3a--3b for Fe and Tables 4a--4b for all other elements.  Similarly, the
chemical abundance ratios for all elements are given in Table 5, and the number
of lines measured for each element per star along with the $\sigma$/$\surd$(N)
values are available in Tables 6a--6b.  The log gf values listed for La and Eu 
in Table 4b represent the \emph{total} gf values instead of an individual 
hyperfine component.  The interested reader can find the full linelists for 
these elements in the references given above.  

\subsubsection{Spectrum Synthesis Analysis}

The abundances of O, Al, and La were determined by spectrum synthesis rather
than the EW fitting method described in $\S$3.2.1.  The primary motivation for
using synthesis instead of an EW analysis for these elements is that the 
available lines suffer from varying degrees of contamination with either 
nearby metal lines or molecular CN.  The 6300.31 \AA\ [O I] line is blended 
with the nearby 6300.70 Sc II feature, and is also moderately sensitive to the
C+N abundance.  Similarly, the 6696/6698 \AA\ Al I lines are both moderately 
blended with nearby Fe I and CN features, and the 6262 \AA\ La II line is 
lightly blended with both CN and a Co I line at 6262.81 \AA.  

Spectrum synthesis modeling was carried out using the \emph{synth} driver in 
MOOG.  The atomic linelist was generated primarily from the Kurucz online
database\footnote{The online database can be found 
at: http://kurucz.harvard.edu/LINELISTS/GF100/} with updated log gf values
provided by C. Sneden (2008, private communication).  The atomic linelist was 
merged with a molecular CN linelist that was created through a combination of 
the Kurucz molecular linelist\footnote{The molecular linelist can be found
at: http://kurucz.harvard.edu/LINELISTS/LINESMOL/} and one provided by
B. Plez (2007, private communication; see also Hill et al. 2002).  Individual 
log gf values for lines of interest were verified through spectrum synthesis 
of both the solar and Arcturus atlases.  As was mentioned in $\S$3.2.1, the 
La II hyperfine structure linelist was taken from Lawler et al. (2001a).

Since most stars in our sample do not have published [C/Fe], [N/Fe], and/or
\iso{12}{C}/\iso{13}{C} ratios, we set [C/Fe]=--0.5, \iso{12}{C}/\iso{13}{C}=4,
and treated the N abundance as a free parameter to fit the available CN 
features.  Previous work on evolved RGB stars in $\omega$ Cen (e.g., Norris \& 
Da Costa 1995; Smith et al. 2002; Origlia et al. 2003; Stanford et al. 2010) 
has shown that our set values for [C/Fe] and \iso{12}{C}/\iso{13}{C} are a 
reasonable approximation given that all of the stars in our sample will have 
already undergone first dredge--up and are above the RGB luminosity bump.  With
these assumptions, values of $+$0.80$\la$[N/Fe]$\la$$+$1.50 tended to provide 
the best fits to the CN lines.

\begin{figure} 
\plotone{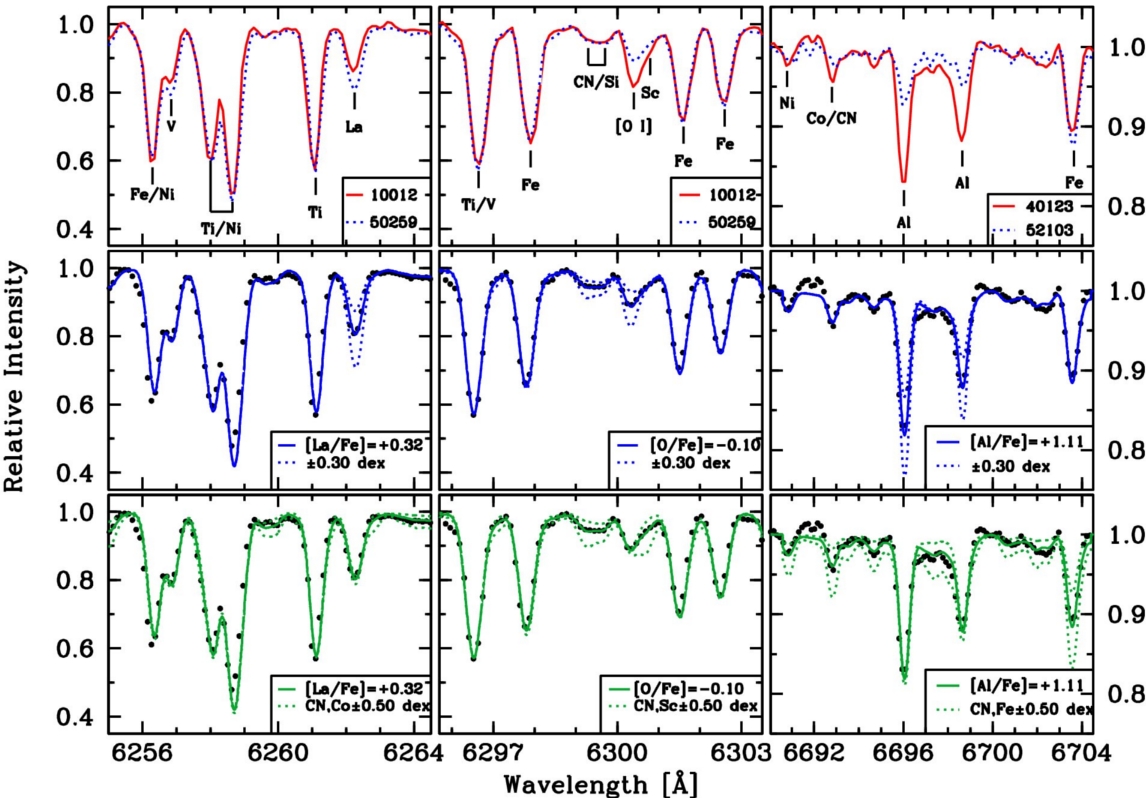}
\caption{The top three panels overplot sample spectra with nearly identical
atmospheric parameters and metallicity, but which exhibit significant 
differences in their derived [La/Fe], [O/Fe], and [Al/Fe] abundances.  The 
middle set of panels show sample spectrum syntheses of La and O for star 50259
in the left and middle panels and Al for star 40123 in the right panel.  In 
all three of the middle panels the solid line indicates the best fit to the
spectra and the dotted lines indicate changes of $\pm$0.30 dex for the 
indicated elements.  The bottom set of panels show sample spectrum syntheses,
but the dotted lines in these cases illustrate changes to the synthetic spectra
when the blended features of CN, Co, Sc, or Fe are altered by $\pm$0.50 dex.
Note that the right panels have a different intensity scale than the left and
middle panels.}
\label{f4}
\end{figure}

Figure \ref{f4} shows sample spectra of four moderately metal--poor 
([Fe/H]$\approx$--1.45) program stars along with synthetic spectrum fits to 
the O, La, and Al regions.  The bottom panels of Figure \ref{f4} indicate the
uncertainty introduced when the abundances of CN and other nearby, blended
metals are altered by $\pm$0.50 dex.  In warmer stars and those that are 
moderately metal--poor, the CN contamination does not provide a significant 
change in the derived abundance.  However, cooler and more metal--rich stars
have O, La, and Al abundances that can deviate by at least 0.10--0.20 dex 
compared to an analysis that does not properly account for molecular blends.
For the Al lines, the nearby Fe lines are generally not much of an issue in 
cool giants because the Fe transitions have excitation potentials $\ga$5 eV.
The O and La lines are also not significantly affected by blends from 
neighboring Fe--peak element features unless the [Fe--peak/Fe] abundance 
exceeds roughly $+$0.3 dex.  However, the O and Sc lines are blended strongly
enough at this resolution to warrant spectrum synthesis regardless of the 
[Sc/Fe] abundance.

\subsubsection{Comparison to Other Studies}

\begin{figure}
\plotone{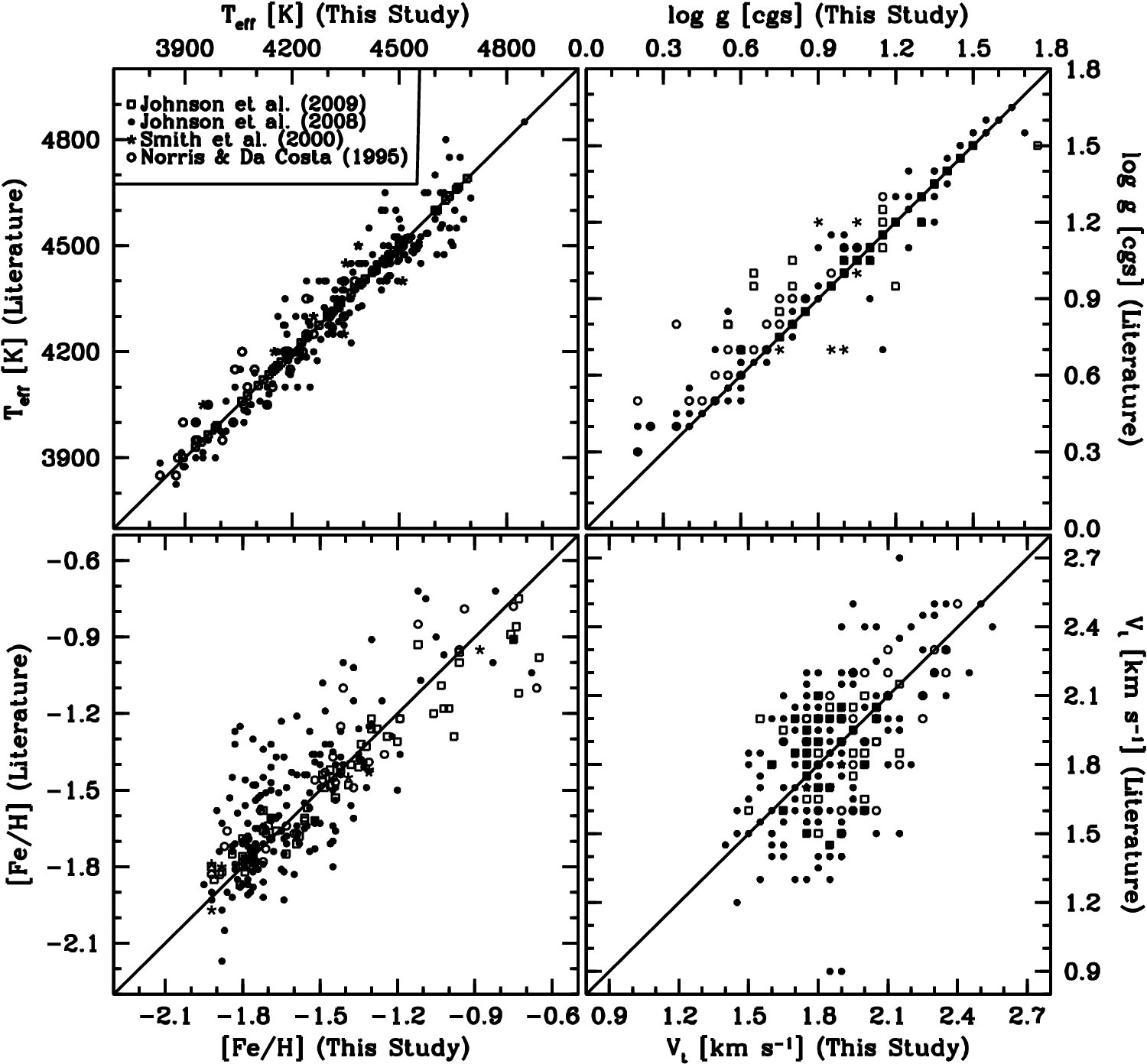}
\caption{The four panels show comparisons of our adopted model atmosphere 
parameters versus those in the literature.  In all panels the solid straight
line indicates perfect agreement.}
\label{f5}
\end{figure}

\begin{figure}
\plotone{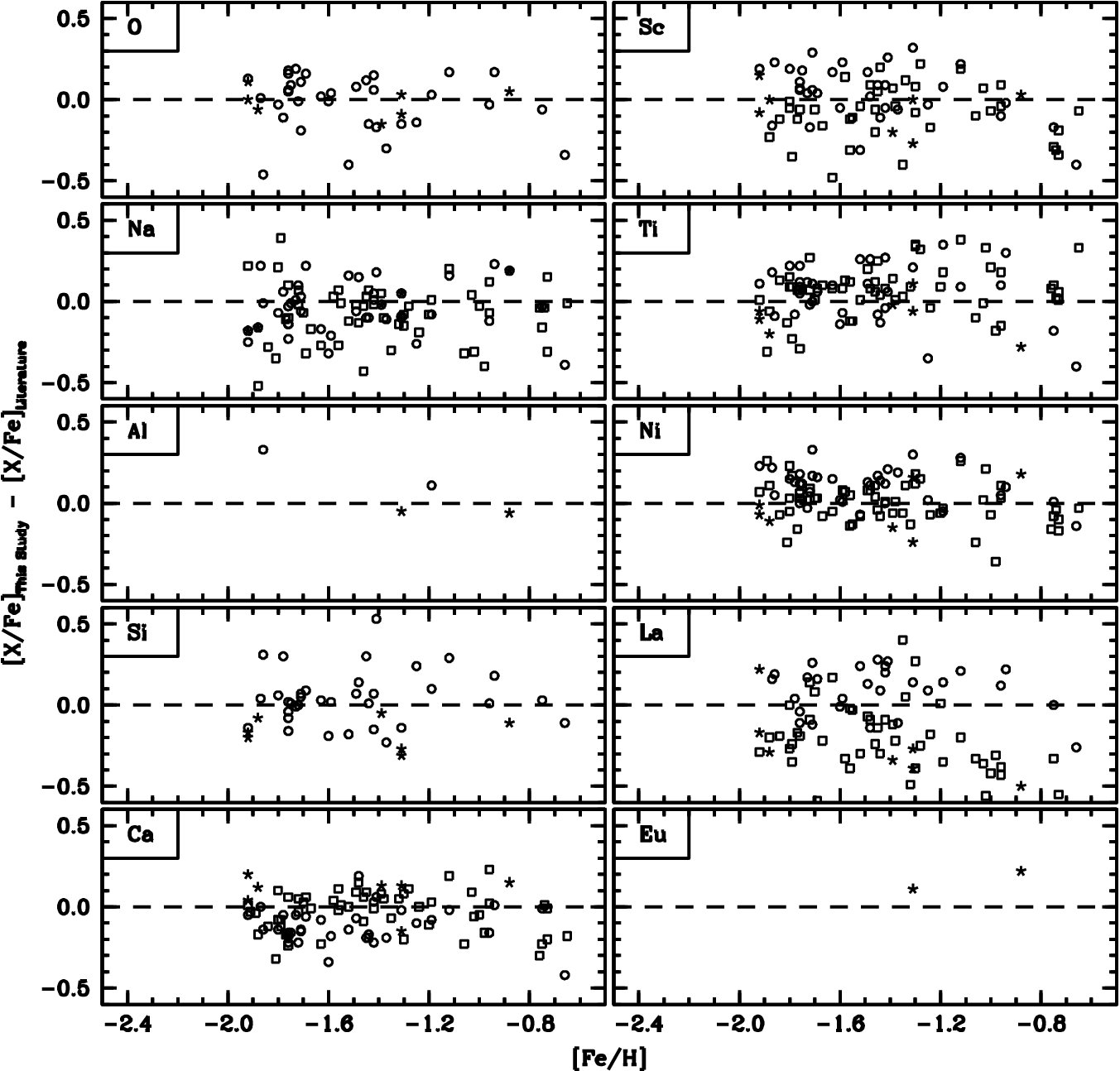}
\caption{The different panels show the [X/Fe] abundance comparisons between
this study and those in the literature.  The dashed line indicates perfect
agreement, and the symbols are the same as those in Figure \ref{f5}.}
\label{f6}
\end{figure}

As described in $\S$1, the chemical composition of $\omega$ Cen has been 
extensively studied using a variety of abundance indicators.  However, there
are only four high resolution spectroscopic studies for which we have more
than five stars in common: Norris \& Da Costa (1995; 35 stars), Smith et al.
(2000; 7 stars), Johnson et al. (2008; 171 stars), and Johnson et al. (2009; 59
stars).  Figure \ref{f5} illustrates the differences between our adopted model 
atmosphere parameters and those found in the literature.  The average 
differences in T$_{\rm eff}$, log g, [Fe/H], and v$_{\rm t}$, in the sense 
present minus literature values, are 0 K ($\sigma$=61 K), 
--0.02 cgs ($\sigma$=0.09), --0.03 dex ($\sigma$=0.17 dex), and
0.02 km s$^{\rm -1}$ ($\sigma$=0.24 km s$^{\rm -1}$), respectively.  We 
conclude from these results that there are no strong systematic offsets among
the studies with regard to the adopted model atmosphere parameters.  This 
conclusion is in agreement with the [X/Fe] abundances comparisons shown in 
Figure \ref{f6}.  The average differences in the chemical abundances between 
this study and those in the literature tend to be $<$0.10 dex 
($\sigma$$\la$0.20 dex).  

The paucity of Al and Eu comparisons shown in Figure \ref{f6} is 
due to two effects: (1) we only obtained about $\sim$40$\%$ as many
spectra in the spectral region that contains the Al and Eu lines and (2) we 
purposely chose to observe stars in the Al/Eu region for which Al and/or Eu
abundances were not already available in the literature.  Further examination
of Figure \ref{f6} indicates that La is the only element showing a systematic
abundance offset.  We tend to find systematically lower [La/Fe] ratios, 
especially at [Fe/H]$\ga$--1.7, because of our inclusion of hyperfine 
structure for the 6262 \AA\ La II line.  Norris \& Da Costa (1995), Smith et 
al. (2000), and Johnson et al. (2009) base all or part of their La abundances 
on the 6774 \AA\ La II line, which suffers from hyperfine broadening.  However,
there are no hyperfine linelists available in the literature for this 
transition.  Since our present data, combined with that from Johnson et al. 
(2009), include both the 6262 and 6774 \AA\ lines, we have derived an 
empirical hyperfine structure correction factor for the 6774 \AA\ line that
is described in Appendix A.

In addition to the studies mentioned above, we also have five stars in common
(ROA 211, 300, 371, WFI 618854, and WFI 222068) with the Pancino et al. (2002)
work that measured [Fe/H], [Si/Fe], [Ca/Fe], and [Cu/Fe] in six relatively
metal--rich ([Fe/H]$\geq$--1.2) $\omega$ Cen giants.  However, despite sharing 
small differences in our derived T$_{\rm eff}$, log g, [Fe/H], and v$_{\rm t}$
values, we find noticeably different [$\alpha$/Fe] abundances for two of the 
most metal--rich stars (ROA 300 and WFI 222068).  This is important because
the Pancino et al. (2002) result is one of the primary studies suggesting that
Type Ia SNe may have significantly affected $\omega$ Cen's chemical 
enrichment.  Origlia et al. (2003) also find a decrease in [$\alpha$/Fe] at
[Fe/H]$>$--1, and we note similar discrepancies in our derived abundances for
the most metal--rich stars.  However, their abundances are based on low
resolution, infrared spectra and may be subject to systematic offsets with
our data.  

\begin{figure}
\plotone{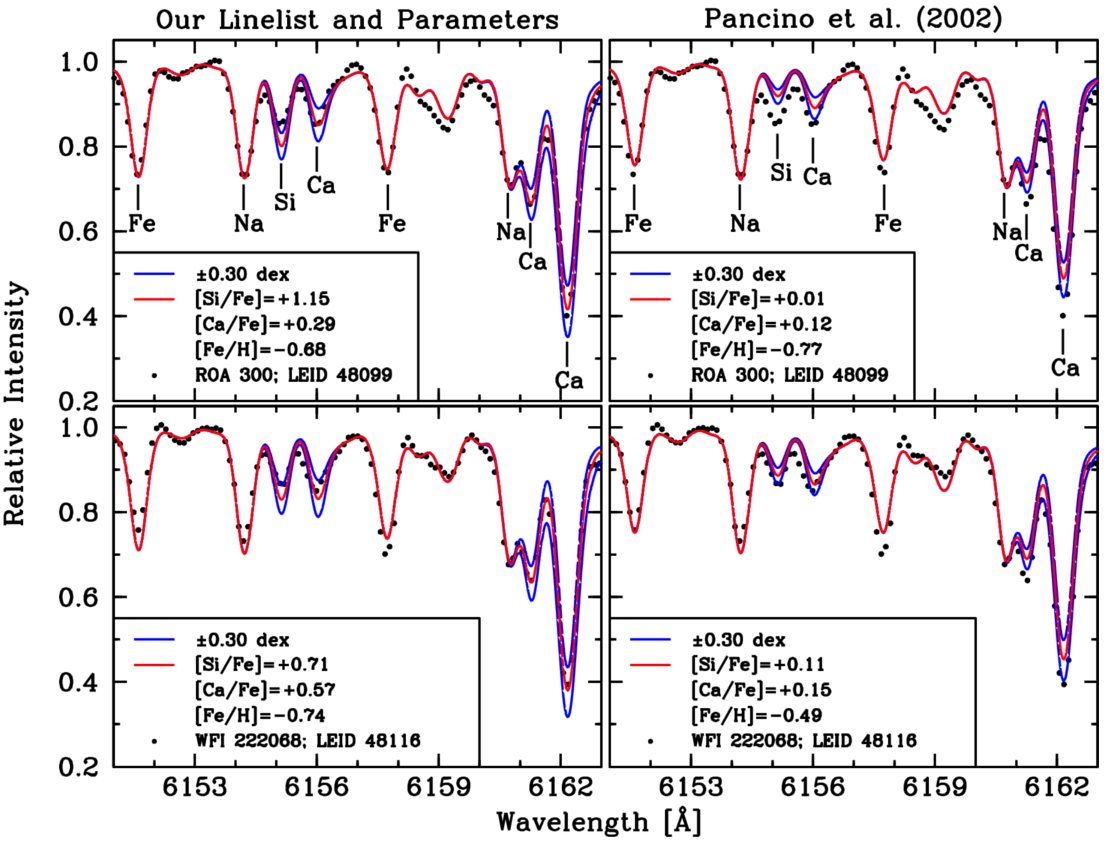}
\caption{Spectrum synthesis fits to the RGB--a stars LEID 48099 (ROA 300) and
LEID 48116 (WFI 222068).  The two left panels show the synthetic spectrum
fits using our measured abundances, model atmosphere parameters, and linelist.
The two right panels show the synthetic spectrum fits using the Pancino et
al. (2002) abundances and model atmosphere parameters \emph{but our linelist}.
In all panels, the solid red line shows the synthesis results using the
predetermined abundances, and the solid blue lines show changes of $\pm$0.3
dex to [Si/Fe] and [Ca/Fe].  Note that the abundances of elements other than
silicon and calcium were set to the values listed in Table 5.}
\label{f7}
\end{figure}

For a direct comparison with the Pancino et al. (2002) data, in star ROA 300
we find Si and Ca offsets of $\Delta$[Si/Fe]=+1.14 and $\Delta$[Ca/Fe]=+0.17.
Similarly, WFI 222068 exhibits differences of $\Delta$[Si/Fe]=+0.60 and 
$\Delta$[Ca/Fe]=+0.42.  To investigate this discrepancy, we ran spectrum 
syntheses for the 6155 \AA\ Si I line and 6156, 6161, and 6162 \AA\ Ca I lines 
(see Figure \ref{f7}).  The results shown in Figure \ref{f7} indicate that the 
Pancino et al. (2002) [Si/Fe] and [Ca/Fe] abundances are too low to match the 
observed spectra \emph{using our linelist and model atmospheres}.  Instead, we 
find better agreement by using the upper limits on the error bars given by 
Pancino et al. (2002; their table 3), which results in increasing their 
[Si/Fe] and [Ca/Fe] abundances by $\sim$+0.2 dex.  

Further inspection of Figure \ref{f7} shows that our EW--based [Si/Fe] 
abundances may have overestimated the true [Si/Fe] abundances for these two
stars by $\sim$+0.3 dex.  We did not find any clear reason for this discrepancy
because the EW--based abundances for calcium and all other elements were in 
agreement with the spectrum synthesis fits, but it is possible that an 
unaccounted for (probably CN) blend is present near the silicon line in 
these very cool (T$_{\rm eff}$$<$4000 K), relatively metal--rich 
([Fe/H]$\sim$--0.7) giants.  It is worth noting that the [Si/Fe] and [Ca/Fe]
abundance values are in much better agreement for two of the warmer, more 
metal--poor stars where the differences between EW-- and synthesis--based 
abundances are negligible.  In the remaining star (ROA 371), the difference
between our derived [Ca/Fe] abundance and that from Pancino et al. (2002) is
mostly negligible, but the [Si/Fe] abundance offset is noticeably larger at
$\Delta$[Si/Fe]=+0.48.  However, this star was also analyzed by both Paltoglou 
\& Norris (1989) and Norris \& Da Costa (1995), and we find in agreement with 
those two studies that ROA 371 is Si--rich with [Si/Fe]$\ga$+0.5.  It seems 
likely that most, if not all, of the discrepancy between our derived abundance
values and Pancino et al. (2002) are the result of differences in adopted log 
gf values, model atmospheres, and line choice.

\subsection{Abundance Sensitivity to Model Atmosphere Parameters}

Table 7 shows the sensitivity of our derived log $\epsilon$(X) abundances 
to changes in the adopted model atmosphere parameters.  The tests were 
conducted at T$_{\rm eff}$=4200 K and T$_{\rm eff}$=4600 K, values
typical of stars in our sample, and metallicities ranging from 
[Fe/H]=--2.0 to --0.50.  The analyses for each test star were run by 
adjusting T$_{\rm eff}$$\pm$100 K, log g$\pm$0.30 cgs, 
[Fe/H]$\pm$0.30 dex, and v$_{\rm t}$$\pm$0.30 km s$^{\rm -1}$ individually 
while holding the other parameters constant.

We find that the chemical abundances derived from subordinate ionization
state transitions (e.g., most neutral metals) are most sensitive to 
changes in T$_{\rm eff}$.  However, abundances derived from dominant ionization
state transitions (e.g., neutral oxygen; singly ionized transition metals and 
heavy elements) are more sensitive to uncertainties in surface gravity and 
metallicity because of their stronger dependence on electron pressure and 
H$^{\rm -}$ opacity.  For stars with [Fe/H]$<$--1, microturbulence was found to
have a negligible effect on abundances derived from all transitions except 
Fe I and Ca I.  Abundances derived from Fe I and Ca I lines were more sensitive
to microturbulence uncertainties because of their typically larger EWs than 
other lines at a given metallicity.  In stars with [Fe/H]$>$--1, most 
abundances were affected at the 0.05--0.10 dex level due to the increased line 
strengths.  Similarly, the abundances of most elements in warmer stars were 
less sensitive to changes in microturbulence because of the generally
weaker line strengths.  It seems likely that our derived log $\epsilon$(X) 
abundance uncertainties do not exceed $\sim$0.20 dex based on our choices of 
model atmosphere parameters.  Additionally, the [X/Fe] abundance ratios for 
most elements are expected to exhibit an even weaker dependence on model 
atmosphere parameter uncertainties because of their similar behavior to Fe I.

In addition to the parameters shown in Table 7, we also tested the abundance
uncertainties based on changes to CN and He.  Since CN lines are strongest in
the O--poor stars, it is possible that standard EW and spectrum synthesis
analyses may not give the same abundances for lines significantly blended with
CN.  However, we find that none of the lines chosen for this study that were
analyzed via a standard EW approach were significantly affected by continuum
suppression or blending from CN.  The robust agreement between the synthesis 
and EW--based analyses for elements other than O, Al, and La is demonstrated 
in Figure \ref{f4}, where the abundances of all other elements studied here 
were preset to those values obtained from a standard EW analysis.  

Since the current interpretation of $\omega$ Cen's blue main sequence is that 
stars belonging to that population are He--rich (Y$\sim$0.38), we investigated
the effects helium enrichment might have on our analyses.  To test this, we
ran both EW and spectrum synthesis analyses using He--normal (Y=0.27) and 
He--rich (Y=0.35) ATLAS9 models\footnote{The He--rich models can be downloaded
at http://wwwuser.oat.ts.astro.it/castelli/grids.html.}.  We find that the 
He--rich model does not result in a significantly different abundance 
($\Delta$log $\epsilon$(X)$<$0.1 dex), and the effects on our derived 
[X/Fe]\footnote{Also note that the decrease in N(H) for He--rich stars will
not affect abundances reported as [X/Fe] ratios because [X/H] and [Fe/H] both
increase by the same amount.} ratios are further mitigated for the low 
ionization potential metals.  These results are in agreement with helium 
enrichment predictions by Boehm--Vitense (1979), and are consistent with 
similar tests on $\omega$ Cen stars in Piotto et al. (2005), Johnson et al. 
(2009), and Cunha et al. (2010).  Furthermore, Girardi et al. (2007) conclude 
that increasing the He abundance to the extreme values predicted in some 
$\omega$ Cen stars should not significantly alter either the bolometric 
correction or V--K color--temperature relation.  We therefore believe that our 
adopted atmospheric parameters are reliable even for He--rich giants.

\section{RESULTS}

\subsection{Iron and the Metallicity Distribution Function}

As discussed in $\S$1, $\omega$ Cen's large metallicity spread has been 
previously verified in many photometric and spectroscopic analyses.  However,
the results presented here are based on direct measurements from high 
resolution, high S/N spectra in a nearly complete sample of $\omega$ Cen 
giants with V$\leq$13.5.  These new data cover the cluster's full metallicity
regime, and are also nearly complete out to $\sim$50$\%$ of the tidal radius.
The data presented here, along with that from Johnson et al. (2008; 2009), 
yield spectroscopic [Fe/H] measurements for 867 giants.  

\begin{figure}
\plotone{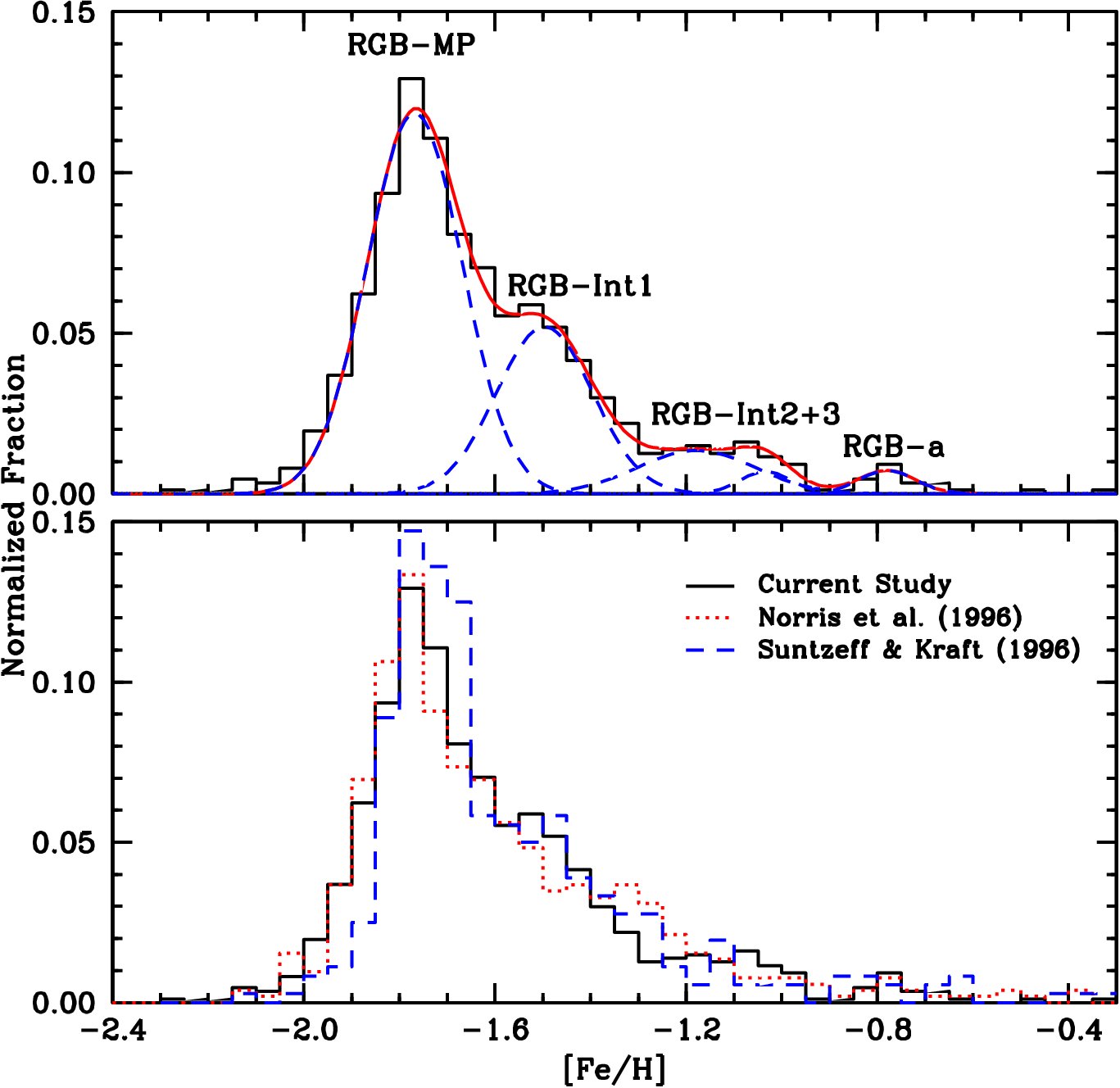}
\caption{The top panel shows the metallicity distribution function for our
complete sample, including data from Johnson et al. (2008; 2009).  For this
panel, the solid red line shows a least--squares fit to the distribution using 
five Gaussian profiles, and the dashed blue lines illustrate the individual
Gaussian component fits.  The bottom panel compares our distribution 
function (solid black line) to those of Norris et al. (1996; dotted red line)
and Suntzeff \& Kraft (1996; dashed blue line).}
\label{f8}
\end{figure}

In Figure \ref{f8}, we plot our derived metallicity distribution function and 
compare with the results of two other large spectroscopic surveys that spanned 
the upper RGB and SGB (Norris et al. 1996; Suntzeff \& Kraft 1996).  The 
general trend among all studies is that a dominant, metal--poor stellar 
population exists at [Fe/H]$\approx$--1.7 along with a higher metallicity tail 
terminating around [Fe/H]$\approx$--0.5.  Our data confirm this result, and 
also support previous observations that found multiple peaks in the metallicity
distribution function but a paucity of stars with [Fe/H]$<$--2.  The full range
of iron abundances in our sample extends from [Fe/H]=--2.26 to --0.32, and in 
Figure \ref{f8} we find five peaks in the metallicity distribution function 
located at [Fe/H]$\approx$--1.75, --1.50, --1.15, --1.05, and --0.75.  These 
peaks correspond to the RGB--MP, RGB--MInt, and RGB--a populations identified 
by Pancino et al. (2000) and Sollima et al. (2005), and also generally agree 
with Str{\"o}mgren photometry estimates (Hilker \& Richtler 2000; Hughes \&
Wallerstein 2000; Calamida et al. 2009).  It is difficult to accurately deblend
the two populations near [Fe/H]=--1.15 and --1.05 because the separation is 
comparable to the line--to--line dispersion of Fe abundance measurements in 
individual stars.  Instead, we will combine these two populations during 
further analyses.  Taking this into account, the (now four) stellar populations
make up roughly 61$\%$, 27$\%$, 10$\%$, and 2$\%$ of our sample, respectively.
For brevity, we will follow a similar naming scheme used by Sollima et al. 
(2005) when referring to the different metallicity populations:  RGB--MP 
([Fe/H]$\leq$--1.6), RGB--Int1 (--1.6$<$[Fe/H]$\leq$--1.3), RGB--Int2+3 
(--1.3$<$[Fe/H]$\leq$--0.9), and RGB--a ([Fe/H]$>$--0.9).

The most metal--poor stars ([Fe/H]$\leq$--2) make up about 2$\%$ (17/867) of 
the full sample and only about 3$\%$ (17/541) of the RGB--MP stellar 
population.  However, the RGB--a stars are slightly underrepresented 
because of our V magnitude cutoff.  To test for any selection effects, we 
rebinned the data to only include stars within $\sim$1 mag of each giant 
branch's RGB tip, which is the approximate magnitude range over which we 
sampled the RGB--a.  We did not find any significant differences 
in the relative population mix, and different magnitude cutoffs only raised the
RGB--a population fraction to $\sim$5$\%$.  These estimates are consistent
with those derived from number counts in photometric analyses (Pancino et al. 
2000; Sollima et al. 2005; Villanova et al. 2007; Calamida et al. 2009).  It 
should also be noted that AGB contamination may affect the number counts of 
each population differently.  Lee et al. (2005a) found that if the intermediate 
metallicity and most metal--rich stars are in fact He--rich then these stars 
will populate the ``extreme" horizontal branch.  Furthermore, D'Cruz et al. 
(2000) estimate that as much as 30$\%$ of the cluster's horizontal branch
population may reside on the ``extreme" horizontal branch, and it is likely 
that these stars evolve directly to white dwarfs rather than first ascending 
the AGB (e.g., Sweigart et al. 1974).  Since it is difficult to differentiate 
between RGB and AGB stars in $\omega$ Cen's color--magnitude diagram, it is 
possible that the number counts for the two most metal--poor populations 
contain a disproportionate number of AGB stars compared to the more metal--rich 
populations.  However, our estimated population fractions are consistent with 
those found along the main sequence and subgiant branch (e.g., Villanova et al.
2007) where AGB contamination is not an issue.

\begin{figure}
\plotone{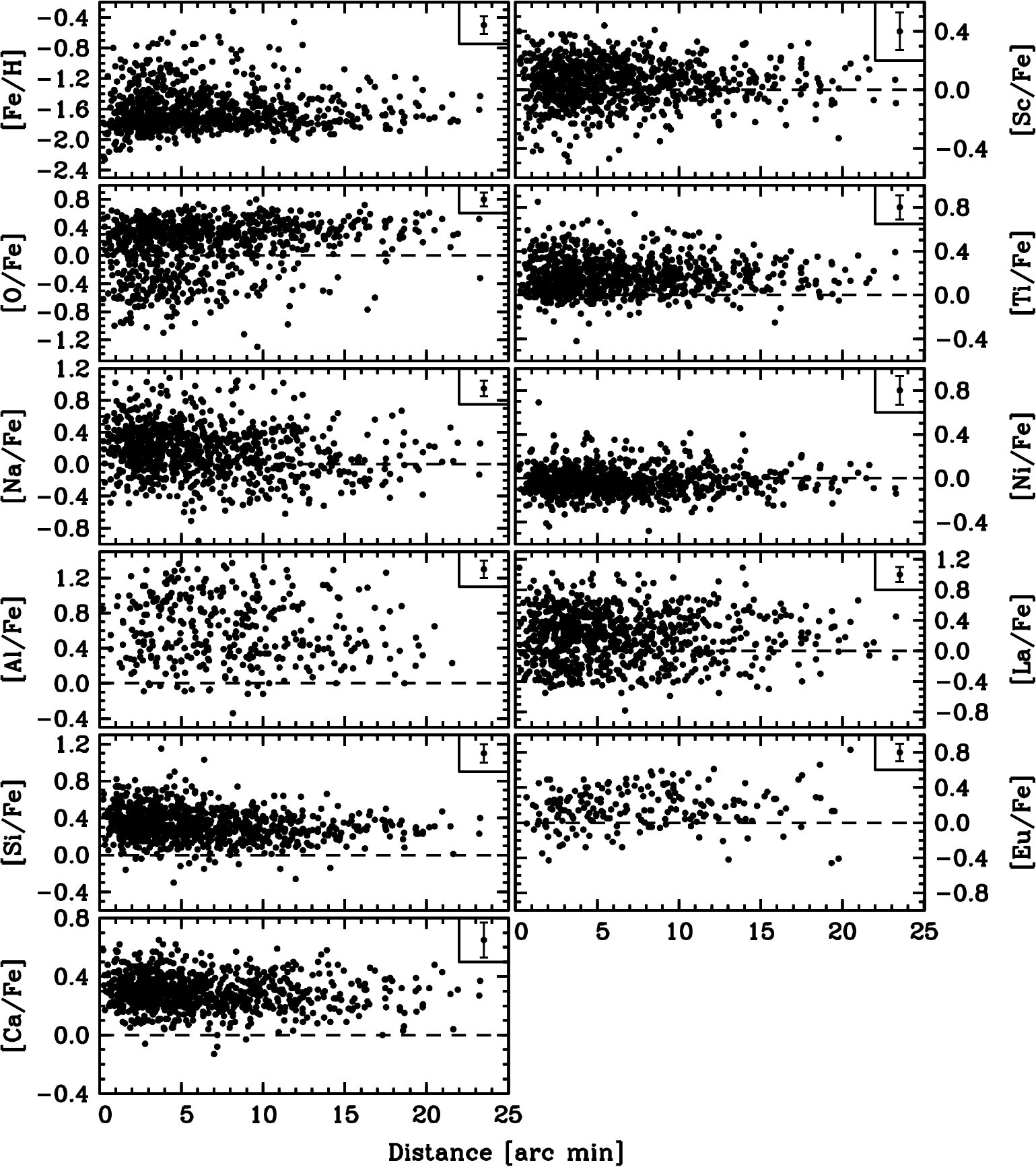}
\caption{Chemical abundance ratios of all elements are plotted as a function
of distance from the cluster center.  The defined cluster center is the same
as that used in Figure \ref{f3}.  The plotted abundances contain the results
from this study and Johnson et al. (2008; 2009).  In all panels, the black
dashed line indicates the solar--scaled abundance values.}
\label{f9}
\end{figure}

In addition to the existence of multiple, discrete stellar populations in 
$\omega$ Cen, there is some evidence that the metal--rich stars are more
centrally located than the more metal--poor populations (Norris et al. 1996;
Suntzeff \& Kraft 1996; Pancino et al. 2000; Hilker \& Richtler 2000; Pancino 
et al. 2003; Rey et al. 2004; Sollima et al. 2005; Johnson et al. 2008; Bellini
et al. 2009b; Johnson et al. 2009).  In Figure \ref{f9}, we plot our derived
abundances as a function of projected distance from the cluster center.  A 
two--sided Kolmogorov--Smirnov (K--S) test (Press et al. 1992) confirms that 
the metal--rich stars ([Fe/H]$>$--1.3) are more centrally located than the
metal--poor stars at the 96$\%$ level\footnote{We adopt the notion that the 
null hypothesis (i.e., that the two distributions are the same) can be rejected
if the p value is ``small" ($<$0.05).}.  Additionally, all of the stars with 
[Fe/H]$\geq$--0.9 are located within 13$\arcmin$ of the cluster center, with 
most of those residing inside 10$\arcmin$.  

Further inspection of Figure \ref{f9} reveals another interesting radial 
distribution trend; all stars with [Fe/H]$\leq$--2 are located within 
12$\arcmin$ of the cluster core, and 88$\%$ (15/17) of these stars reside 
inside 5$\arcmin$.  A two--sided K--S test comparing the radial distribution of
stars with --2.0$<$[Fe/H]$\leq$--1.60 versus those with [Fe/H]$\leq$--2 
indicates that the two distributions are drawn from different parent 
populations at the 99$\%$ level.  Additionally, the star--to--star metallicity 
dispersion decreases with increasing distance from the cluster center, but this
is mostly driven by the metallicity gradient and paucity of stars with 
[Fe/H]$\geq$--1.3 outside $\sim$15$\arcmin$ from the cluster center.  If one 
only considers stars with [Fe/H]$<$--1.3, the standard deviation in [Fe/H] 
between 0--10$\arcmin$ and 10--20$\arcmin$ differs by less than 0.02 dex.  This
indicates that the two most metal--poor stellar populations are well mixed 
inside the cluster.

\subsection{Oxygen}

The chemical evolution of oxygen in $\omega$ Cen has previously been analyzed
via high resolution spectroscopy in several studies containing sample sizes 
ranging from $\sim$5--40 RGB stars (e.g., Cohen 1981; Paltoglou \& Norris 
1989; Brown \& Wallerstein 1993; Norris \& Da Costa 1995; Zucker et al. 1996; 
Smith et al. 2000), and more recently in a sample of $\sim$200 RGB stars 
(Marino et al. 2010).  The main results from past studies indicate that: 
(1) $\omega$ Cen giants exhibit large star--to--star dispersions in [O/Fe] 
abundance, (2) many of the intermediate metallicity stars have [O/Fe]$<$0, 
(3) the majority of metal--poor stars are O--rich with [O/Fe]$\sim$+0.3, and 
(4) oxygen is anticorrelated with both sodium and aluminum.  The results 
presented here add 848 new [O/Fe] abundance measurements.

Figure \ref{f9} includes a plot of our derived [O/Fe] abundances as a function 
of projected distance from the cluster center.  Compared to the other elements 
in Figure \ref{f9}, oxygen appears to exhibit a unique radial distribution.  
Stars with [O/Fe]$\leq$0, and especially those with [O/Fe]$<$--0.4, are more 
centrally concentrated than the bulk of stars with [O/Fe]$>$0.  In our sample, 
62$\%$ (145/233) of stars with [O/Fe]$\leq$0 are located inside 5$\arcmin$ from
the core and 91$\%$ (213/233) are inside 10$\arcmin$.  This is compared to just
42$\%$ (261/615) and 77$\%$ (472/615) for the stars with [O/Fe]$>$0, 
respectively.  A two--sided K--S test reveals that the O--poor stars exhibit a 
different spatial distribution than the O--rich stars at the 99$\%$ level.  
This result may have important implications regarding the origin of the blue
main--sequence, and will be discussed further in $\S$5.2.2.

Figure \ref{f10} shows the chemical evolution of [O/Fe] plotted as a function
of [Fe/H].  This plot reveals that the O--poor stars, in addition to being 
preferentially located near the cluster core, are also well separated from 
the O--rich stars over a large metallicity range.  Figure \ref{f11} shows the 
[O/Fe] data binned in 0.10 dex increments and separated into the population 
subclasses defined in $\S$4.1.  The resultant histograms support the existence 
of two subpopulations, one O--rich ([O/Fe]$>$0) and the other O--poor 
([O/Fe]$<$0), residing inside the RGB--Int1 and RGB--Int2+3 populations.  
Interestingly, neither the RGB--MP nor the RGB--a populations appear to exhibit
this bimodal behavior.  Instead, the RGB--MP stars are predominantly O--rich 
with a median [O/Fe]=+0.32, and the RGB--a stars are moderately O--poor with a 
median [O/Fe]=--0.15.  In the RGB--MP population, the percentages of O--poor 
and O--rich stars are 13$\%$ (71/535) and 87$\%$ (464/535), respectively.  The 
two intermediate metallicity populations show quite different distributions, 
with the percentages being 46$\%$ (100/218) to 54$\%$ (118/218) in the 
RGB--Int1 group and 64$\%$ (47/74) to 36$\%$ (27/74) in the RGB--Int2+3 group.
The relative distribution in the RGB--a stars is 71$\%$ (15/21) O--poor to 
29$\%$ (6/21) O--rich, respectively.

Examining the bulk properties of the [O/Fe] abundances reveals that, in
all but the most metal--poor and metal--rich stars, a significant 
star--to--star dispersion is present with $\Delta$[O/Fe]$>$2 over a large 
metallicity range.  The full range of [O/Fe] abundances found in our sample 
spans from [O/Fe]=--1.30 to +0.80.  The stars with [Fe/H]$\leq$--2 are 
overwhelmingly O--rich with 94$\%$ (15/16) having [O/Fe]$>$0 and 
$\langle$[O/Fe]$\rangle$=+0.38, and the single O--poor star is only moderately 
depleted at [O/Fe]=--0.13.  The [O/Fe] abundance ``ceiling" decreases for 
stars with [Fe/H]$\ga$--1.3, dropping from [O/Fe]$\approx$+0.6 at [Fe/H]=--1.3
to [O/Fe]$\approx$+0.0 at [Fe/H]=--0.3.  Additionally, the super O--poor stars 
([O/Fe]$\leq$--0.4) are only found in the range --1.9$\la$[Fe/H]$\la$--1.0.  
When considering all stars in our sample, the relative percentages of O--rich 
([O/Fe]$>$0), O--poor (--0.4$<$[O/Fe]$\leq$0.0), and super O--poor 
([O/Fe]$\leq$--0.4) stars are 73$\%$ (615/848), 14$\%$ (118/848), and 13$\%$ 
(115/848), respectively.  We also find that in stars with [Fe/H]$\la$--1, 
[O/Fe] is anticorrelated with both [Na/Fe] and [Al/Fe].  The implications of 
these anticorrelations, along with the possible significance of the super 
O--poor stars, will be discussed further in $\S$5.

\subsection{Sodium}

Previous sodium abundance measurements support the idea that $\omega$ Cen 
experienced a significantly different chemical evolutionary path than any other
stellar system (Cohen 1981; Paltoglou \& Norris 1989; Brown \& Wallerstein 
1993; Norris \& Da Costa 1995; Zucker et al. 1996; Smith et al. 2000; Johnson 
et al. 2009; Villanova et al. 2009; Marino et al. 2010).  The results from 
these studies have shown that:  (1) [Na/Fe] appears to increase as a function 
of increasing [Fe/H], (2) $\Delta$[Na/Fe]$>$1 for most values of [Fe/H] in the 
cluster, (3) no strong [Na/Fe] abundance gradient is observed, and (4) [Na/Fe] 
is correlated with [Al/Fe] and anticorrelated with [O/Fe].  Our new results, 
combined with those from Johnson et al. (2009), give [Na/Fe] abundances for 
848 cluster giants.  Although it is likely that our derived sodium abundances 
suffer from moderate non--LTE (NLTE) effects, abundances derived from the 
6154/6160 \AA\ doublet used here are expected to have NLTE offsets $<$0.2 dex 
for giants in our metallicity regime (e.g., Gratton et al. 1999; Mashonkina et 
al. 2000; Gehren et al. 2004).  Since no standard NLTE corrections are 
available in the literature, the abundances reported in Table 5 and shown in 
the figures do not include NLTE corrections.

\begin{figure}
\plotone{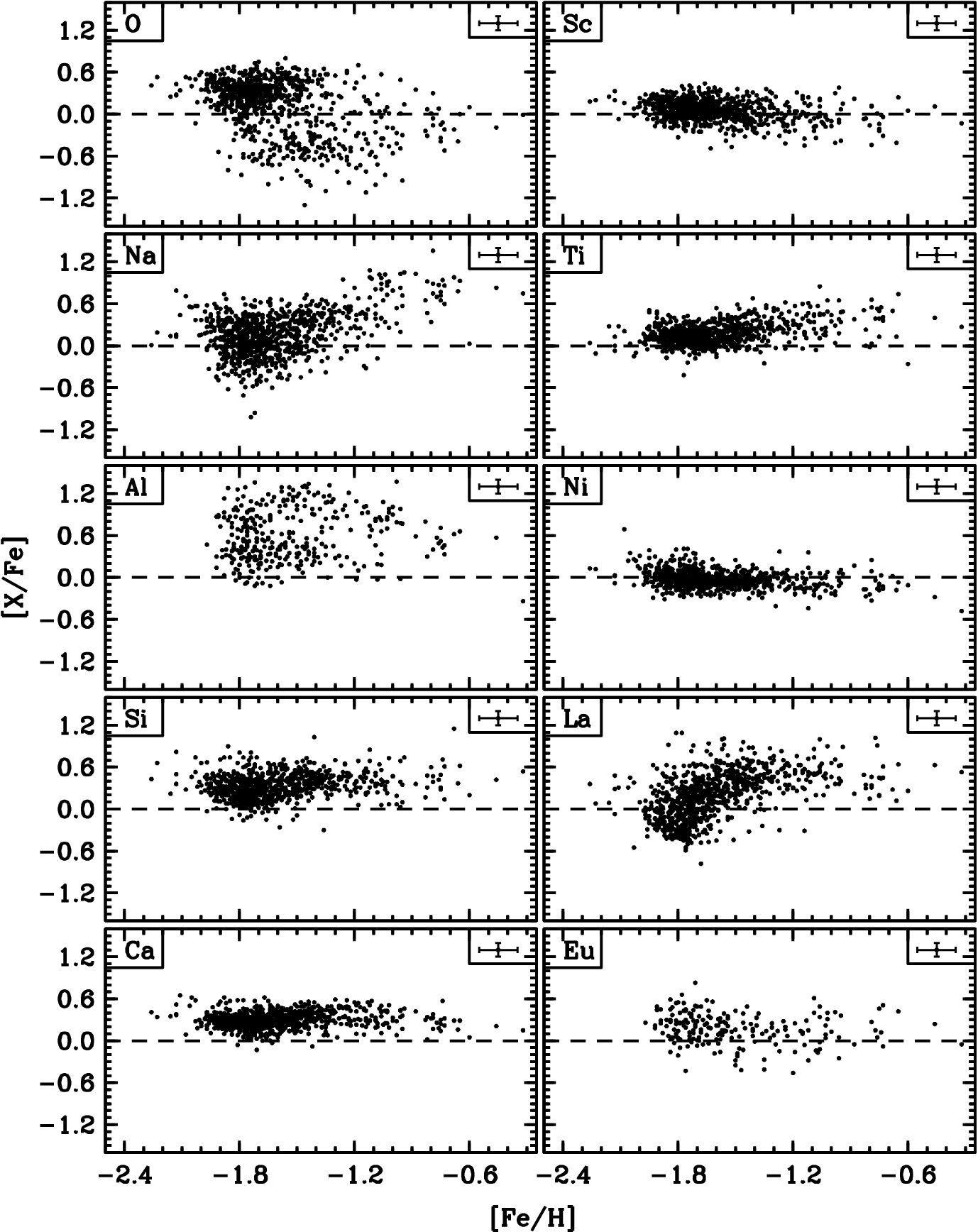}
\caption{[X/Fe] abundances are plotted as a function of [Fe/H] for all elements
analyzed in this study, including non--repeat stars from Johnson et al. (2008; 
2009).  The ordinate axis in all panels spans the same range, and the dashed
line indicates the solar--scaled abundance values.}
\label{f10}
\end{figure}

\begin{figure}
\plotone{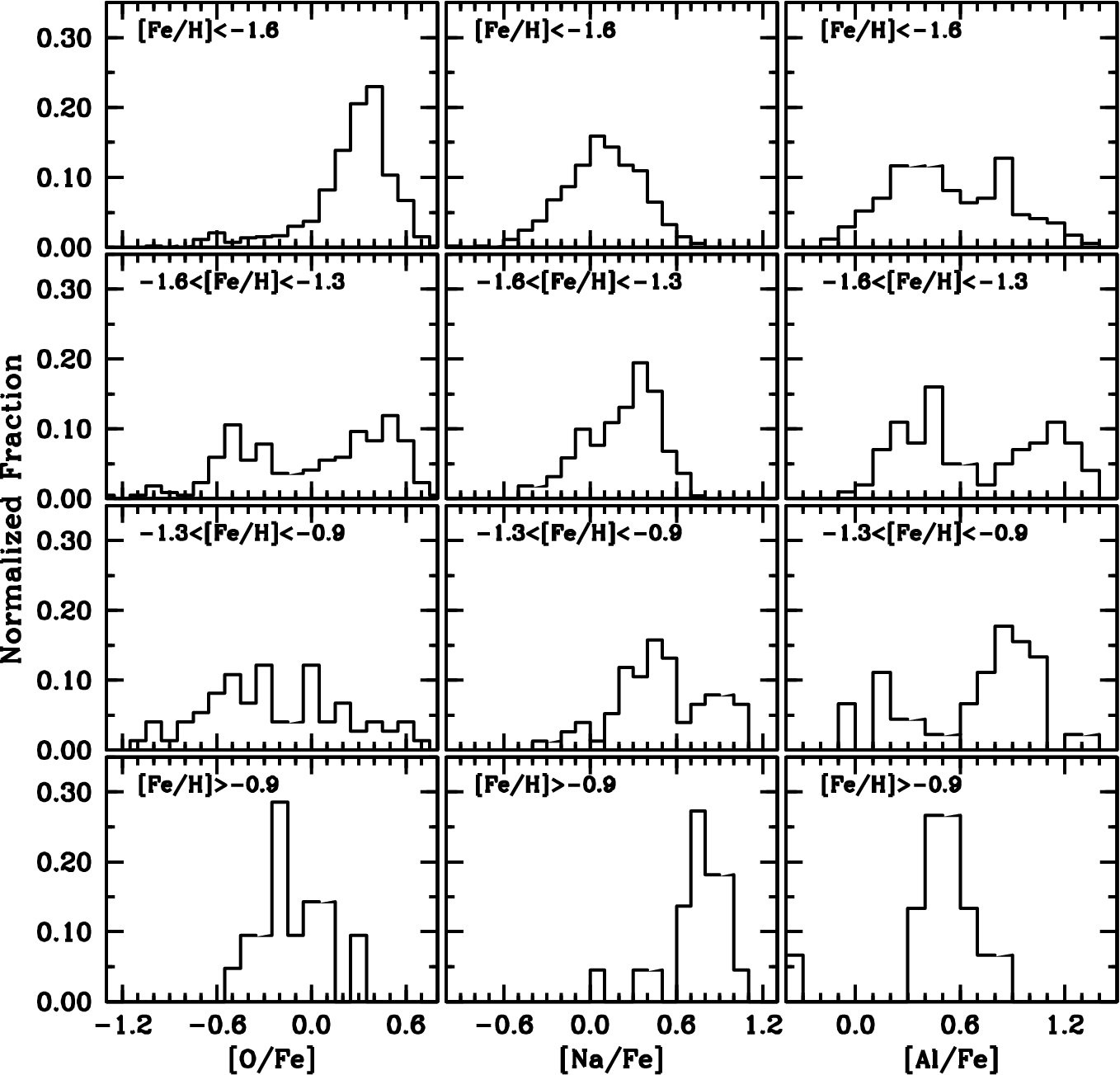}
\caption{Histograms showing the abundance distributions of [O/Fe], [Na/Fe], and
[Al/Fe].  Each histogram is binned in 0.10 dex increments, and the panels are
broken down by the metallicity subclasses described in $\S$4.1.}
\label{f11}
\end{figure}

Inspection of Figure \ref{f11} indicates that [Na/Fe] exhibits a similar 
bimodal abundance pattern as shown by [O/Fe].  That is, the most metal--poor 
and metal--rich stellar populations show a single primary peak in the [Na/Fe] 
distribution function, and the two intermediate metallicity populations may be 
best described as having two peaks in the [Na/Fe] distribution function.  
However, unlike the case with oxygen, we do not find an obvious centrally 
concentrated population that correlates with any [Na/Fe] abundance range.
We do find that the most Na--rich stars in our sample ([Na/Fe]$\geq$+0.6) are
all found inside 13$\arcmin$ from the cluster center, but this observation is
unlikely to be significant because (1) the two most metal--rich stellar 
populations contain 69$\%$ (44/64) of the most Na--rich stars (see 
Figures \ref{f10}--\ref{f11}) and (2) these stellar populations are already 
known to be centrally concentrated.  This is in contrast to the O--poor radial 
trend that is found in stars with --1.6$<$[Fe/H]$\leq$--1.3, which do not 
exhibit a preferred radial location.  However, Figure \ref{f9} shows that a 
weak, declining [Na/Fe] gradient may exist such that the median [Na/Fe] values 
for 0--5$\arcmin$, 5--10$\arcmin$, 10--15$\arcmin$, and 15--20$\arcmin$ are
+0.22, +0.14, +0.08, and --0.03, respectively.

Figure \ref{f10} highlights the chemical evolution of [Na/Fe] as a function of 
[Fe/H].  We find that a large star--to--star dispersion is present at all
metallicities, and that the full range extends from [Na/Fe]=--1.02 to +1.36.
Since we have many stars of the same temperature, surface gravity, and 
metallicity, the line strength differences confirm that the observed abundance 
spread is a real effect and not due to possible underlying NLTE effects.  In
addition to displaying a significant star--to--star dispersion, the sodium 
abundances also exhibit a strong metallicity dependence such that the median 
[Na/Fe] value increases with increasing [Fe/H].  The median [Na/Fe]
value rises from +0.08 in the RGB--MP population to +0.78 in the RGB--a 
population.  As mentioned above, the two most metal--rich populations contain 
the most Na--rich stars in the cluster.  Despite the complex nature of 
sodium's evolution in $\omega$ Cen, the O--Na anticorrelation and Na--Al
correlation are present in all but the most metal--rich stars.

\subsection{Aluminum}

Except for iron and calcium, aluminum has been the most highly studied element
in $\omega$ Cen.  Previous high resolution spectroscopic work has targeted
more than 200 RGB stars (Cohen 1981; Brown \& Wallerstein 1993; Norris \& 
Da Costa 1995; Zucker et al. 1996; Smith et al. 2000; Johnson et al. 2008; 
Johnson et al. 2009) and shown: (1) $\Delta$[Al/Fe]$>$0.5 at all metallicities 
and exceeds more than a factor of ten in the most metal--poor stars, (2) the 
range of observed [Al/Fe] abundances decreases at [Fe/H]$>$--1.3, (3) there 
is a paucity of stars with [Al/Fe]$<$+0.3 at intermediate and high 
metallicities, and (4) a Na--Al correlation and O--Al anticorrelation are 
present in most, if not all, cluster stars.  In this paper we present 133 new 
[Al/Fe] abundance measurements, and when combined with the data from Johnson 
et al. (2008; 2009) provide [Al/Fe] values for 332 $\omega$ Cen giants.  As 
with sodium (see $\S$4.3), we have not applied any NLTE corrections to our 
derived aluminum abundances.  However, all aluminum abundances determined here 
utilized the non--resonance 6696/6698 \AA\ lines, which are not expected to 
have large NLTE offsets in the temperature, gravity, and metallicity range of 
stars in our sample (e.g., Gehren et al. 2004; Andrievsky et al. 2008).

Unlike oxygen, and to a lesser extent sodium, aluminum does not show any 
obvious correlation between [Al/Fe] abundance and radial location.  However,
aluminum does show the same bimodal abundance distribution for the RGB--Int1
and RGB--Int2+3 populations (see Figure \ref{f11}).  By dividing the samples at
[Al/Fe]=+0.6, we find that the percentage of ``Al--enhanced" 
([Al/Fe]$\geq$+0.6) stars in the RGB--Int1 population is 50$\%$ (50/100) 
compared to 50$\%$ (50/100) as well for the ``Al--normal" stars 
([Al/Fe]$<$+0.6).  Similarly, the RGB--Int2+3 stars are distributed as 
69$\%$ (31/45) enhanced and 31$\%$ (14/45) normal, respectively.  
Interestingly, the [Al/Fe] distribution also shows complex substructure in the 
RGB--MP population, which is not observed in the [O/Fe] and [Na/Fe] data.  In 
this population, 41$\%$ (70/172) of the stars are Al--enhanced and 59$\%$ 
(102/172) are Al--normal.  Furthermore, this is the only $\omega$ Cen 
population that contains a significant number of stars over the full [Al/Fe] 
range.  For the RGB--a population, a single peak is observed at [Al/Fe]=+0.5 
in the [Al/Fe] distribution function.

The full range of [Al/Fe] abundances observed here spans from --0.34 to +1.37,
but only 4$\%$ (12/332) of the stars have [Al/Fe]$<$0.  Similarly, we find that
$\Delta$[Al/Fe]$\sim$1.5 dex for [Fe/H]$<$--1.3.  However, the star--to--star
dispersion decreases noticeably at higher metallicities.  Inspection of 
Figure \ref{f10} shows that [Al/Fe] exhibits an interesting trend as a function
of [Fe/H].  The maximum value reached for stars with [Fe/H]$\la$--1.3 remains
steady near [Al/Fe]$\approx$+1.3, but above [Fe/H]$\sim$--1.3 the maximum
abundance decreases to only [Al/Fe]$\approx$+0.6 in the RGB--a stars.  
Furthermore, the number of stars with [Al/Fe]$<$+0.3 strongly decreases at 
[Fe/H]$>$--1.3.  In the RGB--MP and RGB-Int1 populations, stars with 
[Al/Fe]$<$+0.3 constitute 25$\%$ (69/272) of the distribution, but this 
decreases to only 7$\%$ (1/15) of the RGB--a population.  

\subsection{Silicon}

Previous analyses (Cohen 1981; Paltoglou \& Norris 1989; Brown \& Wallerstein 
1993; Norris \& Da Costa 1995; Smith et al. 2000; Pancino et al. 2002; 
Villanova et al. 2009) have used the heavy $\alpha$ element (Si, Ca, and Ti) 
abundances to assess the dominance of Type II versus Type Ia supernovae in 
$\omega$ Cen and other clusters.  In terms of silicon abundances, it has 
been shown that: (1) silicon is enhanced with [Si/Fe]$>$+0.3 in nearly all 
cluster stars, (2) the star--to--star dispersion in [Si/Fe] is significantly 
smaller than for the lighter $\alpha$ and odd--Z elements, and (3) the most 
metal--rich stars may have appreciably lower [Si/Fe] abundances compared to 
the more metal--poor populations.  From this study, we add 821 new [Si/Fe] 
measurements over $\omega$ Cen's full metallicity range.

While we find that the lighter $\alpha$ element oxygen shows a distinctly
unique distribution versus distance from the cluster center, [Si/Fe]
does not show the same trend.  Figure \ref{f9} suggests that a weak [Si/Fe]
gradient may be present such that the stars inside 5$\arcmin$ have a higher
average silicon abundance than those outside 5$\arcmin$.  We find that stars
inside 5$\arcmin$ have $\langle$[Si/Fe]$\rangle$=+0.37, which is noticeably
higher than the $\langle$[Si/Fe]$\rangle$=+0.29 for those at r$>$5$\arcmin$.  
This result does not change even if we limit examination to stars only between 
0--5$\arcmin$ and 5--10$\arcmin$.  Except near the cluster core, the average
[Si/Fe]$\approx$+0.3 at all radii.  It should be noted that Villanova et al.
(2009) find $\langle$[Si/Fe]$\rangle$=+0.5 in the outer 20--30$\arcmin$ of
$\omega$ Cen, which is larger by about 0.2 dex than we find in the same region.
However, we do not presently have sufficient data to assess whether the average
[Si/Fe] ratio increases at larger radii or if this merely reflects a systematic 
offset.

The full range of [Si/Fe] abundances in our data span from --0.30 to +1.15, but
the average over all stars is [Si/Fe]=+0.33 ($\sigma$=0.17).  While we do 
find a few Si--poor stars ([Si/Fe]$<$0), these stars comprise only 
2$\%$ (16/821) of the total sample.  Similarly, the very Si--rich stars 
([Si/Fe]$>$+0.6) only represent 6$\%$ (52/821) of the total sample.  
Figure \ref{f10} reveals that [Si/Fe] may have a more complex morphology as a 
function of [Fe/H] than previously thought.  The average [Si/Fe] ratio 
decreases from $\langle$[Si/Fe]$\rangle$=+0.46 ($\sigma$=0.19) in stars with
[Fe/H]$\leq$--2 to $\langle$[Si/Fe]$\rangle$=+0.29 ($\sigma$=0.16) in the 
stars that comprise the majority of the RGB--MP population
(--2.0$<$[Fe/H]$\leq$--1.6).  In the subsequent populations, the average
[Si/Fe] abundance monotonically increases with [Fe/H] to 
$\langle$[Si/Fe]$\rangle$=+0.45 ($\sigma$=0.23) in the RGB--a population.  This
is in agreement with Norris \& Da Costa (1995) and Smith et al. (2000), but 
contrasts with the claims by Pancino et al. (2002) and Origlia et al. (2003) 
that stars with [Fe/H]$>$--1 have lower [$\alpha$/Fe] abundances (see $\S$3.2.3
for a brief discussion).

\subsection{Calcium}

In addition to iron, calcium abundances have been analyzed in great detail for
$\omega$ Cen stars.  Previous analyses have used calcium as a proxy metallicity
indicator (Freeman \& Rodgers 1975; Cohen 1981; Norris et al. 1996; 
Suntzeff \& Kraft 1996; Rey et al. 2004; Sollima et al. 2005; Stanford et al. 
2006; Lee et al. 2009) and as an $\alpha$ element tracer (Paltoglou \& Norris 
1989; Norris \& Da Costa 1995; Smith et al. 2000; Pancino et al. 2002; Kayser 
et al. 2006; Villanova et al. 2007; Johnson et al. 2009; Villanova et al. 
2009).  These studies have shown: (1) there is a large spread of at least 1 
dex in [Ca/H] with multiple peaks in the distribution function (i.e., confirms 
the different populations found when using [Fe/H] as a metallicity tracer), 
(2) nearly all stars have enhanced [Ca/Fe]$\approx$+0.3 at all metallicities, 
(3) the star--to--star dispersion is significantly smaller than for the lighter
elements, and (4) there may be a downturn in [Ca/Fe] at [Fe/H]$>$--1.  
Combining our new data with that of Johnson et al. (2009), we add 857 [Ca/Fe] 
abundance measurements.

\begin{figure}
\plotone{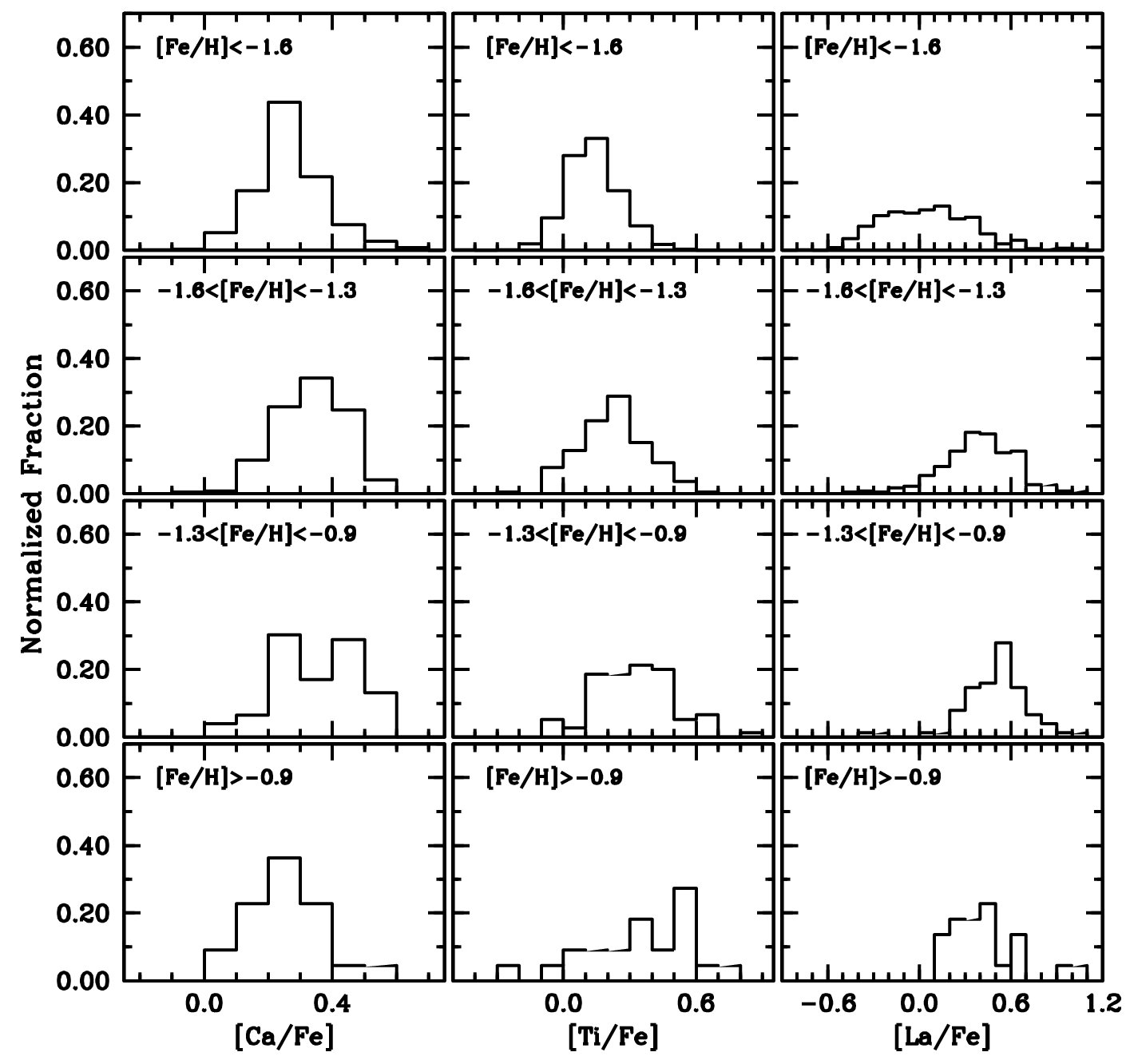}
\caption{Histograms showing the abundance distributions of [Ca/Fe], [Ti/Fe], 
and [La/Fe].  Each histogram is binned in 0.10 dex increments, and the panels 
are broken down by the metallicity subclasses described in $\S$4.1.}
\label{f12}
\end{figure}

Unlike silicon, which provides some evidence for a weak radial abundance 
gradient, [Ca/Fe] does not vary ostensibly between the inner and outer regions
of the cluster.  When considering all stars in our sample, the majority are
Ca--rich with $\langle$[Ca/Fe]$\rangle$=+0.29 ($\sigma$=0.12).  However, the 
full range of observed [Ca/Fe] abundances is smaller than for [Si/Fe], with 
[Ca/Fe] varying between --0.13 and +0.65.  Figure \ref{f10} shows that [Ca/Fe] 
displays a similar morphology to [Si/Fe] when plotted as a function of [Fe/H].
That is, stars with [Fe/H]$\leq$--2 tend to be more Ca--rich with 
$\langle$[Ca/Fe]$\rangle$=+0.37 ($\sigma$=0.16) compared to the majority of 
stars in the RGB--MP population with $\langle$[Ca/Fe]$\rangle$=+0.26
($\sigma$=0.11).  Similarly, the average [Ca/Fe] abundance rises for the 
RGB--Int1 and RGB--Int2+3 populations to $\langle$[Ca/Fe]$\rangle$=+0.34
($\sigma$=0.11; see also Figure \ref{f12}).  However, unlike the case for 
[Si/Fe], the average [Ca/Fe] abundance decreases for [Fe/H]$\ga$--1, and the 
RGB--a stars have $\langle$[Ca/Fe]$\rangle$=+0.26 ($\sigma$=0.12).

Further inspection of Figure \ref{f10} reveals that the distribution of [Ca/Fe]
among the RGB--Int2+3 stars may be bimodal.  Figure \ref{f12} also suggests 
that the RGB--Int2+3 stars may exhibit a bimodal distribution, and shows that
the other populations appear to exhibit a mostly unimodal [Ca/Fe] distribution.
Interestingly, the two RGB--Int2+3 subsets occur in nearly equal proportions
with the stars peaked near [Ca/Fe]=+0.45 constituting 47$\%$ (36/76) of the
subsample and the stars peaked near [Ca/Fe]=+0.25 making up 53$\%$ (40/76) of
the subsample.  However, a two--sided K--S test does not rule out that the 
[Ca/Fe] distributions for the RGB--Int1 and RGB--Int2+3 are different at more
than the 95$\%$ level.  While we caution the reader that the apparent 
bimodality may be a product of small number statistics, it would be interesting
to investigate this possible trend further with additional calcium abundance 
indicators (e.g., HK index).

\subsection{Scandium}

Scandium is typically used as a tracer of Fe--peak element production in 
stellar populations, and Galactic halo and globular cluster stars with 
[Fe/H]$>$--2.5 tend to exhibit solar--scaled [Sc/Fe] abundances.  Although
scandium has been analyzed in only a handful of studies for $\omega$ Cen 
stars (Cohen 1981; Paltoglou \& Norris 1989; Norris \& Da Costa 1995; Zucker 
et al. 1996; Smith et al. 2000; Johnson et al. 2009), the results typically 
show that: (1) the observed star--to--star scatter in [Sc/Fe] is significantly 
smaller than for lighter elements and (2) $\langle$[Sc/Fe]$\rangle$$\approx$0 
at all metallicities.  Combined with the results from Johnson et al. (2009), 
we are able to add 821 [Sc/Fe] abundance measurements.

As can be seen in Figure \ref{f9}, we do not find any evidence for a radial
[Sc/Fe] abundance gradient.  Similarly, Figure \ref{f10} indicates that the 
[Sc/Fe] ratio is approximately constant over the full metallicity regime.  
However, a weak metallicity dependence may be present such that the average
[Sc/Fe] abundance decreases from $\langle$[Sc/Fe]$\rangle$=+0.08 
($\sigma$=0.13) in the RGB--MP population to $\langle$[Sc/Fe]$\rangle$=--0.07 
($\sigma$=0.19) in the RGB--a stars.  The full range of observed [Sc/Fe] 
abundances spans from --0.49 to +0.44, but most stars exhibit a solar--scaled 
[Sc/Fe] ratio.  When considering the entire sample, we find 
$\langle$[Sc/Fe]$\rangle$=+0.05 ($\sigma$=0.15).

\subsection{Titanium}

Titanium is generally considered either the heaviest $\alpha$ element or one
of the lightest Fe--peak elements.  Previous titanium 
abundance measurements for $\omega$ Cen stars (Cohen 1981; Paltoglou \& Norris 
1989; Brown \& Wallerstein 1993; Norris \& Da Costa 1995; Smith et al. 2000; 
Villanova et al. 2007; Johnson et al. 2009; Villanova et al. 2009) have shown: 
(1) the star--to--star dispersion in [Ti/Fe] is comparable to that found in 
[Si/Fe] and [Ca/Fe], (2) the titanium abundance is generally enhanced at 
[Ti/Fe]$\sim$+0.3, and (3) there may be evidence for an increase in [Ti/Fe] 
with increasing [Fe/H].  Our new results, combined with Johnson et al. (2009),
provide 826 [Ti/Fe] measurements.

Inspection of Figure \ref{f9} confirms that we do not find any correlation
between our determined [Ti/Fe] abundance and a star's radial location.  In a 
similar fashion to the behavior of silicon and calcium, Figure \ref{f10} shows 
that titanium also exhibits a metallicity dependent morphology.  The average
[Ti/Fe] ratio is roughly constant across the RGB--MP population's full 
metallicity range ([Fe/H]$\leq$--1.6) at $\langle$[Ti/Fe]$\rangle$=+0.13
($\sigma$=0.12), which is $\sim$0.2 dex lower than the [Si/Fe] and 
[Ca/Fe] ratios in those same stars.  However, the average [Ti/Fe] abundance 
rises monotonically to $\langle$[Ti/Fe]$\rangle$=+0.34 ($\sigma$=0.25) in the 
RGB--a population (see also Figure \ref{f12}).  The full range of abundances in
our sample spans from [Ti/Fe]=--0.42 to +0.85, but most stars are at least 
moderately Ti--enhanced with $\langle$[Ti/Fe]$\rangle$=+0.18 ($\sigma$=0.16).

\subsection{Nickel}

Aside from iron, nickel is the only other ``true" Fe--peak element analyzed 
here.  The chemical evolution of nickel in a stellar population often tracks
very closely to iron, and $\omega$ Cen appears to follow that trend (Cohen 
1981; Paltoglou\& Norris 1989; Norris \& Da Costa 1995; Smith et al. 2000; 
Johnson et al. 2009; Villanova et al. 2009).  Previous studies agree that: 
(1) the derived [Ni/Fe] abundances show the smallest intrinsic dispersion of 
any element and (2) the average [Ni/Fe] abundance is nearly solar at all 
metallicities and locations in the cluster.  We add to these results 806 new 
[Ni/Fe] abundance determinations.

Figure \ref{f9} shows that, like the other transition metals, [Ni/Fe] 
abundances do not exhibit any signs of a radial gradient.  Similarly, 
Figure \ref{f10} indicates that the distribution of [Ni/Fe] is essentially 
constant as a function of [Fe/H] with a small intrinsic scatter, but there
may be a slight decrease in [Ni/Fe] at [Fe/H]$\ga$--1.3.  The full 
spread of [Ni/Fe] values found in our sample ranges from --0.48 to +0.69, and 
the cluster as a whole gives $\langle$[Ni/Fe]$\rangle$=--0.03 ($\sigma$=0.12).

\subsection{Lanthanum}

The heavy element lanthanum is often used as a tracer of the slow 
neutron--capture process (s--process), and its evolution has proved to be 
particularly interesting in $\omega$ Cen.  Previous analyses (Cohen 1981; 
Paltoglou \& Norris 1989; Norris \& Da Costa 1995; Smith et al. 2000; Johnson 
et al. 2009; Marino et al. 2010) have examined the [La/Fe] ratios in $\sim$100 
RGB stars and found:  (1) the most metal--poor stars tend to have [La/Fe] 
abundances consistent with those found in monometallic globular clusters, 
(2) a large increase in [La/Fe] is seen between [Fe/H]$\approx$--1.7 and --1.4,
(3) the intermediate metallicity stars are almost exclusively La--rich, and 
(4) the average [La/Fe] ratio remains super--solar in the most metal--rich 
stars.  When combined with the data from Johnson et al. 2009, we add to these 
past results 810 new [La/Fe] abundances.

As can be seen in Figure \ref{f9}, we find no evidence supporting the existence
of a radial [La/Fe] gradient, and the star--to--star dispersion remains 
approximately constant across all radii sampled here.  On the other hand, 
our data shown in Figure \ref{f10} support previous claims that [La/Fe] 
abundances exhibit an unusual morphology when plotted as a function of [Fe/H].
There is a strong increase in [La/Fe] for stars with [Fe/H]$\ga$--1.7, and 
a large intrinsic scatter of $\Delta$[La/Fe]$\ge$1 is present an nearly all
metallicities.  Furthermore, the average [La/Fe] abundance monotonically 
increases from +0.05 in the RGB--MP population to +0.49 in the RGB--Int2+3
population (see also Figure \ref{f12}).  However, the RGB--a stars have 
$\langle$[La/Fe]$\rangle$=+0.43, which suggests either a leveling off or 
slight decline in [La/Fe] at [Fe/H]$\ga$--1.  

The full range of [La/Fe] abundances observed here spans from --0.78 to +1.17,
and it is worth noting that the proper accounting of hyperfine structure in 
the [La/Fe] derivations has decreased the maximum abundance values found 
in Johnson et al. (2009) from [La/Fe]$\sim$+2 to [La/Fe]$\sim$+1.2.  These 
lower abundance ratios suggest that a large fraction of binary transfer systems
may not be required to account for the significant lanthanum 
enhancements.  However, we still find that only 29$\%$ (232/810) of the stars 
in our sample have [La/Fe]$<$0, and 94$\%$ (217/232) of those stars reside
in the RGB--MP population.  Interestingly, the stars with [Fe/H]$\leq$--2 tend 
to exhibit rather high [La/Fe] abundances.  These stars have 
$\langle$[La/Fe]$\rangle$=+0.19, which is distinctly larger than the 
$\langle$[La/Fe]$\rangle$=+0.05 found for the full sample of RGB--MP stars.
Unfortunately, a two--sided K--S test indicates that the data are insufficient 
to reject the null hypothesis with more than 94$\%$ confidence.

\subsection{Europium}

In an analogous fashion to lanthanum, the heavy element europium is often used 
as an indicator of the rapid neutron--capture process (r--process).  However, 
europium has been analyzed in far fewer stars than lanthanum (Norris \& Da 
Costa 1995; Zucker et al. 1996; Smith et al. 2000; Johnson et al. 2009).  
The primary results from these studies are:  (1) [Eu/Fe] tends to be 
somewhat underabundant relative to monometallic globular clusters of similar 
metallicity, (2) a significant intrinsic scatter is observed, but it is smaller
than that found in [La/Fe], and (3) [Eu/Fe] remains relatively constant as a
function of [Fe/H].  Combined with the data from Johnson et al. (2009), we 
provide [Eu/Fe] abundances for 194 stars.

Given the significantly smaller sample for europium compared to the other 
elements analyzed here, it is difficult to assess whether any true radial
trends exist.  Figure \ref{f9} provides weak evidence that the average [Eu/Fe]
abundance may increase away from the cluster center.  The available data 
support this by showing an increase from $\langle$[Eu/Fe]$\rangle$=+0.12 
for stars between 0--5$\arcmin$ to $\langle$[Eu/Fe]$\rangle$=+0.23 for stars 
between 5--10$\arcmin$ from the core.  Unfortunately, the sample size becomes
too small outside $\sim$10$\arcmin$ to conclude whether this trend continues.

Figure \ref{f10} reveals that [Eu/Fe] exhibits a significantly different 
behavior than [La/Fe] when plotted as a function of [Fe/H].  The full range is 
somewhat smaller with [Eu/Fe] spanning --0.46 to +0.83, and the average
[Eu/Fe] abundance appears to \emph{decrease} in the metallicity range where
[La/Fe] shows its greatest \emph{increase}.  While the intermediate metallicity
populations generally contain the lowest [Eu/Fe] abundances, the average 
[Eu/Fe] ratios differ by only $\sim$+0.1 dex among the different stellar 
populations.

\section{DISCUSSION}

The results of our analyses support previous observations that $\omega$ 
Cen hosts multiple stellar populations exhibiting a complex history of chemical
enrichment.  To briefly summarize, we have confirmed five peaks in the 
metallicity distribution function located at [Fe/H]$\approx$--1.75, --1.50, 
--1.15, --1.05, and --0.75; however, for discussion purposes the [Fe/H]=--1.15 
and --1.05 populations are treated as a single group.  The RGB--MP, RGB--Int1, 
RGB--Int2+3, and RGB--a populations constitute 61$\%$, 27$\%$, 10$\%$, and 
2$\%$ of stars in our sample, respectively.  We also find large intrinsic 
abundance dispersions for O, Na, and Al, and, except for perhaps in the most 
metal--rich stars, these elements exhibit the well--known abundance 
correlations and anticorrelations found in ``normal" globular clusters.  
Additionally, the O--poor ([O/Fe]$\leq$0) stars are located almost exclusively 
within $\sim$5--10$\arcmin$ of the cluster center, but the O--rich 
([O/Fe]$\sim$+0.3) stars are rather evenly distributed at all cluster radii.
The heavier $\alpha$ elements Si, Ca, and Ti exhibit smaller star--to--star
dispersions than the lighter elements and are generally enhanced by about
a factor of two.  The average [$\alpha$/Fe] ratio tends to increase with 
metallicity up to [Fe/H]$\approx$--1, and above this metallicity the average 
[Ca/Fe] ratio begins to decline while the average [Si/Fe] and [Ti/Fe] 
abundances remain roughly constant.  The two Fe--peak elements scandium and 
nickel exhibit little star--to--star dispersion and their [X/Fe] ratios are 
nearly constant as a function of metallicity.  We find a strong increase in the 
[La/Fe] abundances when comparing stars in the RGB--MP and RGB--Int1
populations, but the average [La/Fe] ratios for stars in the RGB--Int2+3 and
RGB--a populations remain roughly the same.  In contrast, [Eu/Fe] does not vary
strongly with metallicity and is only modestly enhanced.  We now aim to 
interpret what these results reveal about $\omega$ Cen's complex evolutionary 
history.

\subsection{Supernova Nucleosynthesis: Evidence from Heavy $\alpha$ and 
Fe--peak Elements}

The standard theory of Galactic chemical evolution suggests that massive stars
($\ga$10 M$_{\sun}$) produce the majority of elements up to the Fe--peak 
during various hydrostatic and/or explosive burning stages, and return the 
newly synthesized material to the interstellar medium (ISM) primarily through 
Type II SN explosions (e.g., Arnett \& Thielemann 1985; Thielemann \& Arnett 
1985; Woosley \& Weaver 1995; Nomoto et al. 2006).  Theoretical yields indicate
that stellar populations where Type II SNe have played the dominant role in 
polluting the ISM should produce future generations of stars with [$\alpha$/Fe]
ratios that are about 0.3--0.5 dex larger than the solar--scaled value, and 
exhibit abundance ratios in the range --0.5$\la$[X/Fe]$\la$+0.3 for other 
elements lighter than about zinc.  The massive stars provide chemical 
enrichment on time scales of $\sim$2$\times$10$^{\rm 7}$ years or less, and are
believed to be the dominant production sources of most elements in the Galactic
halo and disk up to [Fe/H]$\approx$--1 (e.g., Timmes et al. 1995; Samland 
1998).  In contrast, Type Ia SNe primarily produce Fe--peak elements, and can 
contribute to a stellar population's ISM about 5$\times$10$^{\rm 8}$ to 
3$\times$10$^{\rm 9}$ years after the onset of star formation (e.g., Yoshii et 
al. 1996; Nomoto et al. 1997).  Significant contributions from Type Ia SNe are 
believed to drive the observed decrease in the Galactic [$\alpha$/Fe] abundance
trend at [Fe/H]$>$--1.

Figure \ref{f13} shows our measured [X/Fe] ratios as a function of [Fe/H], and 
overplots the expected abundance trends if (1) Type II SNe are responsible for 
all of $\omega$ Cen's chemical enrichment and (2) Type Ia ejecta are mixed with
Type II ejecta in a 75/25$\%$ ratio.  For consistency, we show only the
supernova yields from Nomoto et al. (1997; Type Ia) and Nomoto et al. (2006; 
Type II), but the theoretical yields from other groups (e.g., Woosley \& 
Weaver 1995) follow approximately the same trends.  We find that the $\alpha$ 
and Fe--peak abundance distributions are generally well described by pollution 
from Type II SNe.  However, Figures \ref{f10} and \ref{f13} indicate that 
the behavior of [Si/Fe], [Ca/Fe], and [Ti/Fe] as a function of increasing 
[Fe/H] is more complex than for [Sc/Fe] and [Ni/Fe].  For all three $\alpha$
elements, the average [$\alpha$/Fe] abundance noticeably increases between the
most metal--poor and intermediate metallicity populations.  Additionally, the
stars with [Fe/H]$<$--2 tend to exhibit larger [Si,Ca/Fe] ratios than the rest
of the RGB--MP stars, but the [Ti/Fe] abundances are mostly uniform across
the full RGB--MP metallicity range.  

Some of this behavior may be at least qualitatively explained by examining the 
mass and/or metallicity dependent yields of massive stars.  In 
Figure \ref{f14}, we plot the predicted production factors from Woosley \& 
Weaver (1995) for various elements as a function of progenitor mass.  The 
increase in the average [Si/Fe] and [Ca/Fe] abundances for $\omega$ Cen stars 
at [Fe/H]$>$--1.6 may be explained by the metallicity dependence of the Si and 
Ca yields, especially for stars more massive than about 18--20 M$_{\sun}$.  As 
can be seen in Figure \ref{f14}, the most massive stars are predicted to 
produce higher yields as the metallicity increases from [Fe/H]=--2 to --1, but 
the \emph{difference} between the Si and Ca yields are expected to remain 
roughly constant.  This means that as long as $\omega$ Cen was able to retain 
and mix the ejecta of $\ga$18 M$_{\sun}$ stars, we should expect (1) that the 
average [Si/Fe] and [Ca/Fe] abundances should increase with [Fe/H] and (2) that
both Si and Ca should exhibit the same general morphology until at least 
[Fe/H]$\approx$--1.  Both of these predictions are seen in Figures \ref{f10} 
and \ref{f13}.  However, the similar increase found for [Ti/Fe] may not be due 
to Type II SNe.  The theoretical yields do not predict a significant increase 
in [Ti/Fe] as a function of either progenitor mass or metallicity, and the 
situation does not improve if $>$25 M$_{\sun}$ stars are included (e.g., 
McWilliam 1997).  Instead, it seems likely that titanium has additional 
production sources.  We should note that this all follows the assumption that 
the observed abundances trace \iso{48}{Ti}, in addition to \iso{28}{Si} and
\iso{40}{Ca}, but an increase in the production of other stable isotopes could
alter this scenario.

Mass dependent yields may also be responsible for explaining the discrepancy 
in [Si,Ca/Fe] between the stars with [Fe/H]$<$--2 and the rest of the 
RGB--MP population.  As noted in $\S$4, the average [Si/Fe] and [Ca/Fe] 
abundances are 0.17 and 0.11 dex larger for the [Fe/H]$<$--2 stars.  This trend
can be reconciled if the most metal--poor stars in the cluster, which 
represent only 3$\%$ of the RGB--MP population, preferentially formed from
the ejecta of $\ga$20 M$_{\sun}$ stars.  However, this would require very rapid 
enrichment of the early $\omega$ Cen environment because $>$20 M$_{\sun}$ stars
live $\la$10$^{\rm 7}$ years (e.g., Schaller et al. 1992).  Note that
this scenario is compatible with the observation that the [Fe/H]$<$--2 stars 
have the same mean [Ti/Fe] abundance as the rest of the RGB--MP population
because, as mentioned above, the titanium yields from $>$20 M$_{\sun}$ stars 
are comparable to those of lower mass stars.  Additionally, if a monotonic
relationship between [Fe/H] and formation time exists for at least the RGB--MP
stars, then the mass dependent yields may also explain the apparent decrease
in [Si,Ca/Fe] as [Fe/H] increases from $\sim$--2 to --1.6, as well as, the 
steeper decline for [Si/Fe] compared to [Ca/Fe].  As indicated by Figure 
\ref{f14}, the decline in Si yield is a stronger function of progenitor mass 
between 18--25 M$_{\sun}$ than for Ca.  Therefore, forming stars from gas
polluted by progressively less massive SNe should qualitatively 
reproduce the observed trend.  The sudden increase in [Si,Ca/Fe] in the 
RGB--Int1 population would then make sense if a new round of star formation 
began with $>$20 M$_{\sun}$ stars contributing once again.  

\subsubsection{Are Type Ia SNe Required?}

Since previous analyses have estimated that the age spread among the various
$\omega$ Cen populations is $\sim$2--4 Gyr (e.g., Stanford et al. 2006), it 
would seem reasonable to assume that Type Ia SNe could have contributed to
the cluster's chemical enrichment.  However, the consistently elevated 
[$\alpha$/Fe] ratios observed for nearly all stars in the cluster suggests
that Type Ia enrichment has been limited.  Pancino et al. (2002) and Origlia
et al. (2003) found in a small sample of $\omega$ Cen giants that the RGB--a 
stars had noticeably lower [$\alpha$/Fe] and higher [Cu/Fe] abundances than the
lower metallicity stars, and attributed these trends to the onset of Type Ia 
SNe at [Fe/H]$>$--1.  On the other hand, Cunha et al. (2002) analyzed [Cu/Fe] 
abundances in a larger sample spanning [Fe/H]$\sim$--2 to --0.8, and did not 
find evidence for an increase in [Cu/Fe].  Similarly, Norris \& Da Costa (1995)
and Smith et al. (2000) did not find evidence for a decrease in [$\alpha$/Fe] 
or an increase in [Cu/Fe].

\begin{figure}
\plotone{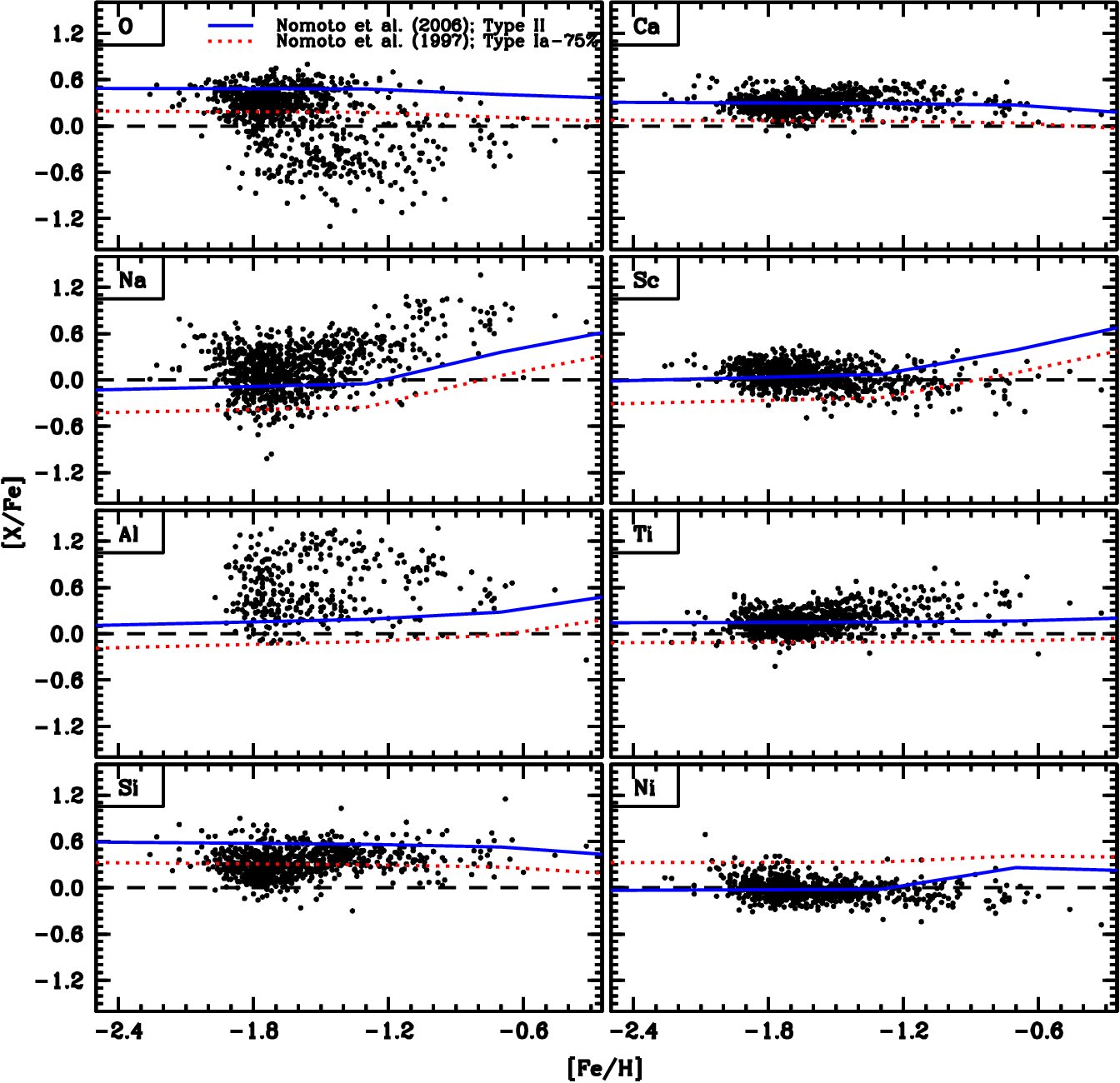}
\caption{[X/Fe] abundances for $\omega$ Cen stars plotted as a function of 
[Fe/H].  The solid blue lines indicate the combined Type II SNe yields of 
Nomoto et al. (2006) weighted by a standard IMF integrated from 0.07--50 
M$_{\sun}$, including contributions from hypernovae.  The dotted red lines 
represent the expected abundance trends if the Type Ia SN yields from Nomoto et
al. (1997) are mixed with the Type II yields in a with a 75$\%$ Type Ia and 
25$\%$ Type II ratio.  Note that the Type II yields have been systematically 
adjusted for [Sc/Fe], [Ti/Fe], and [Ni/Fe] to match the average abundances of 
the RGB--MP stars.}
\label{f13}
\end{figure}

\begin{figure}
\plotone{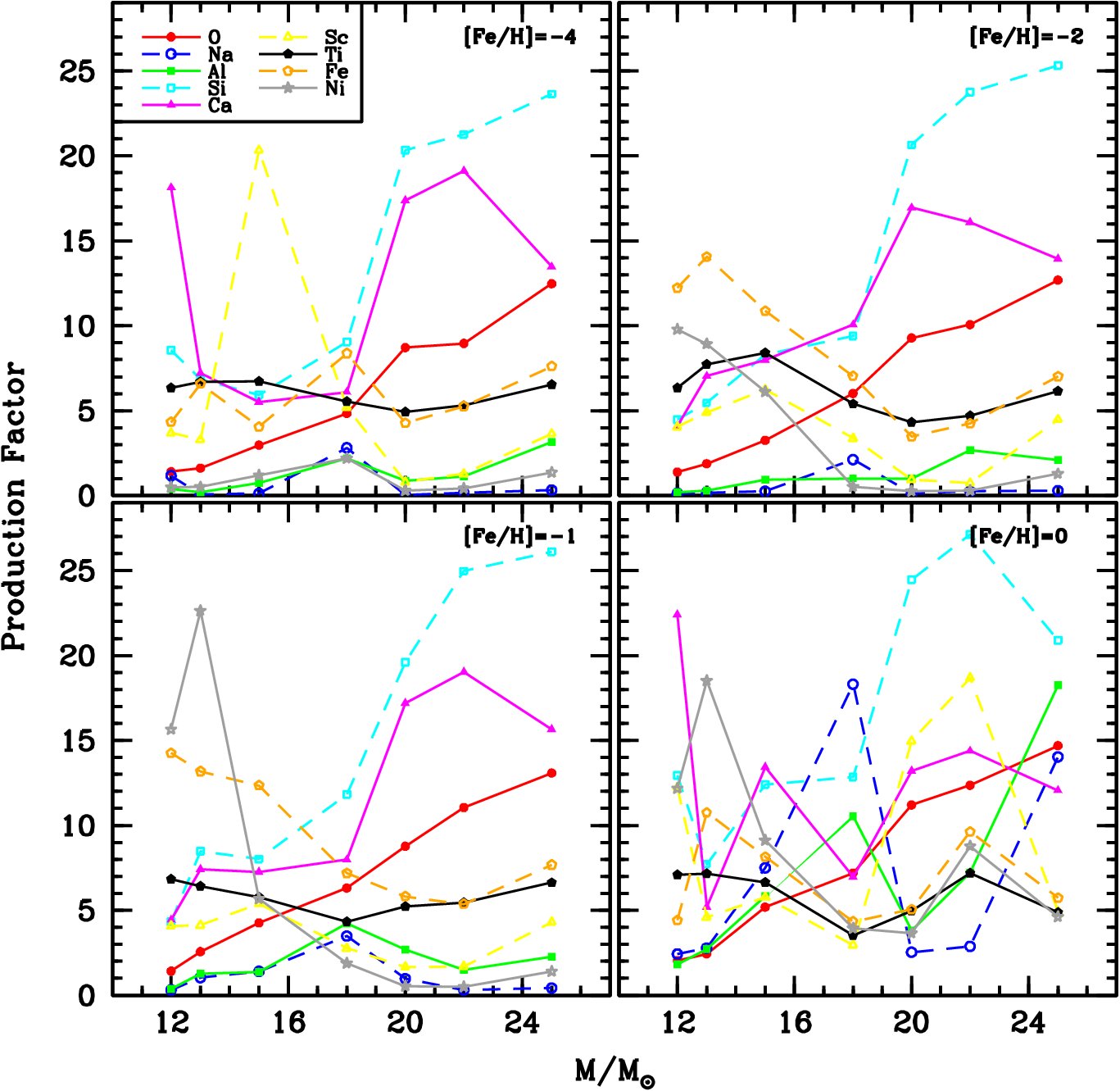}
\caption{Production factors for 12--25 M$_{\sun}$ Type II SNe from Woosley \& 
Weaver (1995).  The production factors are calculated as the ratio of an 
isotope's mass fraction in the ejecta compared to its mass fraction in the Sun.
Note that the isotopes plotted here are \iso{16}{O}, \iso{23}{Na}, 
\iso{27}{Al}, \iso{28}{Si}, \iso{40}{Ca}, \iso{45}{Sc}, \iso{48}{Ti}, 
\iso{56}{Fe}, and \iso{58}{Ni}.}
\label{f14}
\end{figure}

While the primary production source of Cu is uncertain (e.g., Sneden et al. 
1991b; Matteucci et al. 1993), it is clear that ambiguity remains regarding 
the significance of Type Ia SNe to $\omega$ Cen's chemical evolution.  Our data
are generally inconsistent with the rather extreme 75$\%$ Type Ia to 25$\%$ 
Type II mixture plotted in Figure \ref{f13}, especially at [Fe/H]$<$--1.
Although we find a slight decrease in [Ca/Fe] at [Fe/H]$>$--0.7, at least part
of this decrease may be explained by a reduction in calcium yields from more 
metal--rich Type II SNe (e.g., see Figures \ref{f13}--\ref{f14}).  
Interestingly, the [Si/Fe] and [Ti/Fe] ratios do not exhibit similar decreases 
at [Fe/H]$>$--0.7.  However, the larger measurement error for silicon compared 
to calcium may be masking any subtle trends, and although titanium is often 
enhanced like other $\alpha$ elements in globular cluster stars its dominant 
isotope \iso{48}{Ti} is not an $\alpha$ isotope.  Additionally, analyzing 
different mixtures of Type II versus Ia ejecta requires inherent assumptions 
about the massive star IMF and the source of Type Ia SNe, which in Figure 
\ref{f13} is the ``standard" white dwarf deflagration model.  The model values 
shown in Figure \ref{f13} could easily be changed by using different 
assumptions and adjusting the aforementioned parameters.

\begin{figure}
\plotone{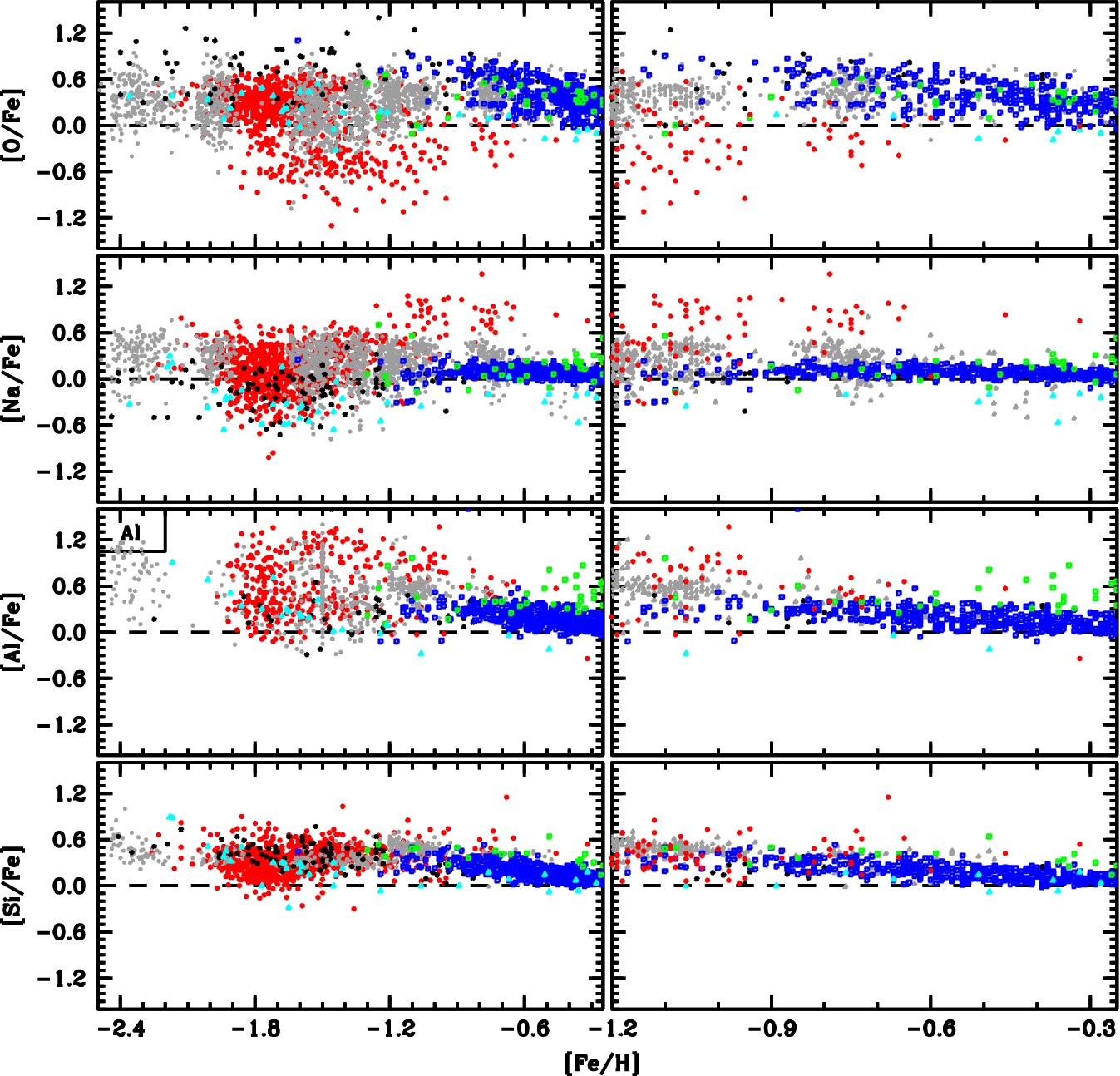}
\caption{[O/Fe], [Na/Fe], [Al/Fe], and [Si/Fe] abundances for individual stars 
in $\omega$ Cen (filled red circles), other globular clusters (filled grey 
circles), Galactic halo (black stars), thin/thick disk (open blue boxes), bulge
(open green boxes), and dwarf galaxies (open cyan triangles).  The left panels
show the [X/Fe] abundances as a function of [Fe/H] for $\omega$ Cen's full
metallicity range.  The right panels show the same abundance trends for 
[Fe/H]$\geq$--1.2, but the $\omega$ Cen points are plotted on top.  The 
literature references are listed in Table 8.}
\label{f15}
\end{figure}

\begin{figure}
\plotone{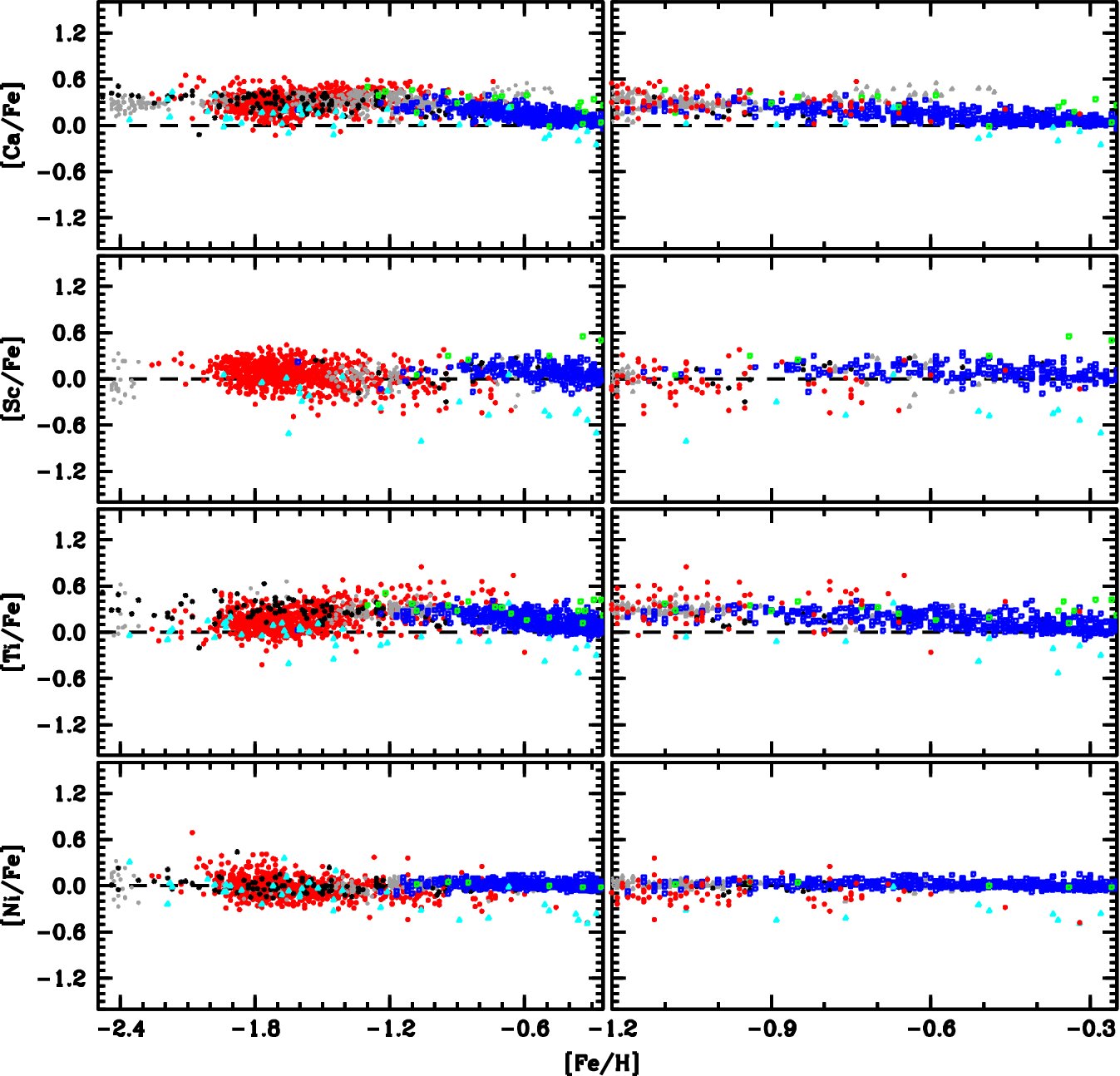}
\caption{Similar plot to Figure \ref{f15} showing [Ca/Fe], [Sc/Fe], [Ti/Fe],
and [Ni/Fe] abundances.  The symbols are the same as those in 
Figure \ref{f15}.}
\label{f16}
\end{figure}

A more empirical approach is to compare the evolution of $\alpha$ and Fe--peak
elements with other stellar populations exhibiting different levels of Type Ia
enrichment.  In Figures \ref{f15}--\ref{f17} we plot our derived abundances
for $\omega$ Cen stars as a function of [Fe/H], and compare with data from the
literature tracing the chemical evolution of other globular clusters, the
Galactic thin/thick disk, halo, bulge, and nearby dwarf galaxies (see Table 8
for literature references).  Focusing on the heavy $\alpha$ and Fe--peak
elements at the metal--rich end of the distribution shows that, at least for
stars with [Fe/H]$<$--0.7, $\omega$ Cen generally follows a morphology similar 
to that found in monometallic globular clusters, the Galactic halo, and the 
Galactic bulge.  In contrast, the most metal--rich $\omega$ Cen stars 
([Fe/H]$>$--0.7) exhibit [Ca/Fe] ratios that are more similar to those found in
Galactic thick disk stars (e.g., see Brewer \& Carney 2006).  Additionally, the
most metal--rich $\omega$ Cen stars tend to exhibit [Ca/Fe] ratios that are, on
average, at least 0.1--0.2 dex lower than those found in the more metal--poor 
stars.  This may indicate that the level of Type Ia enrichment in the most
metal--rich $\omega$ Cen stars and the thick disk were comparable.  However, 
at [Fe/H]$>$--0.7 the [Ni/Fe] ratios are noticeably low in the $\omega$ Cen
stars, and as mentioned previously the [Si/Fe] and [Ti/Fe] data do not 
exhibit similar abundance decreases in concert with [Ca/Fe].  Although $\omega$
Cen is widely believed to be the remnant core of a dwarf spheroidal galaxy, the
heavy $\alpha$ elements are enhanced in $\omega$ Cen stars by a factor of 2--3
compared with other dwarf galaxies, at least for [Fe/H]$\ga$--1.5.

In addition to the heavy $\alpha$ and Fe--peak elements, the lighter elements 
O, Na, and Al are also inconsistent with significant contributions from Type Ia
SNe.  Figure \ref{f13} shows that nearly all of the stars with [Fe/H]$>$--1 
have [Na/Fe] and [Al/Fe] abundances that are well above even the levels 
predicted by Type II SNe, but [O/Fe] is abnormally low.  The abundance patterns
expected from Type Ia production should lead to an overall decrease in the 
average abundance of all three elements as [Fe/H] increases.  However, these 
elements can be altered by either \emph{in situ} mixing or pollution from other
sources, and therefore may not be reliable indicators of a star's original 
composition.  While the heavy $\alpha$ element data, in particular [Ca/Fe],
provide some evidence for Type Ia SN contributions in the most metal--rich
stars, the light element data are in better agreement with a Type II SN 
pollution model that includes an additional proton--capture production 
mechanism.  The apparent suppression of Type Ia SNe in $\omega$ Cen 
remains an open problem, but it may be at least partially tied to the cluster's
several Gyr relaxation time scale (e.g., van de Ven et al. 2006) and low 
($\sim$3--4$\%$) binary frequency (Mayor et al. 1996).

\subsection{Proton--Capture Processing: Light Element Variations}

The light elements oxygen through aluminum provide sensitive diagnostics for
determining the chemical enrichment history of stellar populations.  These
elements are primarily produced in the hydrostatic helium, carbon, and/or
neon burning stages of massive ($\ga$10 M$_{\rm \sun}$) stars (e.g.,
Arnett \& Thielemann 1985; Thielemann \& Arnett 1985; Woosley \& Weaver 1995).
Stars forming out of gas that has been primarily polluted by Type II SNe should
have [O/Fe]$\sim$+0.4 and exhibit increasing [Na/Fe] and [Al/Fe] abundances 
with increasing metallicity.  However, these elements can also be produced (or 
destroyed) in lower mass stars that reach internal temperatures high enough to 
activate the proton--capture ON, NeNa, and MgAl cycles.  If this processed 
material is mixed to the surface, then some stars may return gas to the 
ISM that is O--poor and Na/Al--rich compared to the material ejected by Type II
SNe.  This scenario is believed to occur in the RGB and AGB phases of low and 
intermediate mass ($\la$8 M$_{\rm \sun}$) stars (e.g., Sweigart \& Mengel 1979;
Cottrell \& Da Costa 1981; Denisenkov \& Denisenkova 1990; Langer et al. 1993; 
Ventura \& D'Antona 2009; Karakas 2010), but also in the cores of massive, 
rapidly rotating main sequence stars (e.g., Decressin et al. 2007).

Figures \ref{f15}--\ref{f18} highlight the distinct light element abundance
patterns found in several different stellar populations.  Examination of these
trends indicates that although $\omega$ Cen shares some abundance patterns
with other globular cluster, Galactic disk, halo, bulge, and nearby dwarf
galaxy stars, it differs from all of these both in the extent of its 
star--to--star abundance variations and its individual abundance ratios.  
Approximately half of the RGB--MP stars have O, Na, and Al abundances that are 
consistent with those found in similar metallicity halo, and to a lesser 
extent, dwarf galaxy stars.  The chemical composition of these stars is 
believed to be primarily a result of Type II SN enrichment, and the chemical 
similarities among these populations is not unexpected.  It seems likely that 
$\omega$ Cen would have had considerable interaction with the primordial gas 
that formed the Galactic halo, and it has been shown that reproducing the 
cluster's metallicity distribution function is only possible in an open box 
scenario (e.g., Ikuta \& Arimoto 2000; Romano et al. 2007).  However, the 
remaining RGB--MP stars exhibit [O/Fe], [Na/Fe], and [Al/Fe] abundances that 
are significantly different than those found in metal--poor halo and dwarf 
galaxy stars.  In particular, the ``enhanced"  RGB--MP stars are O--poor and 
Na/Al--rich.  Similar chemical compositions are only found in some monometallic
globular cluster stars (e.g., see reviews by Kraft 1994; Gratton et al. 2004).  
Interestingly, the number of stars in $\omega$ Cen that are O--poor and 
Na/Al--rich increases to 60--95$\%$ at higher metallicities.  The RGB--Int2+3, 
and especially the RGB--a, stars have [O/Fe], [Na/Fe], and [Al/Fe] ratios that 
differ significantly even from individual globular clusters by at least a 
factor of two.  Figure \ref{f18} shows that this is true even when considering 
[O/Na], [O/Al], and [Na/Al] ratios instead of [X/Fe].  Our data indicate that 
the $\omega$ Cen stars at [Fe/H]$\ga$--1.3 experienced an additional enrichment
process that is not observed in any other stellar system studied so far, but
the combined populations of M54 and the Sagittarius dwarf galaxy may share some 
similar trends (Carretta et al. 2010).

The light element abundance patterns in other globular clusters are typically
believed to be the result of high temperature proton--capture nucleosynthesis
operating in an environment where a combination of the ON, NeNa, and MgAl
cycles are or were active.  Material that has been processed through these 
proton--capture cycles is expected to exhibit a deficiency in [O/Fe] concurrent
with supersolar [Na/Fe] and [Al/Fe] ratios, which should naturally lead to 
O--Na and O--Al anticorrelations along with a Na--Al correlation.  In Figures 
\ref{f19}--\ref{f21}, we plot [O/Fe], [Na/Fe], and [Al/Fe] against each other 
for the major $\omega$ Cen populations described in $\S$4.1.  We find that the 
O, Na, and Al abundance relations found in the RGB--MP, RGB--Int1, and 
RGB--Int2+3 populations are consistent with the abundance patterns that are 
characteristic of high temperature proton--capture processing.  Furthermore,
the impact of proton--capture nucleosynthesis appears to increase as a function
of increasing metallicity.  Both the extent of the light element variations and
the percentage of stars that are O--poor and Na/Al--rich increases 
monotonically with [Fe/H].  However, the same O--Na, O--Al, and Na--Al 
relations are not observed in the RGB--a population.  Instead, the RGB--a, as
well as a few RGB--Int2+3, stars exhibit a rather uniform composition that is 
moderately O--poor ([O/Fe]$\sim$--0.15), very Na--rich ([Na/Fe]$\sim$+0.78), 
and is unlike any of the more metal--poor $\omega$ Cen stars.

A common interpretation of the light element abundance trends in monometallic
globular clusters is that the O--rich, Na/Al--poor stars represent the first 
generation of stars formed from the ejecta of Type II SNe, and the O--poor, 
Na/Al--rich stars represent a subsequent generation formed from gas that had
been chemically enriched by intermediate mass AGB stars or some other polluting
source in which the ON, NeNa, and/or MgAl cycles were active (e.g., D'Ercole 
et al. 2008; Carretta et al. 2009b).  The first generation stars are often
referred to as ``primordial" stars, and the enriched populations are referred
to as either ``intermediate" or ``extreme", depending on the level of
O--depletion and Na--enrichment (e.g., Carretta et al. 2009a; but see also
Lee 2010 for a different interpretation).

The $\omega$ Cen data can be divided into similar subpopulations.  Here we 
follow a similar definition to that used in Carretta et al. (2009a) where the
primordial component is defined as having [O/Fe]$\geq$0 and [Na/Fe]$\leq$+0.1,
the intermediate component includes stars with [O/Fe] ratios satisfying the 
relation [O/Fe]$\geq$[0.62([Na/Fe])--0.65], and the extreme component consists 
of the remaining most O--poor stars.  Monometallic globular clusters typically 
consist of $\sim$20--40$\%$ of stars belonging to the primordial component, 
$\sim$30--80$\%$ in the intermediate component, and $\la$20$\%$ in the extreme 
component (e.g., Carretta et al. 2009a).  As can be seen in Figures 
\ref{f19}--\ref{f21}, the RGB--MP stars follow the general trend observed in 
monometallic globular clusters with a primordial:intermediate:extreme 
distribution of 50$\%$:43$\%$:7$\%$, respectively.  The RGB--Int1 population 
contains roughly an equal proportion of primordial, intermediate, and extreme 
abundance stars with a distribution of 30$\%$:32$\%$:38$\%$.  However, the 
RGB--Int2+3 and RGB--a stars contain far more extreme abundance stars than are 
found in any globular cluster with distributions of 11$\%$:15$\%$:74$\%$ and 
5$\%$:14$\%$:81$\%$, respectively.  The large number of intermediate and 
extreme abundance stars indicates that $\omega$ Cen likely experienced a
similar enrichment process to that in monometallic globular clusters during 
each round of star formation, and it is interesting to note that the 
populations expected to be He--rich exhibit the largest fraction of extreme 
abundance stars.  While there is a clear delay in the onset of whichever 
mechanism drives the O--poor, Na/Al--rich abundance phenomenon, it is worth 
noting that we find a very low incidence of carbon stars\footnote{While we 
do not provide explicit carbon abundance measurements in this paper, the 
possible carbon stars listed in Figure \ref{f1}, Figure \ref{f3}, and Table
2 were identified by visual inspection of their spectra.  However, all three 
of the possible carbon stars identified here that also overlap with the van 
Loon et al. (2007) survey (LEID 32059, 41071, and 52030) are confirmed carbon
stars based on the presence of strong C$_{\rm 2}$ bands in their spectra.} 
($<$2$\%$; see Figure \ref{f1}) despite the large population of O--poor stars. 
Unfortunately, we cannot distinguish between \emph{in situ} carbon stars and 
those formed from mass transfer, but the frequency of carbon stars on the giant
branch is consistent with the expected binary fraction of $\sim$3--4$\%$ (Mayor
et al. 1996).

\subsubsection{Enrichment by Pollution and \emph{in situ} Processing}

Although we have identified the major light element abundance trends for
$\omega$ Cen, the information so far has only led us to conclude that
proton--capture nucleosynthesis has likely played a significant role in the
cluster's chemical enrichment.  Further examination is required in order to
understand the possible location(s) where these processes are or were active.  
The comparatively small star--to--star dispersion in [X/Fe] exhibited by the
heavy $\alpha$ and Fe--peak elements (see Figure \ref{f10}) indicates that the
$>$1 dex variations observed for [O/Fe], [Na/Fe], and [Al/Fe] are not due to
incomplete mixing of SN ejecta, as is suspected for [Fe/H]$<$--3 halo stars
(e.g., McWilliam 1997).  Previous studies have found that many of the light
element abundance patterns exhibited by monometallic globular cluster stars,
which are subsequently shared by many $\omega$ Cen stars, may be
best explained by proton--capture nucleosynthesis operating at temperatures
near 70$\times$10$^{\rm 6}$ K (e.g., Langer et al. 1997; Prantzos et al. 2007).
If at least part of the abundance patterns found in $\omega$ Cen and other 
globular cluster stars are due to pollution from external sources, then the 
currently favored production mechanisms are: (1) hot bottom burning in 
$>$5 M$_{\sun}$ AGB stars (e.g., Ventura \& D'Antona 2009; Karakas 2010), 
(2) hydrogen shell burning in now extinct but slightly more massive RGB stars 
(Denissenkov \& Weiss 2004), and (3) core hydrogen burning in rapidly rotating 
massive stars (Decressin et al. 2007).  

While massive, rapidly rotating stars and extinct $\sim$0.9--2 M$_{\sun}$ RGB 
stars may also reproduce many of the observed light element trends, presently
there are no detailed theoretical yields spanning a fine grid of metallicities 
similar to those available for intermediate mass AGB stars.  Furthermore, the
time scale of pollution from extinct low mass RGB stars is at least 2--3 times 
longer than the estimated age spread among the different $\omega$ Cen 
populations, but this does not rule out possible mass transfer pollution from
these objects.  Additionally, the massive, rapidly rotating star scenario is
expected to produce a continuum of polluted stars with varying He abundances
(Renzini 2008), which is inconsistent with the singular Y=0.38 value that
seems required to fit the blue main sequence (e.g., Piotto et al. 2005).  
Romano et al. (2010) also point out that if the winds from massive main 
sequence stars are also responsible for the anomalous light element abundance
variations in the current generations of $\omega$ Cen stars, it is not clear
why the He enrichment was delayed until higher metallicities.  However, Renzini
(2008) and Romano et al. (2010) find that intermediate mass AGB stars may 
provide a reasonable explanation for the high He content in some stars, in 
addition to the \emph{average} behavior of [Na/Fe], and to a lesser extent 
[O/Fe], in $\omega$ Cen.  Therefore, we will only consider the AGB pollution
scenario here, but we caution the reader that several qualitative and 
quantitative hurdles remain in order for AGB pollution to be a viable 
explanation of light element variations in globular clusters (e.g., Denissenkov
\& Herwig 2004; Denissenkov \& Weiss 2004; Fenner et al. 2004; Ventura \& 
D'Antona 2005; Bekki et al. 2007; Izzard et al. 2007; Choi \& Yi 2008).

In Figure \ref{f22}, we plot our derived O, Na, and Al abundances as a function
of [Fe/H], and overplot the metallicity dependent theoretical yields from
Type II SNe, as well as, 3--6 M$_{\sun}$ AGB stars.  While the $\omega$ Cen
stars with chemical compositions similar to the Galactic disk and halo appear
well bounded by production from Type II SNe, the enhanced stars at least
qualitatively follow the general trends predicted by production from $>$5
M$_{\sun}$ AGB stars.  In particular, the depletion of oxygen concurrent with
the rise in sodium and decline in the \emph{maximum} [Al/Fe] ratio with
increasing metallicity are all consistent with the predicted patterns exhibited
by material that has been processed via hot bottom burning in $>$5 M$_{\sun}$
AGB stars.  However, the theoretical AGB yield curves shown in Figure \ref{f22}
do not include lifetime estimates for the polluting AGB stars, and one could 
envision sliding the various curves along the abscissa to account for age
differences among the different populations.  In other words, plots similar
to Figure \ref{f22} lend insight into whether the abundance trends are possibly
consistent with AGB pollution, but numerical chemical evolution models are 
required to fully constrain which mass ranges have contributed to the chemical 
composition of stars in a given population.

Despite this limitation, we can use Figure \ref{f22} to elicit some 
constraints.  We find that while 5--6 M$_{\sun}$ AGB ejecta are generally
consistent with the abundance trends observed at all metallicities, 
3--4 M$_{\sun}$ AGB stars likely did not contribute significantly to $\omega$ 
Cen's chemical enrichment until about [Fe/H]=--1.3.  This is most evident by
examining the [O/Al] and [Na/Al] ratios in Figure \ref{f22}.  The 
$<$5 M$_{\sun}$ AGB stars produce [O/Fe] and [Na/Fe] ratios that are too high 
and [Al/Fe] ratios that are too low to fit the data, even when diluted
with SN or $>$5 M$_{\sun}$ AGB ejecta.  Figures \ref{f19}--\ref{f21} also 
support the rejection of 3--4 M$_{\sun}$ AGB ejecta, \emph{which originate from
AGB stars of comparable metallicity}, from contributing significantly to the 
chemical composition of stars with [Fe/H]$<$--1.3.  However, Figures 
\ref{f19}--\ref{f22} do not rule out that $<$5 M$_{\sun}$ AGB stars with 
[Fe/H]$\la$--1.5 impacted enrichment of the RGB--Int2+3 and RGB--a populations.
The [Na/Fe] and [Al/Fe] yields from metal--poor AGB stars are mostly consistent
with the trends observed in the intermediate and most metal--rich $\omega$ Cen
giants, but it seems that an additional mechanism may be required to explain
the [O/Fe] abundances.  Note that our conclusions are not drastically altered 
if we adopt the theoretical AGB yields from Karakas (2010)\footnote{This 
statement is based on using the average mass fraction data from Tables A2--A6 
in Karakas (2010).}, which uses mixing length theory for convection, instead 
of the Ventura \& D'Antona (2009) yields, which use the full spectrum of 
turbulence theory for convection and are shown in Figures \ref{f19}--\ref{f22}.
Unfortunately, Karakas (2010) does not provide yield information for 
metallicities between [Fe/H]=--2.3 and --0.7, which makes direct comparison 
with $\omega$ Cen difficult because most stars fall in the missing range.

One of the most puzzling aspects concerning the abundance patterns of light 
elements in $\omega$ Cen is the strongly bimodal distribution at intermediate
metallicities (see Figure \ref{f11}).  If ISM pollution was driven by AGB 
stars, then it is unclear why (1) only the RGB--MP stars exhibit a 
continuous distribution of [O/Fe], [Na/Fe], and [Al/Fe] abundances and (2) more
than 70$\%$ of the more metal--rich stars have envelope material that has 
experienced significant proton--capture processing.  As can be seen in Figures
\ref{f19}--\ref{f22}, the [O/Fe] yields from AGB stars are by far the most 
inconsistent with our data, but the full mass range of AGB stars may reproduce
the [Na/Fe] and [Al/Fe] abundances at nearly all metallicities.  Depleting the 
oxygen abundance from [O/Fe]=+0.4 to [O/Fe]$<$--0.4 via hot bottom burning in 
AGB stars is generally not achieved for any mass or metallicity range.  
However, D'Ercole et al. (2010) showed that including the ejecta of 
``super--AGB" ($>$6.5 M$_{\sun}$) stars may reproduce the super O--poor 
([O/Fe]$<$--0.4) abundances found in some globular clusters under the 
assumption that the massive AGB stars deplete to [O/Fe]$\approx$--1.  Despite 
this, it is unlikely that more massive AGB stars are the culprits behind the 
large contingent of super O--poor $\omega$ Cen stars because one would have to 
assume an IMF strongly weighted toward $\sim$5--9 M$_{\sun}$ stars in order to 
produce so many super O--poor stars.  Note that this is not as much of a 
problem in monometallic globular clusters because the number of super O--poor 
stars is $<$20$\%$ (e.g., Carretta et al. 2009a).  It seems that invoking some 
degree of \emph{in situ} proton--capture processing is required to explain the 
observed abundance patterns of $\omega$ Cen stars with [Fe/H]$\ga$--1.6,
in order to avoid unrealistic requirements such as IMFs strongly weighted 
toward intermediate mass stars or forming a majority of the RGB--Int1, 
RGB--Int2+3, and RGB--a stars almost entirely out of a narrow mass range of 
AGB stars.

A key assumption when considering \emph{in situ} processing in low mass RGB 
stars is that the material being enriched near the hydrogen burning shell must
be able to mix into the convective envelope and be brought to the surface.
In stars with normal helium abundances, it is not believed that this can occur
until the hydrogen burning shell erases the molecular weight barrier left 
behind by the convective envelope after first dredge--up (e.g., see review by 
Salaris et al. 2002).  However, some or all of the intermediate metallicity 
stars in $\omega$ Cen are thought to be quite He--rich, and D'Antona \& 
Ventura (2007) found that stars with Y=0.35--0.40 should contain a much more
shallow molecular weight gradient that might not inhibit deep mixing.  Instead,
deep mixing in He--rich stars might be active over a wide range of 
luminosities on the giant branch, which would be consistent with our 
observation that the degree of light element enrichment is not strongly 
correlated with luminosity.  These authors also find that reproducing the 
abundance patterns exhibited by the super O--poor stars can be achieved by 
\emph{in situ} mixing if the RGB stars are already polluted by the ejecta of 
intermediate mass AGB stars.  In their scenario, \emph{in situ} mixing should
decrease the envelope [O/Fe] ratio by up to a factor of 10 while only 
increasing the [Na/Fe] ratio by about 0.2 dex.  While the evolution of [Al/Fe]
is not reported by D'Antona \& Ventura (2007), we can speculate that the 
enhancement in [Al/Fe] is smaller than that experienced by [Na/Fe] given the 
higher temperatures required to convert Mg to Al.

As mentioned above, the proposed deep mixing scenario only works if the 
intermediate metallicity RGB stars in $\omega$ Cen formed from material that 
was already enriched by hot bottom burning in intermediate mass AGB stars.
Our current data set does not provide direct evidence of this, but we may
look to the behavior of silicon as a proxy indicator because \iso{28}{Si} can 
be produced through leakage from the MgAl cycle at temperatures 
$>$65$\times$10$^{\rm 6}$ K (e.g., Yong et al. 2005; Carretta et al. 2009b).  
In Figure \ref{f23}, we plot our [X/Fe] abundances as a function of [Fe/H]
color coded by the primordial, intermediate, and extreme abundance components
described above.  While most of the $\alpha$ and Fe--peak elements do not 
display any particular dependence on light element abundance, the RGB--MP and
RGB--Int1 extreme component stars exhibit silicon enhancements of nearly 0.3 
dex compared to the primordial and intermediate component stars.  Furthermore, 
in Figure \ref{f24} we plot [O/Fe], [Na/Fe], and [Al/Fe] versus [Si/Fe], 
[Ca/Fe], and [Ti/Fe] and find that only silicon shows any semblance of a 
correlation with O, Na, and Al, as is indicated by the respective Pearson
correlation coefficients shown in Figure \ref{f24}.  This suggests that silicon
may have undergone an additional production process not experienced by the 
heavier $\alpha$ elements.  The existence of an Al--Si correlation concurrent 
with an O--Si anticorrelation suggests that the O--poor stars were likely 
polluted by material that had been processed at temperatures exceeding 
$\sim$65$\times$10$^{\rm 6}$ K.  These conditions are reached during hot 
bottom burning in intermediate mass AGB stars, but not in the hydrogen burning
shells of low mass RGB stars.  

Since the $\omega$ Cen stars likely satisfy the prerequisites needed for 
\emph{in situ} mixing to occur, we may attribute a large portion of the 
[O/Fe], and to a lesser extent the [Na/Fe], variations to this process.  The 
relatively small number of RGB--MP stars that are super O--poor suggests that 
the helium content had not yet been significantly increased in the cluster to 
allow the formation of He--rich stars.  In fact, there are very few super 
O--poor stars at [Fe/H]$<$--1.7.  The radial segregation of O--poor stars 
(see Figure \ref{f9}) is also consistent with the idea that additional time 
was needed to increase the cluster He content, and may indicate that He--rich 
gas was preferentially funneled into the cluster core, as is suggested in the 
models by D'Ercole et al. (2008).  We find that the light element abundance 
trends in the intermediate metallicity and RGB--a stars are consistent with an 
AGB pollution plus \emph{in situ} mixing scenario.  In these stars, the high 
[Na/Fe] and [Al/Fe] abundances are consistent with production in comparable 
metallicity or more metal--poor AGB stars because \emph{in situ} mixing is not 
expected to significantly increase [Na/Fe] or [Al/Fe] in He--rich RGB stars 
that are already O--poor and Na/Al--rich (D'Antona \& Ventura 2007).  
Additionally, the increasing minimum [O/Fe] abundance at [Fe/H]$\ga$--1
is consistent both with the increase in the [O/Fe] yields for $>$5 M$_{\sun}$ 
AGB stars and the fact that \emph{in situ} mixing should produce less advanced 
proton--capture processing at higher metallicities.  This is due primarily to 
the lower temperatures achieved in the interiors of more metal--rich stars, 
but may also occur if the He mass fraction in the RGB--a stars is smaller than 
in the RGB--Int2+3 stars, which could lead to more shallow mixing.  Lastly, we 
note that because the [Na/Fe] and [Al/Fe] abundances do not share the same 
correlation as [O/Fe] with radial location, it may be the case that some stars 
producing high Na and Al yields do not necessarily produce large He yields.

One of the most important effects of including \emph{in situ} mixing in the
chemical enrichment picture is that it reduces the necessity for AGB stars to 
account for all abundance patterns, and also increases the mass range of 
available AGB polluters to more than just those with favorable yields.  The
inferred high He--content of the blue main sequence population and the observed
large increase in s--process enrichment in this cluster indicate that the 
large degree of ISM pollution required for the above scenario to work is not 
unreasonable.  A detailed comparison of the \iso{24}{Mg}, \iso{25}{Mg}, and 
\iso{26}{Mg} isotopes may be particularly illuminating in order to investigate 
whether hot bottom burning, \emph{in situ} processing, or both played an active
role in shaping the abundance patterns of $\omega$ Cen giants.  Note that 
fluorine is also expected to be strongly depleted in the proposed deep mixing
scenario.  Furthermore, a large sample spectroscopic abundance analysis of 
stars at lower luminosities could provide an interesting test for the impact 
of \emph{in situ} mixing.

\subsubsection{Oxygen Abundances and a Possible Connection to the Blue Main 
Sequence}

The discovery and subsequent detailed analyses of $\omega$ Cen's blue main 
sequence (Anderson 1997, 2002; Bedin et al. 2004; Norris 2004; Piotto et al. 
2005; Sollima et al. 2007; Bellini et al. 2009b) have revealed that this 
population represents $\sim$30$\%$ of all main sequence stars, is 
preferentially located near the cluster center, and perhaps most importantly is
more metal--rich than the dominant red main sequence.  As mentioned in $\S$1, 
the commonly accepted reason for the existence of the blue main sequence is 
that these stars are significantly more He--rich than the dominant population 
of more metal--poor stars.  In fact, Norris (2004) and Piotto et al. (2005) 
find that the blue main sequence is best fit with an extreme helium abundance 
of Y$\approx$0.38.  One of the most interesting characteristics of the blue 
main sequence is that it is well detached from the red main sequence in color, 
and almost no stars are found in between the two sequences (e.g., see Bedin et 
al. 2004; their Figure 1).  From this and the information above, we might 
expect the current RGB stars that were once part of the blue main sequence to 
be chemically conspicuous, preferentially located near the cluster center, and 
more metal--rich that the dominant stellar population.

Examination of the abundance patterns in the RGB--Int1 and RGB--Int2+3 stars
reveals that the [O/Fe] ratio stands out as a possible indicator of which stars
once belonged to the blue main sequence.  At intermediate metallicities, a 
majority of the stars have [O/Fe]$\leq$0, and the radial distribution of these
stars shows that more than 90$\%$ are located inside 10$\arcmin$ from the 
cluster center while only 70$\%$ of those with [O/Fe]$>$0 are located in the 
same range (see Figure \ref{f9}).  However, in order for the [O/Fe]
abundance to be considered as a chemical tracer of the blue main sequence it 
must qualitatively and quantitatively agree with the observed trends of blue 
main sequence stars.  In Figure \ref{f25}, we plot the number ratio of 
O--poor to O--rich stars out to $\sim$15$\arcmin$, and also plot the measured 
number ratios of blue to red main sequence stars from Bellini et al.
(2009b).  Although we are plotting an indirect measurement of the ratio of blue
to red main sequence stars with N$_{\rm O-poor}$/N$_{\rm O-rich}$ and the 
Bellini et al. (2009b) data represent direct measurements, we find that the 
two trends are in reasonable agreement.  Both data sets indicate that the 
majority of O--poor (blue main sequence) stars are inside $\sim$5$\arcmin$ of 
the cluster center, and the \emph{relative} ratio of O--poor/O--rich (blue/red
main sequence) stars decreases at larger radii.  Note that Sollima et al. (2007)
come to a similar conclusion when considering stars located at 
$\sim$7--23$\arcmin$ from the cluster center.  

Since the relative ratios of O--poor to O--rich stars follow those observed
for the blue and red main sequences, we may expect the absolute number of 
O--poor stars to also be consistent with that of the blue main sequence stars
due to our high completion percentage (see Figure \ref{f2}).  As mentioned 
previously, it is estimated that the blue main sequence constitutes 
$\sim$25--35$\%$ of all main sequence stars, and we find in agreement with this 
estimate that 27$\%$ of all RGB stars in our sample are O--poor.  Additionally,
Piotto et al. (2005) showed that the blue main sequence is best fit by a 
metallicity similar to that of the RGB--Int1 and RGB--Int2+3 populations.  We
find that at least 65$\%$ of the O--poor stars in our sample are located in 
the appropriate metallicity range.  This percentage may in fact be somewhat 
larger if we consider that (1) very few O--poor stars are found at 
[Fe/H]$<$--1.7, (2) the average [Fe/H] abundance error is roughly $\pm$0.1 dex,
and (3) the boundary between the RGB--MP and RGB--Int1 populations is not 
uniquely defined.  However, we would still find that $\sim$20--30$\%$ of 
intermediate metallicity stars are O--rich.  Note that a significant number
of O--rich, intermediate metallicity stars (i.e., stars in the correct 
metallicity range that would not lie on the blue main sequence) would be 
consistent with the observation by Sollima et al. (2006) that many RR Lyrae 
stars with [Fe/H]$\sim$--1.2 have standard helium abundances.  

In any case, we have demonstrated that the O--poor giants are spatially 
similar, found mostly in the same metallicity range, and are present in nearly
identical proportions to those found on the blue main sequence.  It is not 
entirely clear why the [O/Fe] ratio in the giants shares a similar sensitivity 
to radial location and metallicity with the blue main sequence stars, but we 
speculate that the oxygen deficient stars are connected with the blue main 
sequence through helium enrichment.  That is, the He--rich main sequence stars 
are pushed blueward on the color--magnitude diagram, and the He--rich giants 
experience \emph{in situ} mixing that strongly depletes oxygen without 
similarly large increases in sodium and aluminum.  Comparison between the 
7770 \AA\ oxygen triplet line strengths in blue and red main sequence stars
would provide a direct confirmation of our hypothesis.  Although we have 
invoked \emph{in situ} mixing to explain the very large O--depletion in these
stars, note again that the scenario proposed by D'Antona \& Ventura (2007) 
requires that these star were already somewhat O--poor.

\subsection{Neutron--Capture Processing}

While the isotopes of most elements lighter than about zinc are produced 
primarily through charged particle reactions, the isotopes of elements beyond 
the Fe--peak are mostly produced through neutron--capture reactions.  
Neutron--capture nucleosynthesis is believed to proceed through two main 
channels: (1) the s--process where the neutron--capture rate is slow compared 
to the $\beta$--decay rate of unstable nuclei and (2) the r--process where the 
neutron--capture rate is fast compared to the $\beta$--decay rate of unstable 
nuclei (e.g., see recent review by Sneden et al. 2008).  The large difference 
in neutron fluxes required for the two processes points to different 
operational environments.  The main component of the s--process is widely 
believed to be active in thermally pulsing low and intermediate mass AGB 
stars, but theoretical models indicate that most of the s--process element 
production is probably constrained to stars in the range $\sim$1.3--3 
M$_{\sun}$ (e.g., Busso et al. 1999; Herwig 2005; Straniero et al. 2006).  AGB 
stars $\la$1.3 M$_{\sun}$ have envelope masses that are too small for third 
dredge--up to occur, and more massive AGB stars are only believed to experience 
a few third dredge--up episodes.  Conversely, the exact location(s) where the 
r--process operates is (are) not well defined, but significant circumstantial 
evidence suggests an explosive origin associated with core collapse SNe (e.g., 
Mathews \& Cowan 1990; Cowan et al. 1991; Wheeler et al. 1998; Arnould et al. 
2007; Sneden et al. 2008).

\begin{figure}
\plotone{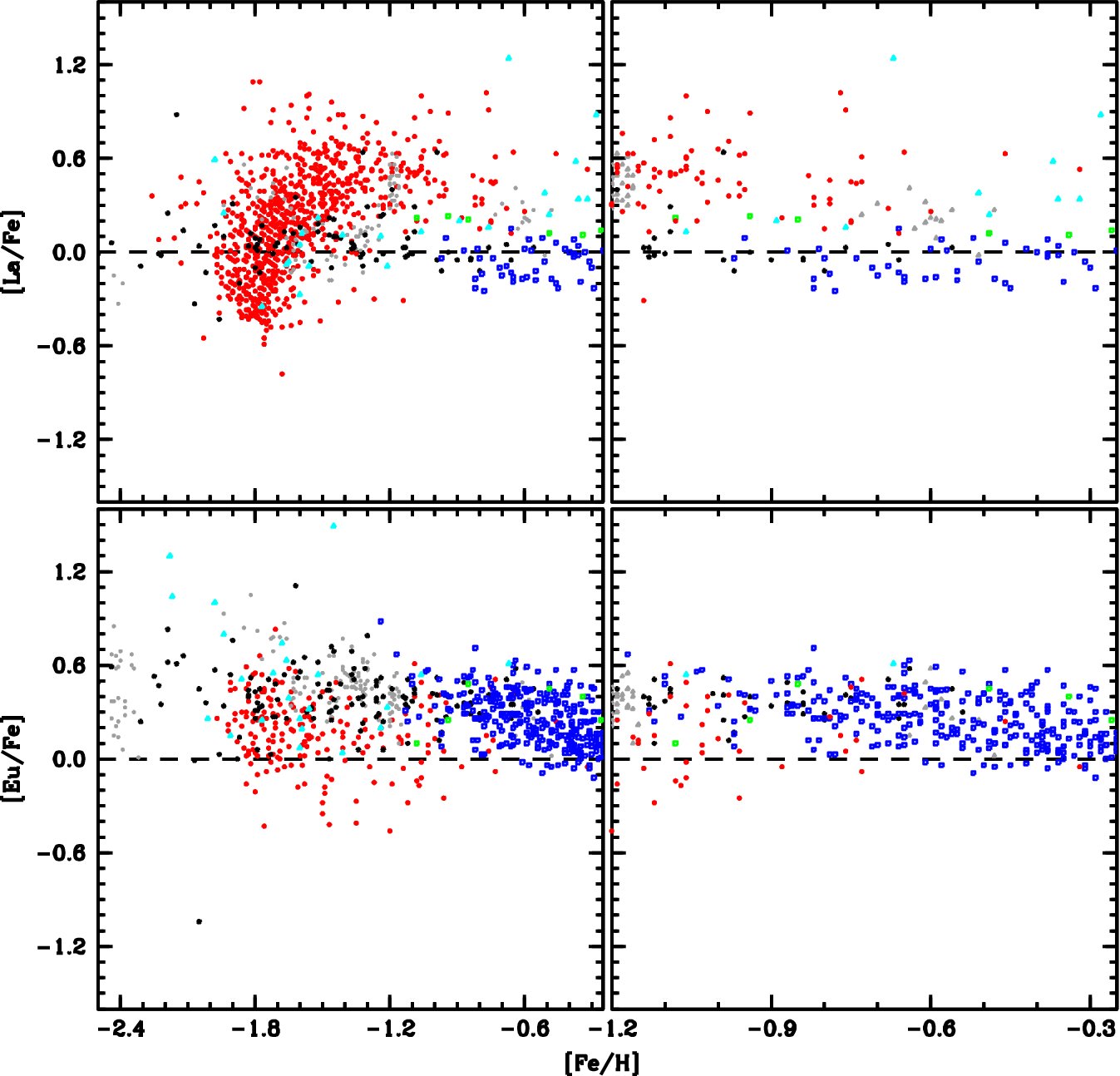}
\caption{Similar plot to Figure \ref{f15} showing [La/Fe] and [Eu/Fe] 
abundances.  The symbols are the same as those in Figure \ref{f15}.}
\label{f17}
\end{figure}

\begin{figure}
\plotone{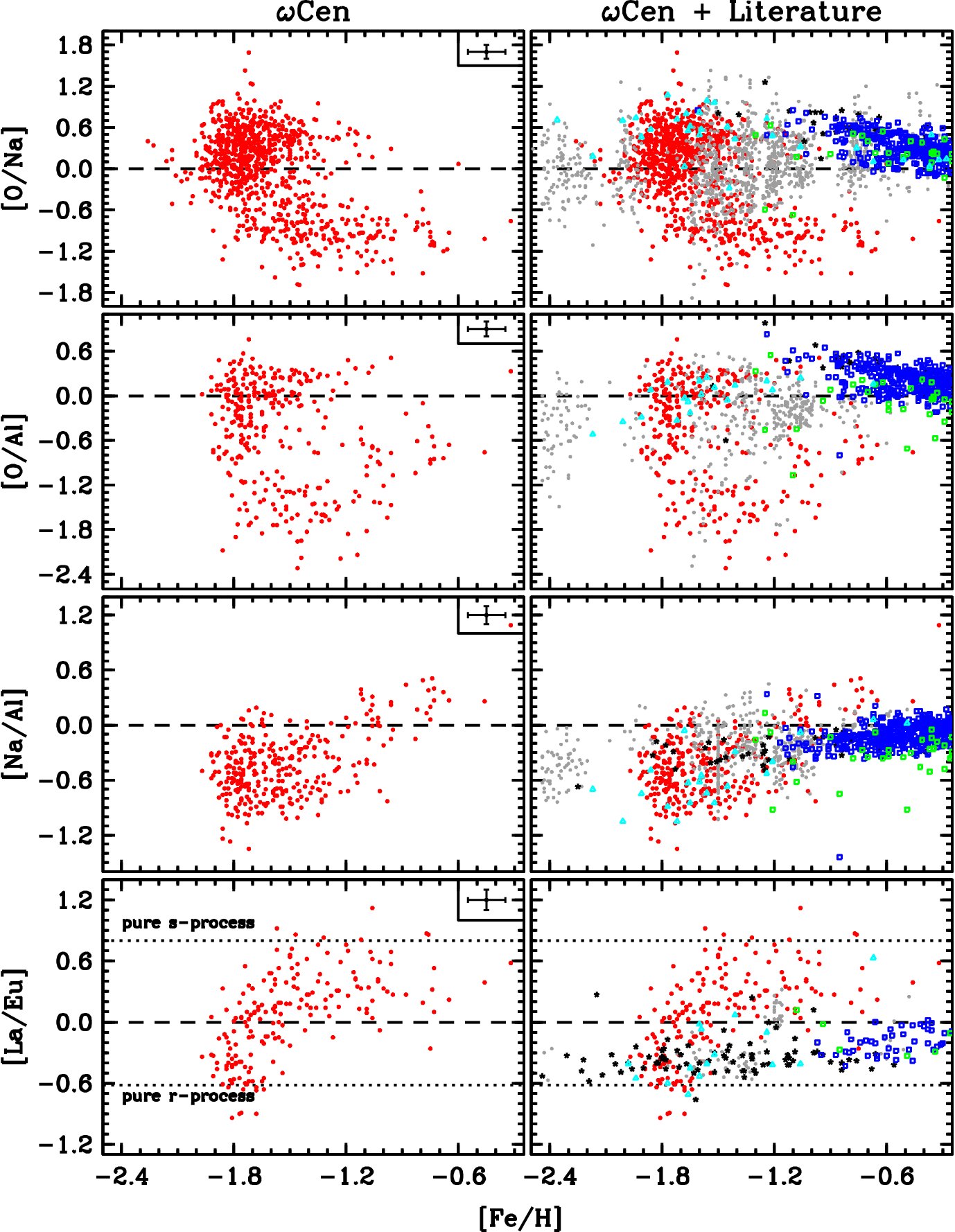}
\caption{[O/Na], [O/Al], [Na/Al], and [La/Eu] abundances are plotted as a
function of [Fe/H] for $\omega$ Cen (left panels) and the literature (right
panels).  The symbols are the same as those in Figure \ref{f15}.  The dotted
lines in the [La/Eu] panels indicate the abundance ratios expected for pure
r-- and s--process enrichment given in McWilliam (1997).}
\label{f18}
\end{figure}

\begin{figure}
\plotone{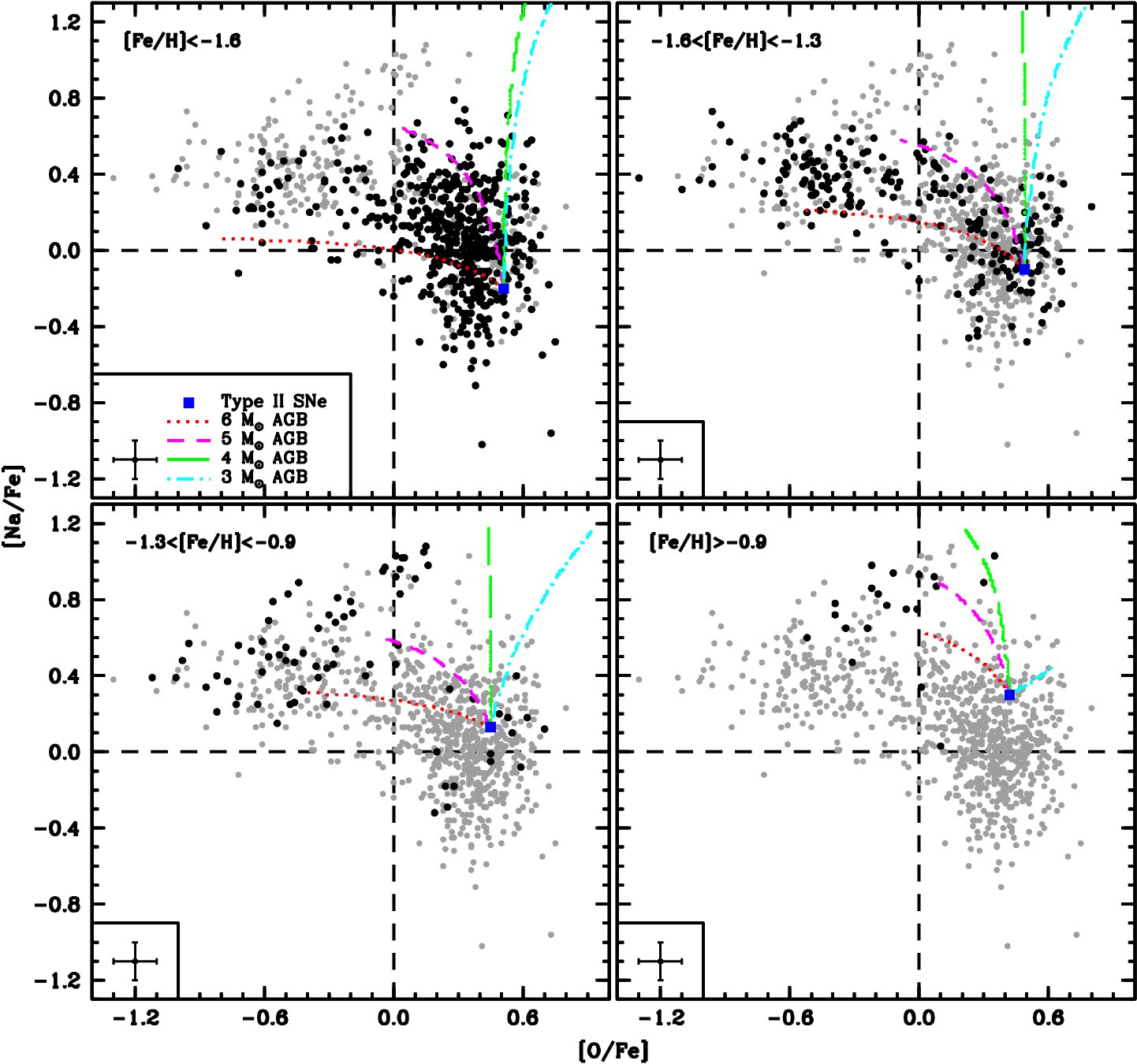}
\caption{[Na/Fe] versus [O/Fe] abundances for the four primary $\omega$ Cen
populations (see $\S$4.1).  The filled grey circles represent the full sample, 
and the filled black circles represent only the stars residing in the 
designated metallicity range.  Individual yields from Ventura \& D'Antona 
(2009) are shown for 3 (dot--dashed cyan lines), 4 (long dashed green lines), 
5 (dashed magenta lines), and 6 M$_{\sun}$ (dotted red lines) AGB stars of 
varying [Fe/H].  The filled blue squares indicate the approximate
Type II SN+Hypernova yields from Nomoto et al. (2006) expected for the given 
metallicity regime.}
\label{f19}
\end{figure}

\begin{figure}
\plotone{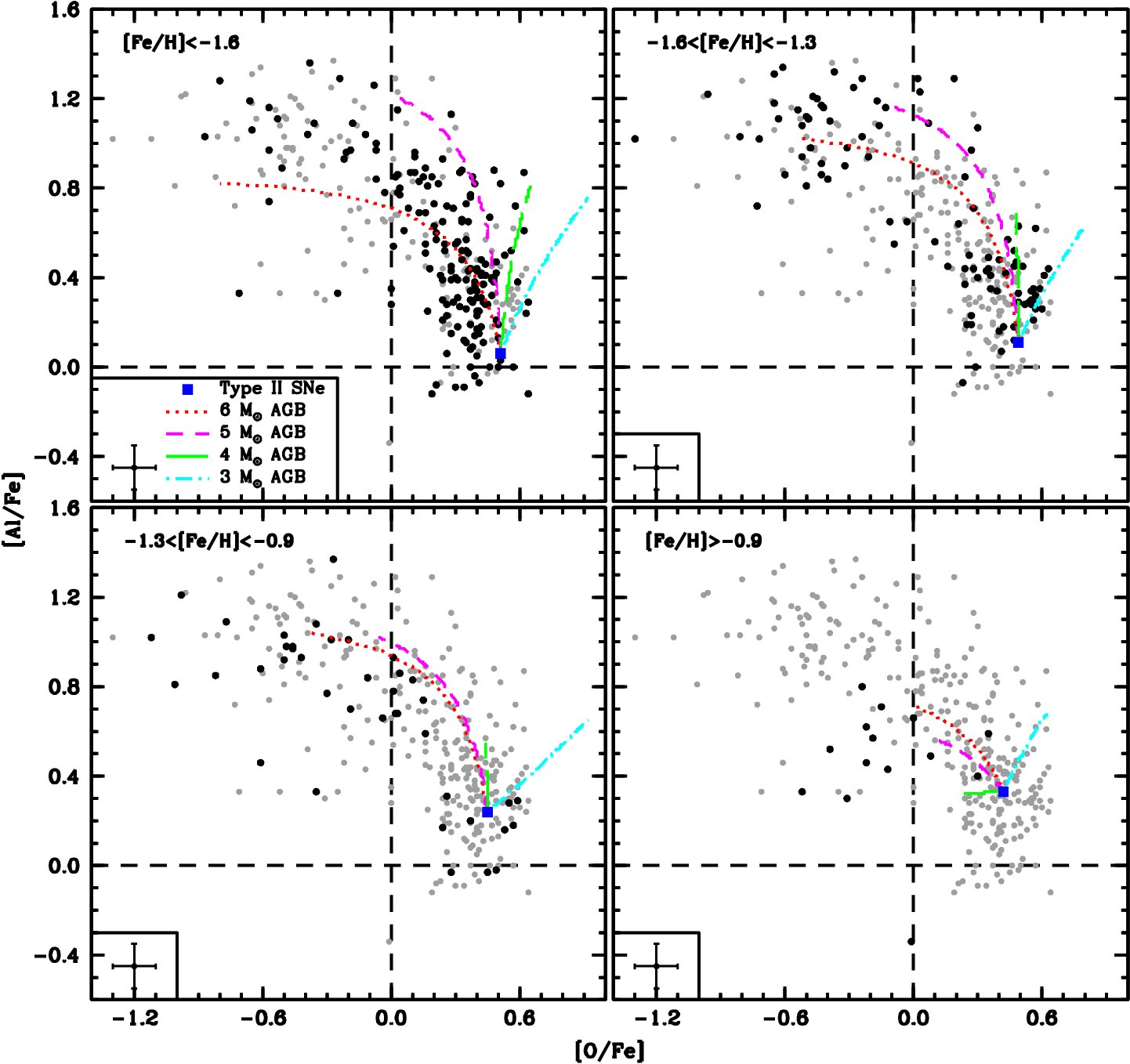}
\caption{Similar plot to Figure \ref{f19} showing the run of [Al/Fe] versus
[O/Fe] abundances for the different populations.  The symbols are the same
as those in Figure \ref{f19}.}
\label{f20}
\end{figure}

\begin{figure}
\plotone{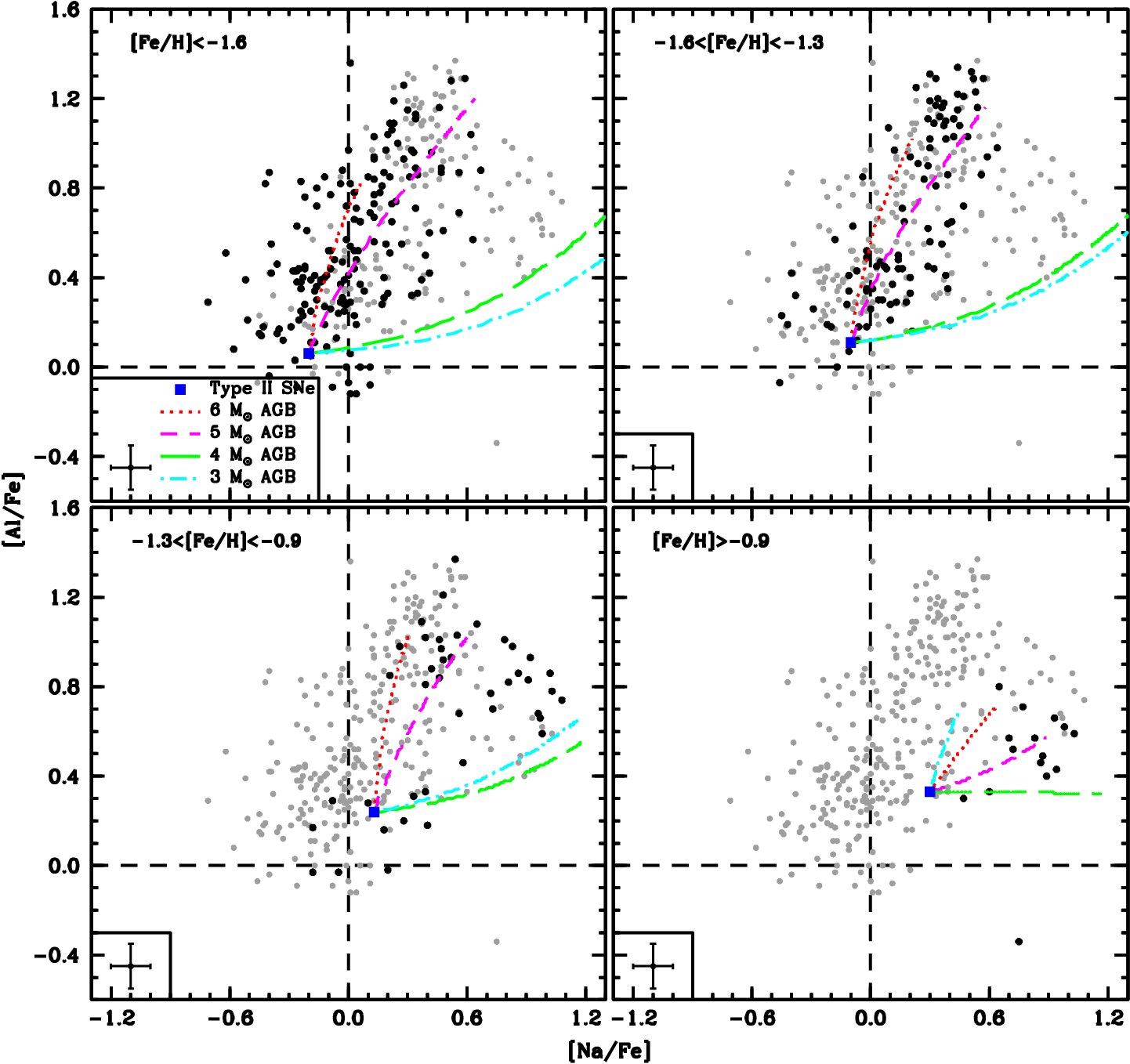}
\caption{Similar plot to Figure \ref{f19} showing the run of [Al/Fe] versus
[Na/Fe] abundances for the different populations.  The symbols are the same
as those in Figure \ref{f19}.}
\label{f21}
\end{figure}

\begin{figure}
\plotone{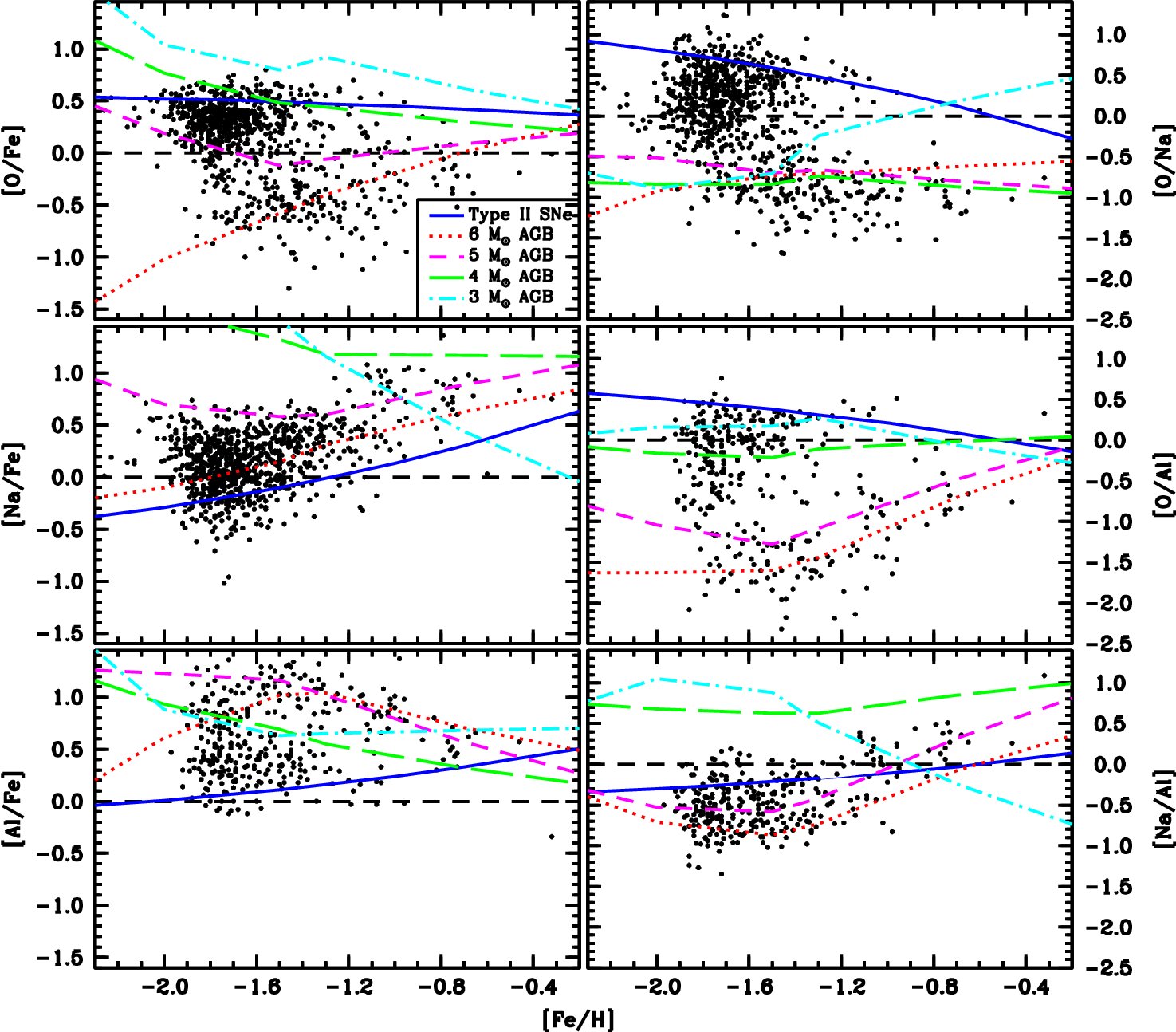}
\caption{The abundances of O, Na, and Al are plotted as a function of [Fe/H]
for the full sample that includes our new results and those from Johnson et al.
(2008; 2009).  The solid blue lines illustrate the Salpeter IMF--weighted Type
II SN yields from Nomoto et al. (2006), and the other curves are the same
as those in Figure \ref{f19}.}
\label{f22}
\end{figure}

\begin{figure}
\plotone{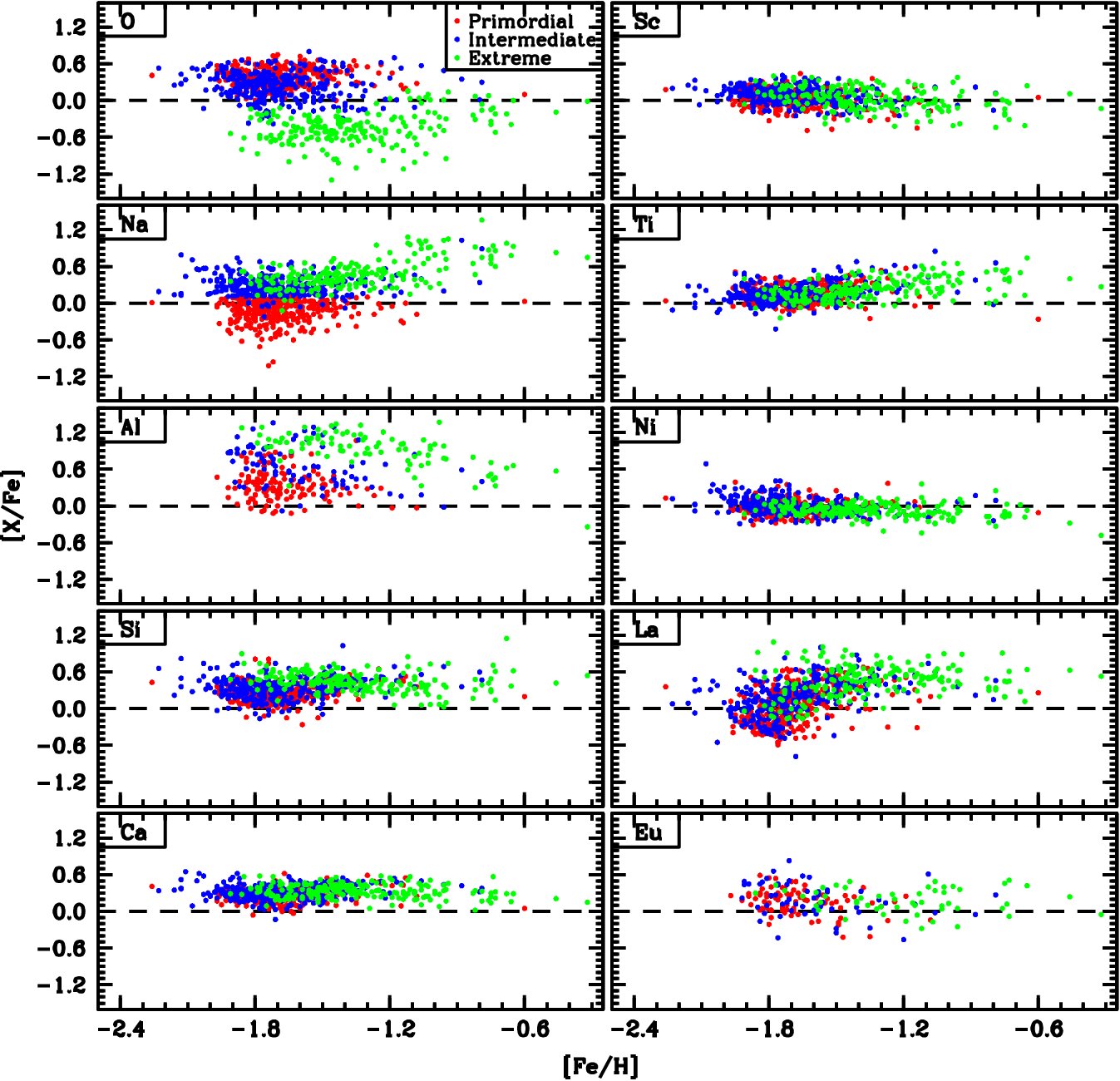}
\caption{Similar plot to Figure \ref{f10} showing the abundance distribution
of all elements as a function of [Fe/H].  The stars are broken down into the
primordial (filled red circles), intermediate (filled blue circles), and
extreme (filled green circles) components defined in $\S$5.}
\label{f23}
\end{figure}

\begin{figure}
\plotone{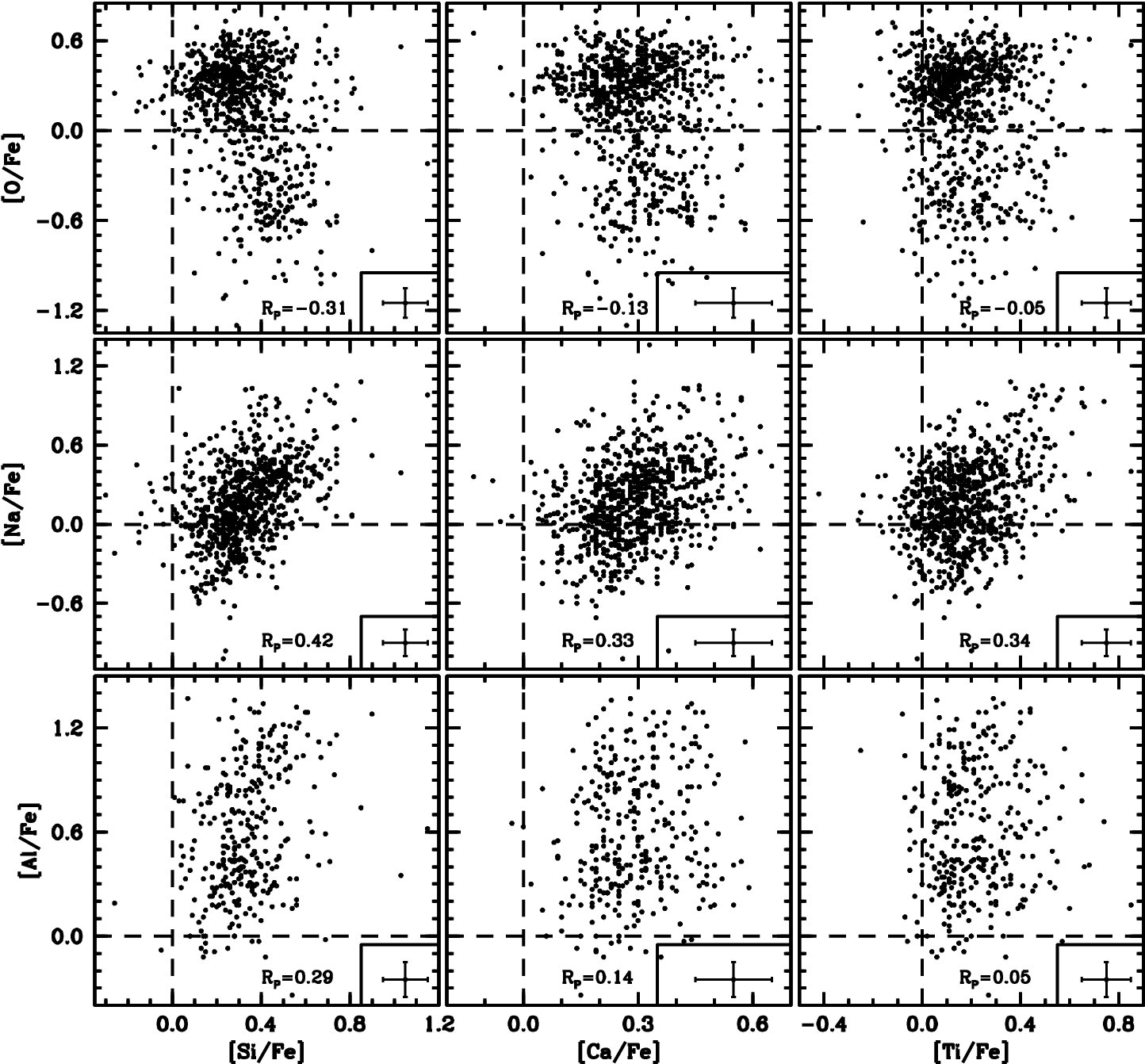}
\caption{The light elements O, Na, and Al plotted as a function of the heavier
$\alpha$ elements Si, Ca, and Ti for all $\omega$ Cen giants.  The Pearson
correlation coefficient (R$_{\rm P}$) is also calculated and displayed for 
each panel.}
\label{f24}
\end{figure}

\begin{figure}
\plotone{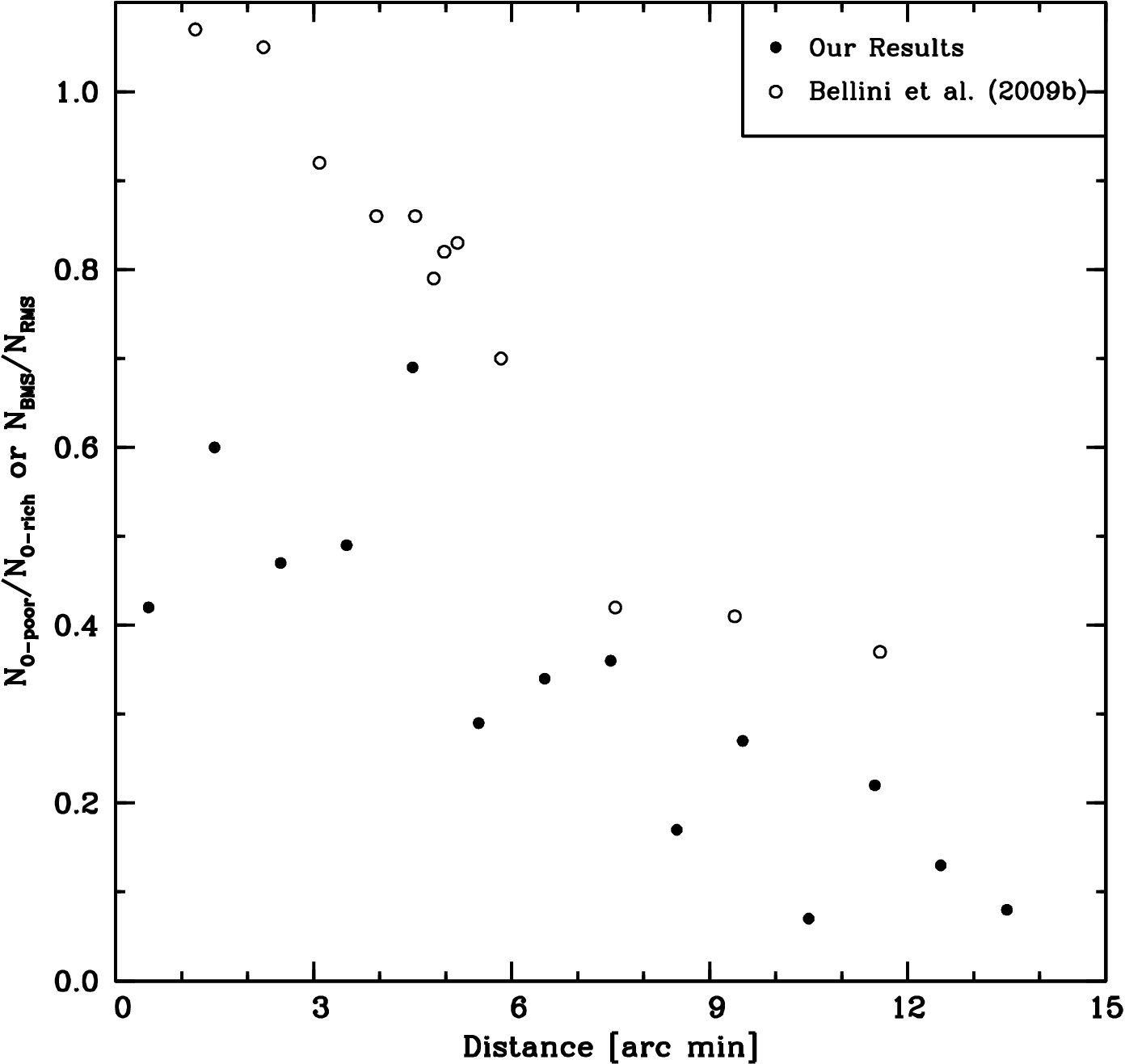}
\caption{A plot illustrating the change in N$_{\rm O-poor}$/N$_{\rm O-rich}$
(our results) or N$_{\rm BMS}$/N$_{\rm RMS}$ (Bellini et al. 2009b) as a 
function of radial distance from the cluster center.  Our results are
shown as the filled circles, and the results from Bellini et al. (2009b) are
shown as the open circles.  Note that N$_{\rm BMS}$ refers to the number of 
blue main sequence stars and N$_{\rm RMS}$ the number of red main sequence 
stars in the Bellini et al. (2009b) sample.}
\label{f25}
\end{figure}

The solar system abundances indicate that $\sim$70--75$\%$ of lanthanum is 
produced via the s--process and more than 95$\%$ of europium is produced 
by the r--process (e.g., Sneden et al. 1996; Bisterzo et al. 2010).  Therefore,
we adopt lanthanum as an s--process indicator and europium as an r--process
indicator.  As can clearly be seen in Figure \ref{f17}, the average [La/Fe] 
ratio increases by more than a factor of three between the RGB--MP and 
intermediate metallicity populations.  Similar increases are not found for any 
other elements in our sample.  This indicates that the s--process has played a 
significant role in the chemical evolution of $\omega$ Cen, and is a dominant 
process at [Fe/H]$\ga$--1.6.  Comparison with the other stellar populations
plotted in Figure \ref{f17} shows that the level of s--process enrichment was
far greater in $\omega$ Cen.  It is interesting to note that the [La/Fe] ratio 
does not continue to increase beyond [Fe/H]$\geq$--1.5 despite the fact that 
s--process production appears to peak in the metallicity range 
--1.5$\la$[Fe/H]$\la$--0.8, at least for the ``standard" \iso{13}{C} pocket 
(e.g., see Bisterzo et al. 2010, their Figure 8).  This may indicate that a 
large fraction of the gas was swept out of the cluster through interaction 
with the Galaxy before low mass AGB stars with [Fe/H]$\ga$--1.5 had a chance 
to contribute to $\omega$ Cen's chemical enrichment.  Comparison between the
RGB--Int2+3 and RGB--a stars shows that the average [La/Fe] ratio decreases by 
$\sim$0.2 dex (see Figure \ref{f12}) for higher metallicities, but is still 
significantly enhanced compared to the Galactic disk and bulge trends.  This 
suggests that s--process production still continued at high metallicities, but 
the rate of production did not exceed that of iron.

For the RGB--MP stars, the average [Eu/Fe] abundances are similar to those 
found in halo, dwarf galaxy, and individual globular cluster stars.  At higher
metallicities, the average [Eu/Fe] abundance of $\omega$ Cen stars actually 
decreases while the average [La/Fe] abundance shows a significant increase.  In
fact, many intermediate metallicity $\omega$ Cen stars have [Eu/Fe] abundances
that are lower than those found in halo and dwarf galaxy stars, and are 
especially Eu--deficient compared to globular cluster stars.  However, the 
average [Eu/Fe] abundance increases again at [Fe/H]$\ga$--1.2 toward values 
similar to those found in the Galactic disk and bulge.  The cause of the 
decrease in [Eu/Fe] at intermediate metallicities, and the low [Eu/Fe] 
abundances in general, is not entirely clear.  It is believed that $\sim$8--10 
M$_{\sun}$ SNe may produce a large portion of the r--process elements (e.g., 
Mathews \& Cowan 1990), but other processes such as neutron star and black hole
mergers may be important as well (e.g., see review by Sneden et al. 2008 and 
references therein).  It may be the case that either the IMF did not favor a 
large number of stars in the 8--10 M$_{\sun}$ range or that one or more of the 
typical r--process production mechanisms was not active at ``normal" levels in 
the intermediate metallicity range.  Interestingly, the metallicity range at 
which the [Eu/Fe] abundance is lowest is also where Cunha et al. (2002; 2010) 
find low [Mn/Fe] and [Cu/Fe] values.  Since Cunha et al. (2010) attributes the 
low [Cu/Fe] and [Mn/Fe] abundances to metallicity dependent SN yields, we can
speculate that the low [Eu/Fe] values might also be due to a related effect.  
Although manganese and europium are produced through different processes, 
their production may be tied to similar progenitor objects and/or environments.

Despite the obvious differences in [La/Fe] and [Eu/Fe] abundances for $\omega$
Cen stars compared to those in other populations, the ratio of these elements
provides a better diagnostic for analyzing the impact of the s-- and 
r--processes.  In the Galactic disk and halo, the [La/Eu] ratio slowly 
increases with metallicity, and this is believed to be primarily due to the
longer time scales required for low and intermediate mass stars to evolve into
AGB stars (e.g., Simmerer et al. 2004).  Dwarf galaxies also tend to exhibit
an increase in s--process elements at higher metallicities, but are typically
more s--process enhanced than similar metallicity halo and disk stars (e.g.,
Geisler et al. 2007).  In contrast, most globular clusters follow the disk/halo
trend and are generally r--process rich (e.g., Gratton et al. 2004).  Figure 
\ref{f18} plots the [La/Eu] ratio as a function of [Fe/H] for these populations
and also illustrates the relatively rapid transition in $\omega$ Cen from being
r--process to s--process dominated.  Many stars in the RGB--MP population 
exhibit [La/Eu] ratios that are identical to those found in halo, dwarf galaxy,
and globular cluster stars.  However, almost all of the more metal--rich stars 
have [La/Eu]$>$0, and many of these stars have [La/Eu] ratios matching those 
expected for pure s--process production.  Note that proper accounting of 
hyperfine structure for both the La and Eu lines has revised our [La/Eu] ratios
downward, at least for the most La--rich stars, from those found in Johnson et 
al. (2009).  The new results are consistent with the more metal--rich stars 
forming from gas that was already heavily polluted with s--process elements, 
but does not require surface pollution from mass transfer.  This is in 
agreement with the results from Stanford et al. (2010), which suggest that the 
strontium abundances (a light s--process element) in several $\omega$ Cen stars
are the result of primordial pollution rather than surface accretion.

Although AGB stars of about 1.3--8 M$_{\sun}$ may be able to produce 
s--process elements, Smith et al. (2000) used the [Rb/Zr] ratio to show that 
AGB stars between $\sim$1.5--3 M$_{\sun}$ were likely the dominant s--process 
enrichment sources in $\omega$ Cen.  Since these stars have lifetimes of 
3$\times$10$^{\rm 8}$--2$\times$10$^{\rm 9}$ years, the time delay between the
formation of RGB--MP stars and subsequent generations had to be at least this 
long.  This delay is consistent with the estimated 2--4 Gyr age range of 
$\omega$ Cen stars (e.g., Stanford et al. 2006), and is also consistent with 
the time required for $>$4--5 M$_{\sun}$ AGB stars to have polluted the ISM, as
seems required to explain at least part of the light element abundance trends.

\begin{figure}
\plotone{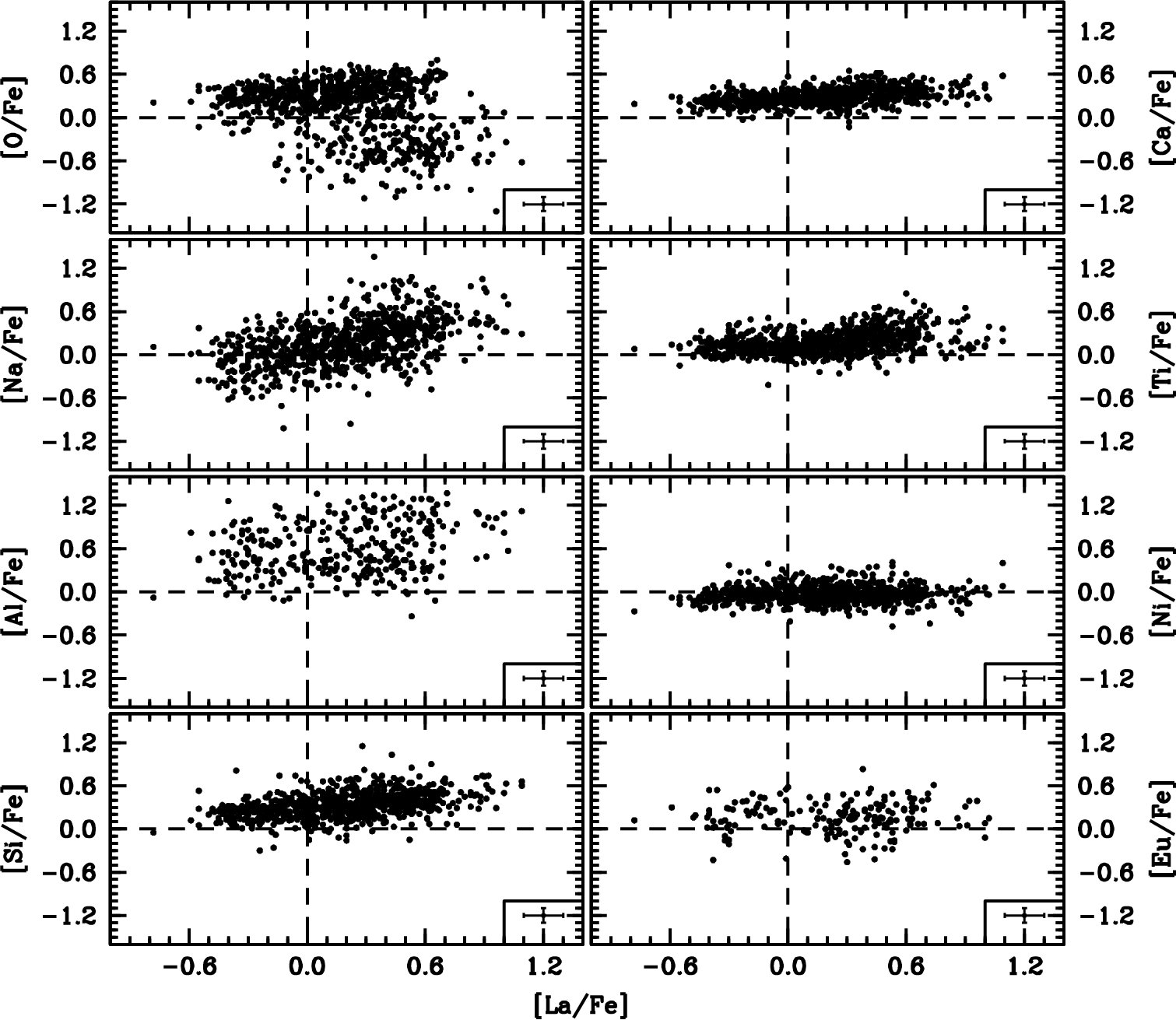}
\caption{A plot of multiple elements as a function of [La/Fe].  The dashed 
lines indicate the solar--scaled abundance values.}
\label{f26}
\end{figure}

In addition to analyzing the behavior of elements produced exclusively through
neutron--capture processes, we can also examine how neutron--capture 
nucleosynthesis may have affected the abundances of lighter elements.  In 
Figure \ref{f26}, we plot multiple elements as a function of lanthanum 
abundance.  As expected, the [Ni/Fe] and [Eu/Fe] ratios do not exhibit any 
correlation with [La/Fe].  This confirms our assumption that europium is 
produced almost exclusively through the r--process, and that nickel, along with
other Fe--peak elements, is not significantly affected by the s--process.  
Additionally, we find that all other elements exhibit a mild correlation with 
[La/Fe].  Given the strong enhancement in lanthanum, it is not surprising that 
the lighter elements might also be mildly affected.  Unfortunately, it is 
difficult to disentangle the production of these elements from other sources.  
We suspect that much of the correlation between the heavy $\alpha$ elements and
lanthanum may be due to the combined effects of Type II SN and AGB s--process 
production overlapping in the same metallicity regime.  In particular, the 
largest increase in [La/Fe] occurs at the transition between the RGB--MP and 
RGB--Int1 populations.  The elevated [Si/Fe] and [Ca/Fe] ratios concurrent
with an increase in [Fe/H] strongly suggests that Type II SNe were the major
producers of these elements.  As mentioned in $\S$5.1, the increase in Si and
Ca abundances may be the result of metallicity dependent Type II SN yields 
rather than additional production from the s--process.  At present, we do not 
have a definitive explanation for the increase in [Ti/Fe] or its correlation 
with [La/Fe].  However, we point out that the stable isotope \iso{50}{Ti} is a 
neutron magic nucleus, and it has been predicted that the helium shell of 
thermally pulsing AGB stars may exhibit a large \iso{50}{Ti}/\iso{48}{Ti} ratio
(e.g., Gallino et al. 1994).  If at least some AGB stars that eject large 
amounts of s--process elements also eject material with a high 
\iso{50}{Ti}/\iso{48}{Ti} ratio, then this may provide an explanation for the 
Ti--La correlation.

It is interesting to note that while the O--rich stars exhibit a correlation 
with [La/Fe], the same relation appears to be mostly absent from the O--poor 
stars.  This supports the idea that the depletion of oxygen is driven by an 
additional process, such as \emph{in situ} mixing, that does not alter the 
[La/Fe] ratio.  Although we find that nearly all of the O--poor stars also 
have [La/Fe]$>$--0.2, we point out that this may be mostly related to the 
fact that the O--poor (He--rich?) stars formed at a time when the average
lanthanum abundance was already becoming significantly enhanced in the cluster 
ISM.  Also note that we do not find any correlation between lanthanum abundance 
and radial location in the cluster, which does not match the observed trend for
the O--poor stars (see Figure \ref{f9}).  We therefore conclude that the 
simultaneous rise in the number of O--poor and La--rich stars are not due to 
the exact same mechanisms.  However, we believe that both phenomena are at 
least in some way related to pollution from low and/or intermediate mass AGB 
stars.

\subsection{Final Remarks}

The data presented here and in previous analyses indicate that $\omega$ Cen
experienced a unique chemical enrichment history.  The occurrence of at least
4--5 discrete star formation episodes spanning $>$1--2$\times$10$^{\rm 9}$ 
years seems required to rectify the breadth of the main sequence turnoff, the
metallicity distribution function, and the large enhancement of s--process
elements.  Despite $\omega$ Cen's rather extensive chemical enrichment, the 
most metal--poor stars ([Fe/H]$<$--2) exhibit abundance patterns and 
star--to--star dispersions that are nearly identical to those found in similar 
metallicity halo and dwarf galaxy stars.  These signatures strongly suggest a 
rapid enrichment time scale in which only massive stars had time to contribute 
to $\omega$ Cen's chemical composition.  Additionally, at least half of the
RGB--MP stars exhibit abundance trends that are consistent with the metal--poor
halo, and the heavy $\alpha$ element trends seem to indicate that the initial
chemical enrichment occurred on a time scale that was sensitive to Type II SNe
of different masses.  However, a significant portion of the RGB--MP stars have
[O/Fe], [Na/Fe], and [Al/Fe] abundances that are unlike stars found in the
halo, and are instead more similar to those found in monometallic globular
clusters.  A clear delay in the presence of O--poor, Na/Al--rich stars until
[Fe/H]$\sim$--1.7 suggests that new generations significantly polluted by the 
ejecta of $\la$8 M$_{\sun}$ stars did not form until about the same time as the
second major episode of star formation.  Furthermore, the neutron--capture data
indicate that at least 1 Gyr had to have elapsed between the formation of the 
RGB--MP and RGB--Int1 populations.  

Since a majority of RGB--MP stars have abundance patterns matching those 
predicted for Type II SN pollution, it seems likely that $\omega$ Cen was 
able to retain and mix a significant percentage of SN ejecta at early times 
in the cluster's evolution.  However, at intermediate metallicities $\omega$ 
Cen's overall chemistry experienced a dramatic shift that strongly deviates
from trends observed in the Galactic halo and most dwarf galaxies.  The 
products of proton-- and neutron--capture nucleosynthesis began to dominate 
the chemical composition of progressively more metal--rich stars, despite
obvious contributions from Type II SNe.  The significant pollution of 
intermediate metallicity stars by O--poor (He--rich?), Na/Al--rich, and 
s--process enhanced gas is undoubtedly the result of the RGB--MP stars evolving
and enriching the cluster ISM.  In order for pollution to occur at the 
levels observed in $\omega$ Cen, the cluster must not have strongly interacted 
with the Galaxy until after at least the formation of the RGB--Int1 population.
Otherwise, it is likely that the gas would have been removed by ram pressure
stripping.  The radial concentration of the RGB--Int2+3 and RGB--a stars near 
the cluster core indicates that enriched gas was funneled toward the cluster 
center and/or the central region was the only location where the escape velocity
was large enough to retain gas ejected by SNe or AGB stars.  The rapid decline
in the relative number of ``primordial" composition stars in the RGB--Int2+3 
and RGB--a populations may be evidence that $\omega$ Cen began to lose mass
at [Fe/H]$\ga$--1.3.  Significant mass loss from the cluster may also help 
explain the minimal impact Type Ia SNe have played in $\omega$ Cen's chemical
enrichment.

\section{SUMMARY}

We have measured chemical abundances of O, Na, Al, Si, Ca, Sc, Ti, Fe, Ni, 
La, and Eu for 855 RGB stars in the globular cluster $\omega$ Cen.  The 
abundances were obtained using moderate resolution (R$\approx$18,000), high
S/N ($>$100) spectra obtained with the Hydra multifiber spectrograph on the 
Blanco 4m telescope at CTIO.  The data set covers more than 80$\%$ of stars 
with V$\leq$13.5, more than 90$\%$ of stars with V$\leq$13.0, and samples the 
full breadth of the giant branch to include the most metal--poor and most 
metal--rich stars in the cluster.  Similarly, we have achieved a completion 
fraction of $\sim$50--100$\%$ at radii extending out to $\sim$24$\arcmin$ from 
the cluster center.  All abundances were determined using either equivalent 
width or spectrum synthesis analyses along with the inclusion of blended 
molecular lines, hyperfine structure, and/or isotope broadening when 
appropriate.  An empirical hyperfine structure correction for the 6774 \AA\ 
La II line is also provided.

We find in agreement with past photometric and spectroscopic studies that 
$\omega$ Cen contains multiple, discrete stellar populations with large
star--to--star abundance variations for all elements.  The metallicity
distribution function contains five peaks centered at [Fe/H]=--1.75,
--1.50, --1.15, --1.05, and --0.75.  However, for the analysis we have combined
the [Fe/H]=--1.15 and --1.05 peaks into a single population.  The (now four)
stellar populations are identified as the RGB--MP ([Fe/H]$\leq$--1.6), 
RGB--Int1 (--1.6$<$[Fe/H]$\leq$--1.3), RGB--Int2+3 (--1.3$<$[Fe/H]$\leq$--0.9),
and RGB--a ([Fe/H]$>$--0.9, which constitute 61$\%$, 27$\%$, 10$\%$, and 2$\%$ 
of our sample, respectively.  The metallicity distribution function also
exhibits a sharp cutoff at the metal--poor end such that only 2$\%$ of the 
stars in our sample have [Fe/H]$<$--2.  The RGB--MP and RGB--Int1 populations 
appear to be uniformly mixed in the cluster, but the RGB--Int2+3 and RGB--a 
stars are preferentially located near the cluster core.  Additionally, almost 
90$\%$ of the most metal--poor stars ([Fe/H]$<$--2) reside within 5$\arcmin$ 
of the cluster center.

The abundance trends exhibited by the heavy $\alpha$ (Si, Ca, and Ti) and 
Fe--peak elements (Sc and Ni) are generally well described by production from 
Type II SNe at all metallicities.  That is, the $\alpha$ elements are 
typically enhanced at [$\alpha$/Fe]$\approx$+0.3 and [Sc,Ni/Fe]$\approx$0.  
While the Fe--peak element [X/Fe] ratios and star--to--star variations remain 
mostly constant over $\omega$ Cen's full metallicity range, the heavy $\alpha$ 
elements show a more complicated morphology.  Over the metallicity range 
spanned by the RGB--MP, there is a noticeable decrease in the average [Si/Fe] 
and [Ca/Fe] abundances with increasing [Fe/H], but the average [Ti/Fe] 
abundance remains essentially constant.  However, the decrease in [Si/Fe] is a
stronger function of [Fe/H] than for [Ca/Fe].  The average
[X/Fe] ratios for all three heavy $\alpha$ elements increase with metallicity
between the RGB--MP and RGB--Int1 populations and remains mostly enhanced at 
higher metallicities.  It seems that many of these abundance trends may be 
driven by mass and/or metallicity dependent Type II SN yields and a new round 
of star formation creating the RGB--Int1 and subsequent populations.  The 
simultaneous rise in [Ti/Fe] at [Fe/H]$\ga$--1.6 may be driven by a different 
production mechanism because theoretical Type II SN yields do not predict a 
large increase in titanium with either progenitor mass or metallicity.

Although some previous analyses have suggested that Type Ia SNe may have become
significant contributors to $\omega$ Cen's chemical enrichment at [Fe/H]$>$--1,
we do not find particularly strong evidence supporting this claim.  In the 
more metal--rich RGB--Int2+3 and RGB--a populations, we find that the average
[$\alpha$/Fe] abundances remain elevated above the level found in disk and 
dwarf galaxy stars of similar metallicity.  While there does appear to be a 
decrease in [Ca/Fe] at [Fe/H]$>$--1, this may be attributed to metallicity
dependent Type II SN yields.  Additionally, the strong rise in the average 
[Na/Fe] ratio for the RGB--Int2+3 and RGB--a stars seems inconsistent with 
Type Ia SNe production.  The maximum [O/Fe] abundance also begins to 
decrease at [Fe/H]$\ga$--1.2; however, this element, as well as Na and Al, may 
be altered by \emph{in situ} mixing or pollution from sources other than Type 
II or Ia SNe.  Therefore, the [X/Fe] ratios for these light elements may not
directly trace SN production or even reflect a star's original composition.
We cannot explicitly rule out that Type Ia SNe have contributed to $\omega$ 
Cen's chemical enrichment, but it seems that their involvement has been mostly
limited.

Unlike the heavy $\alpha$ and Fe--peak elements, the light elements (O, Na, and
Al) exhibit $>$0.5 dex star--to--star abundance variations at all 
metallicities.  Although roughly half of the RGB--MP stars exhibit light 
element abundance patterns that are consistent with those found in similar
metallicity halo and dwarf galaxy stars, the remaining RGB--MP stars, as well
as $>$70$\%$ of more metal--rich stars, show light element abundance patterns
that are more similar to those found in individual globular clusters (i.e.,
O--poor and Na/Al--rich).  Interestingly, the presence of 
these stars is a strong function of metallicity, and the [X/Fe] distribution
functions are bimodal at intermediate metallicities.  While very few 
O--poor, Na/Al--rich stars are found at [Fe/H]$<$--1.7, the majority of stars 
in the RGB--Int1 and subsequent populations exhibit these characteristics.  We 
find that many of the metallicity dependent light element trends can be 
at least qualitatively reproduced by hot bottom burning in intermediate mass
AGB stars.  This is evidenced by the pervasive O--Na and O--Al anticorrelations
and concurrent Na--Al correlation present in all stars with [Fe/H]$\la$--1.
Interestingly, the RGB--a stars no longer exhibit the light element relations
and instead appear to have a roughly uniform composition.  In any case, the
light element trends in stars with [Fe/H]$\la$--1 are similar to what is found 
in monometallic globular clusters, but the relative fraction of O--poor, 
Na/Al--rich stars in $\omega$ Cen at [Fe/H]$>$--1.6 is significantly larger 
than those found in other globular clusters.  Since a wide mass range of AGB
stars seem able to reproduce the observed [Na/Fe] and [Al/Fe] trends but only
a narrow range eject O--poor material, we conclude that the [Na/Fe] and 
[Al/Fe] ratios in the ``enhanced" $\omega$ Cen stars may be explained 
solely by pollution from intermediate mass AGB stars.  However, the strongly
depleted [O/Fe] ratios in many stars appear to require an additional process.
Interestingly, we find a low incidence of carbon stars ($<$2$\%$) in our 
sample despite the large population of O--poor giants.

It seems that some degree of \emph{in situ} processing must be invoked in order
to interpret the large population of O--poor stars.  We find an interesting
parallel between O--poor giants and blue main sequence stars that may explain
at least part of this phenomenon.  The two populations share strikingly
similar radial locations, metallicities, and number fractions.  In particular,
the O--poor and blue main sequence stars are both predominantly found inside 
$\sim$10$\arcmin$ from the cluster center, are mostly found at intermediate 
metallicities, and constitute $\sim$30$\%$ of the RGB and main sequence by 
number.  Since the blue main sequence stars are believed to be He--rich, it 
seems likely that the O--poor stars may also be He--rich.  Previous theoretical
analyses of He--rich, globular cluster RGB stars predict that significant 
\emph{in situ} mixing can occur more easily in He--rich compared to He--normal 
stars.  Furthermore, it is predicted that the surface [O/Fe] abundance may be 
significantly depleted, but [Na/Fe] (and presumably [Al/Fe]) should be mostly
unaffected.  However, this scenario assumes that the O--poor, Na/Al--rich stars
were already polluted by material that was moderately processed by 
proton--capture nucleosynthesis before ascending the RGB.  The observed
O--Si anticorrelation and Al--Si correlation may support this 
scenario.  These relations can naturally arise due to leakage from the MgAl 
cycle at temperatures exceeding $\sim$65$\times$10$^{\rm 6}$ K; temperatures 
this high are achieved in hot bottom burning conditions but not in the 
interiors of low mass RGB stars.  If we assume that the observed O--poor stars 
are also He--rich and therefore more apt to experience \emph{in situ} deep 
mixing, then this may explain why only the [O/Fe] ratio is correlated with 
radial location.  Since the Na and Al abundances do not strongly correlate 
with radial location like O, this may be an indication that the stars 
responsible for producing the high Na and Al abundances do not necessarily 
produce high He yields as well.  

A majority of RGB--MP stars have [La/Fe], [Eu/Fe], and [La/Eu] ratios 
indicating that the r--process was the primary production mechanism early
in $\omega$ Cen's history.  This is similar to what is found in metal--poor 
halo, globular cluster, and dwarf galaxy stars, and is consistent with a 
rapid formation time scale of the RGB--MP population.  However, the 
[La/Fe], [Eu/Fe], and [La/Eu] abundance patterns indicate that the s--process 
became the dominant neutron--capture production mechanism at [Fe/H]$>$--1.6,
and was active at a level above that observed in any other stellar population 
to date.  In fact, almost no stars with [Fe/H]$>$--1.6 have [La/Fe]$<$0, and 
many stars in the intermediate metallicity populations exhibit [La/Eu] ratios 
suggesting pure s--process production.  However, proper accounting of hyperfine
structure in determining both La and Eu abundances has revised our [La/Eu] 
ratios downward from those in Johnson et al. (2009), and we now find that 
surface pollution from mass transfer is not generally required to explain the 
stars with large [La/Eu] ratios.  Interestingly, the typical [Eu/Fe] abundances
in the RGB--Int1 and RGB--Int2+3 stars are well below those observed in 
similar metallicity halo and globular cluster stars.  This suggests that 
typical r--process production mechanisms may have been suppressed in 
$\omega$ Cen.

While we find that both the [Ni/Fe] and [Eu/Fe] ratios are independent of a
star's [La/Fe] abundance, all other elements exhibit a mild correlation with
[La/Fe].  Since the $<$3 M$_{\sun}$ AGB stars believed to produce most of the 
s--process elements in $\omega$ Cen are also predicted to produce some light
elements, the correlation with La is not entirely unexpected.  Interestingly,
the O--rich stars show a correlation with [La/Fe], but the O--poor stars 
do not.  This suggests that the O--depletion phenomenon is driven by an 
additional process, such as \emph{in situ} mixing, that does not alter the 
envelope [La/Fe] ratio.  With regard to the heavy $\alpha$ elements, we 
suspect that the correlation with La may be due to the combined effects of 
Type II SNe producing $\alpha$ elements and low/intermediate--mass AGB stars 
producing s--process elements at approximately the same time.  This is 
supported by the observation that the rise in s--process and $\alpha$ elements 
occurs in the same metallicity range.  We do not have a definitive explanation
for the correlation between [Ti/Fe] and [La/Fe] because the [Ti/Fe] ratio is
not believed to be significantly enhanced in Type II SNe.  However, we point 
out that the stable isotope \iso{50}{Ti} is a neutron magic nucleus and that
the He shell of thermally pulsing AGB stars are predicted to exhibit large 
\iso{50}{Ti}/\iso{48}{Ti} ratios.  Therefore, if at least some AGB stars 
that eject large amounts of s--process elements also eject material with a 
high \iso{50}{Ti}/\iso{48}{Ti} ratio, this may explain both the rise [Ti/Fe]
at [Fe/H]$\ga$--1.6 and the Ti--La correlation.

\acknowledgments

This publication makes use of data products from the Two Micron All Sky Survey,
which is a joint project of the University of Massachusetts and the Infrared 
Processing and Analysis Center/California Institute of Technology, funded by 
the National Aeronautics and Space Administration and the National Science 
Foundation. This research has made use of NASA's Astrophysics Data System 
Bibliographic Services.  Support of the College of Arts and Sciences and the 
Daniel Kirkwood fund at Indiana University Bloomington for CIJ and CAP is 
gratefully acknowledged.  We would like to thank Bob Kraft and Chris Sneden 
for many helpful discussions, Katia Cunha for sending an electronic version of
her paper in advance of publication, and TalaWanda Monroe for her assistance in 
obtaining these observations.  We would also like to thank Frank and Janet 
Winkler and the CTIO staff for their generous hospitality.  We also thank the 
referee for his/her careful reading and thoughtful comments that led to 
improvement of the manuscript.

\appendix

\section{Empirical Lanthanum 6774 \AA\ Hyperfine Broadening Correction}

\begin{figure}
\plotone{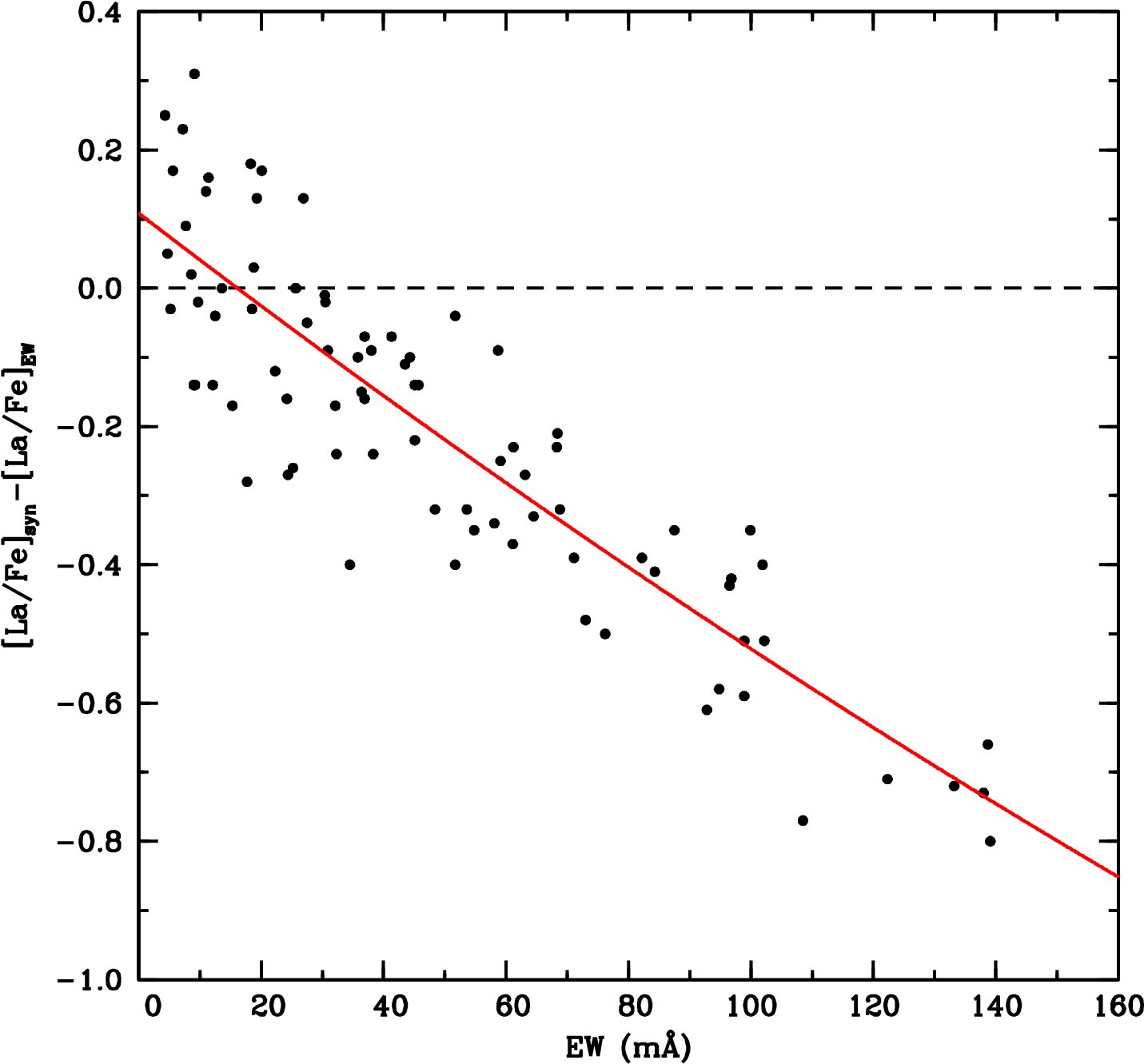}
\caption{A plot of the difference between the [La/Fe] abundances derived using
an EW approach for the 6774 \AA\ La II line and a spectrum synthesis approach
for the 6262 \AA\ La II line.  The solid red line shows the least--squares 
fit to the data, and the horizontal dashed line indicates perfect agreement.}
\label{fappend1}
\end{figure}

The 6774 \AA\ La II line is often measurable in the spectra of [Fe/H]$\ga$--2
RGB and AGB stars, but accurate La abundance determinations from this line
can be hampered by hyperfine broadening if the EW exceeds $\sim$50 m\AA.  
Unfortunately, we are not aware of any publicly available linelists that 
include log gf values for the individual hyperfine components of the 
6774 \AA\ line.  However, the spectra used here and in Johnson et al. (2009)
provide EW measurements of the 6774 \AA\ line and spectrum synthesis
abundance determinations from the 6262 \AA\ line in 85 giants.  Since the 
6262 \AA\ abundance determinations properly account for hyperfine broadening,
we can use these data to derive an empirical correction factor for EW--based
abundance measurements that use the 6774 \AA\ line.  In Figure \ref{fappend1},
we plot [La/Fe]$_{\rm syn}$--[La/Fe]$_{\rm EW}$ ($\Delta$[La/Fe]$_{\rm EW}$) as 
a function of EW.  The least--squares fit to the data gives the empirical 
correction factor as,
\begin{equation}
\Delta[La/Fe]_{EW}=[(5.0\times10^{-6})(EW^{2})]-[(0.0068)(EW)]+0.1084 
(\sigma=0.07),
\end{equation}
where the EW is measured in units of m\AA.  This relation is qualitatively
expected because it shows that a straight--forward EW analysis will
overestimate the La abundance by an increasingly larger amount as one moves
up the curve--of--growth to larger EWs.  Since this is an empirical correction,
it is difficult to predict how the relation might change outside the 
T$_{\rm eff}$ (3800--5000 K), log g ($\la$2), and metallicity 
(--2.5$\la$[Fe/H]$\la$--0.5) regime of our sample.

{\it Facilities:} \facility{CTIO}

\clearpage

\clearpage

\clearpage

\clearpage

\clearpage

\clearpage

\clearpage

\clearpage

\clearpage

\clearpage

\clearpage

\clearpage

\clearpage

\clearpage

\clearpage

\clearpage

\clearpage

\clearpage

\clearpage

\clearpage

\clearpage

\clearpage

\clearpage

\clearpage

\clearpage

\clearpage

\clearpage

\clearpage

\tablenum{1}
\tablecolumns{5}
\tablewidth{0pt}

\begin{deluxetable}{ccccc}
\tablecaption{Hydra Observation Log}
\tablehead{
\colhead{Field}         &
\colhead{UT Date}   &
\colhead{Wavelength Center}       &
\colhead{Filter}     &
\colhead{Exposure}	\\
\colhead{}	&
\colhead{}      &
\colhead{(\AA)}      &
\colhead{}      &
\colhead{(s)} 
}
\startdata
1       &       2008 March 25   &       6250    &       E6257   &       3$\times$3600   \\
2	&	2008 March 25   &       6250    &       E6257   &       2$\times$3600   \\
\nodata	&	2008 March 29   &       6250    &       E6257   &       1$\times$2200   \\
3       &       2008 March 26   &       6250    &       E6257   &       3$\times$3600   \\
4       &       2008 March 26   &       6250    &       E6257   &       3$\times$3600   \\
5       &       2008 March 26   &       6250    &       E6257   &       1$\times$3600   \\
\nodata	&	2008 March 26   &       6250    &       E6257   &       1$\times$1800   \\
\nodata &       2008 March 27   &       6250    &       E6257   &       2$\times$2700   \\
\enddata

\end{deluxetable}

\clearpage

\tablenum{2}
\tablecolumns{13}
\tablewidth{0pt}

\begin{deluxetable}{ccccccccccccc}
\tabletypesize{\scriptsize}
\tablecaption{Star Identifiers, Photometry, and Atmospheric Parameters}
\tablehead{
\colhead{Star}         &
\colhead{Alt. ID}         &
\colhead{V}         &
\colhead{B--V}         &
\colhead{J}         &
\colhead{H}         &
\colhead{K$_{\rm s}$}         &
\colhead{T$_{\rm eff}$}         &
\colhead{log g}         &
\colhead{[Fe/H]}         &
\colhead{v$_{\rm t}$}         &
\colhead{S/N}	&   
\colhead{S/N}	\\
\colhead{LEID\tablenotemark{a}}	&
\colhead{ROA\tablenotemark{b}}      &
\colhead{}      &
\colhead{}      &
\colhead{}      &
\colhead{}      &
\colhead{}      &
\colhead{(K)}      &
\colhead{(cgs)}      &
\colhead{Avg.}      &
\colhead{(km s$^{\rm -1}$)}      &
\colhead{6250 \AA}      &
\colhead{6650 \AA} 
}
\startdata
9	&	370	&	12.529	&	1.250	&	10.382	&	9.755	&	9.627	&	4500	&	1.20	&	$-$1.35	&	1.95	&	250	&	\nodata	\\
5009	&	548	&	12.912	&	1.080	&	10.752	&	10.185	&	10.040	&	4525	&	1.25	&	$-$1.86	&	1.75	&	200	&	\nodata	\\
6017	&	240	&	12.233	&	1.420	&	9.717	&	8.982	&	8.808	&	4145	&	0.85	&	$-$1.19	&	1.90	&	200	&	\nodata	\\
8014	&	6734	&	13.365	&	1.013	&	11.353	&	10.781	&	10.705	&	4700	&	1.75	&	$-$1.73	&	1.60	&	200	&	\nodata	\\
9013	&	6771	&	13.269	&	0.973	&	11.261	&	10.717	&	10.603	&	4700	&	1.50	&	$-$1.67	&	1.80	&	200	&	150	\\
10009	&	\nodata	&	12.573	&	1.234	&	10.331	&	9.664	&	9.520	&	4390	&	1.15	&	$-$1.43	&	1.70	&	150	&	\nodata	\\
10012	&	43	&	11.529	&	1.618	&	8.684	&	7.895	&	7.706	&	3930	&	0.25	&	$-$1.44	&	2.10	&	200	&	\nodata	\\
11019	&	537	&	12.841	&	1.223	&	10.595	&	9.952	&	9.823	&	4405	&	1.25	&	$-$1.54	&	1.65	&	225	&	\nodata	\\
\enddata

\tablecomments{The possible carbon stars shown in Figures \ref{f1} and \ref{f3}
are: LEID 32059, 33062, 35250, 37199, 39105, 41071, 42044, 43199, 43368, 44126,
44262, 44361, 44484, and 52030.  Coordinates and photometry for these stars
can be found in van Leeuwen et al. (2000).}
\tablenotetext{a}{Identifier from van Leeuwen et al. (2000).}
\tablenotetext{b}{Identifier from Woolley (1966).}

\end{deluxetable}

\clearpage
\pagestyle{empty}
\setlength{\voffset}{1in}

\tablenum{3a}
\tablecolumns{15}
\tablewidth{0pt}

\begin{deluxetable}{ccccccccccccccc}
\rotate
\tabletypesize{\scriptsize}
\tablecaption{Fe Atomic Parameters, Equivalent Widths, and Solar Abundances}
\tablehead{
\colhead{Wavelength (\AA)}	&
\colhead{6151.62}       &
\colhead{6157.73}       &
\colhead{6165.36}       &
\colhead{6173.34}       &
\colhead{6180.20}       &
\colhead{6187.99}       &
\colhead{6200.32}       &
\colhead{6219.28}       &
\colhead{6226.74}       &
\colhead{6229.23}       &
\colhead{6232.64}       &
\colhead{6246.32}       &
\colhead{6252.56}       &
\colhead{6265.14} \\
\colhead{Ion}  &
\colhead{Fe I}  &
\colhead{Fe I}  &
\colhead{Fe I}  &
\colhead{Fe I}  &
\colhead{Fe I}  &
\colhead{Fe I}  &
\colhead{Fe I}  &
\colhead{Fe I}  &
\colhead{Fe I}  &
\colhead{Fe I}  &
\colhead{Fe I}  &
\colhead{Fe I}  &
\colhead{Fe I}	&
\colhead{Fe I}	\\
\colhead{E.P. (eV)}  &
\colhead{2.18}  &
\colhead{4.07}  &
\colhead{4.14}  &
\colhead{2.22}  &
\colhead{2.73}  &
\colhead{3.94}  &
\colhead{2.61}  &
\colhead{2.20}  &
\colhead{3.88}  &
\colhead{2.84}  &
\colhead{3.65}  &
\colhead{3.60}  &
\colhead{2.40}	&
\colhead{2.18}	\\
\colhead{Log gf}	&
\colhead{$-$3.33}        &
\colhead{$-$1.22}        &
\colhead{$-$1.51}        &
\colhead{$-$2.89}        &
\colhead{$-$2.66}        &
\colhead{$-$1.69}        &
\colhead{$-$2.41}        &
\colhead{$-$2.42}        &
\colhead{$-$2.19}        &
\colhead{$-$3.00}        &
\colhead{$-$1.23}        &
\colhead{$-$0.85}        &
\colhead{$-$1.71}        &
\colhead{$-$2.56}	\\
\colhead{Log $\epsilon$$_{\rm \odot}$}	&
\colhead{7.52}  &
\colhead{7.52}  &
\colhead{7.52}  &
\colhead{7.52}  &
\colhead{7.52}  &
\colhead{7.52}  &
\colhead{7.52}  &
\colhead{7.52}  &
\colhead{7.52}  &
\colhead{7.52}  &
\colhead{7.52}  &
\colhead{7.52}  &
\colhead{7.52}  &
\colhead{7.52}	\\
\hline
\colhead{Star (LEID)}	&
\multicolumn{14}{c}{Equivalent Widths (m\AA)}
}
\startdata
9	&	73	&	55	&	35	&	107	&	68	&	\nodata	&	99	&	136	&	14	&	37	&	81	&	110	&	162	&	137	\\
5009	&	36	&	22	&	16	&	59	&	30	&	10	&	58	&	90	&	\nodata	&	23	&	45	&	77	&	116	&	95	\\
6017	&	103	&	\nodata	&	46	&	133	&	100	&	46	&	123	&	158	&	\nodata	&	69	&	113	&	130	&	\nodata	&	\nodata	\\
8014	&	30	&	27	&	11	&	51	&	28	&	17	&	58	&	87	&	\nodata	&	18	&	45	&	67	&	113	&	85	\\
9013	&	29	&	26	&	12	&	48	&	27	&	8	&	56	&	88	&	\nodata	&	\nodata	&	40	&	74	&	110	&	88	\\
10009	&	66	&	\nodata	&	31	&	103	&	61	&	31	&	95	&	123	&	15	&	40	&	78	&	107	&	147	&	\nodata	\\
10012	&	119	&	80	&	42	&	148	&	105	&	48	&	140	&	183	&	26	&	81	&	106	&	142	&	192	&	165	\\
11019	&	64	&	55	&	26	&	92	&	59	&	28	&	91	&	119	&	19	&	39	&	73	&	100	&	147	&	123	\\
\enddata

\end{deluxetable}

\clearpage
\pagestyle{empty}
\setlength{\voffset}{1in}

\tablenum{3b}
\tablecolumns{15}
\tablewidth{0pt}

\begin{deluxetable}{ccccccccccccccc}
\rotate
\tabletypesize{\scriptsize}
\tablecaption{Fe Atomic Parameters, Equivalent Widths, and Solar Abundances}
\tablehead{
\colhead{Wavelength (\AA)}	&
\colhead{6270.23}       &
\colhead{6271.28}       &
\colhead{6290.97}       &
\colhead{6297.79}       &
\colhead{6301.50}       &
\colhead{6302.49}       &
\colhead{6311.50}       &
\colhead{6315.81}       &
\colhead{6322.69}       &
\colhead{6335.33}       &
\colhead{6336.83}       &
\colhead{6344.15}       &
\colhead{6149.24}	&
\colhead{6247.56}	\\
\colhead{Ion}  &
\colhead{Fe I}  &
\colhead{Fe I}  &
\colhead{Fe I}  &
\colhead{Fe I}  &
\colhead{Fe I}  &
\colhead{Fe I}  &
\colhead{Fe I}  &
\colhead{Fe I}  &
\colhead{Fe I}  &
\colhead{Fe I}  &
\colhead{Fe I}  &
\colhead{Fe I}  &
\colhead{Fe II}	&
\colhead{Fe II}	\\
\colhead{E.P. (eV)}  &
\colhead{2.86}  &
\colhead{3.33}  &
\colhead{4.73}  &
\colhead{2.22}  &
\colhead{3.65}  &
\colhead{3.69}  &
\colhead{2.83}  &
\colhead{4.07}  &
\colhead{2.59}  &
\colhead{2.20}  &
\colhead{3.69}  &
\colhead{2.43}  &
\colhead{3.89}	&
\colhead{3.89}	\\
\colhead{Log gf}	&
\colhead{$-$2.62} &
\colhead{$-$2.74} &
\colhead{$-$0.72} &
\colhead{$-$2.74} &
\colhead{$-$0.68} &
\colhead{$-$1.11} &
\colhead{$-$3.17} &
\colhead{$-$1.69} &
\colhead{$-$2.41} &
\colhead{$-$2.22} &
\colhead{$-$0.78} &
\colhead{$-$2.92} &
\colhead{$-$2.78}	&
\colhead{$-$2.43}	\\
\colhead{Log $\epsilon$$_{\rm \odot}$}	&
\colhead{7.52}  &
\colhead{7.52}  &
\colhead{7.52}  &
\colhead{7.52}  &
\colhead{7.52}  &
\colhead{7.52}  &
\colhead{7.52}  &
\colhead{7.52}  &
\colhead{7.52}  &
\colhead{7.52}  &
\colhead{7.52}  &
\colhead{7.52}  &
\colhead{7.52}  &
\colhead{7.52}	\\
\hline
\colhead{Star (LEID)}	&
\multicolumn{14}{c}{Equivalent Widths (m\AA)}
}
\startdata
9	&	67	&	\nodata	&	\nodata	&	109	&	\nodata	&	\nodata	&	42	&	29	&	103	&	156	&	109	&	88	&	30	&	49	\\
5009	&	34	&	\nodata	&	\nodata	&	81	&	81	&	44	&	\nodata	&	11	&	57	&	101	&	66	&	43	&	13	&	29	\\
6017	&	101	&	44	&	\nodata	&	142	&	135	&	110	&	\nodata	&	40	&	131	&	172	&	136	&	\nodata	&	34	&	40	\\
8014	&	\nodata	&	\nodata	&	\nodata	&	72	&	81	&	45	&	17	&	9	&	62	&	97	&	63	&	48	&	9	&	29	\\
9013	&	26	&	\nodata	&	\nodata	&	68	&	72	&	37	&	14	&	9	&	55	&	101	&	63	&	39	&	18	&	45	\\
10009	&	68	&	23	&	\nodata	&	111	&	106	&	\nodata	&	34	&	24	&	87	&	134	&	102	&	91	&	24	&	41	\\
10012	&	102	&	31	&	40	&	\nodata	&	132	&	115	&	71	&	42	&	137	&	185	&	129	&	\nodata	&	\nodata	&	35	\\
11019	&	57	&	\nodata	&	\nodata	&	108	&	102	&	71	&	39	&	23	&	90	&	131	&	94	&	85	&	17	&	34	\\
\enddata

\end{deluxetable}

\clearpage
\pagestyle{empty}
\setlength{\voffset}{1in}

\tablenum{4a}
\tablecolumns{15}
\tablewidth{0pt}

\begin{deluxetable}{ccccccccccccccc}
\rotate
\tabletypesize{\scriptsize}
\tablecaption{O--Eu Atomic Parameters, Equivalent Widths, and Solar Abundances}
\tablehead{
\colhead{Wavelength (\AA)}	&
\colhead{6300.31}	&
\colhead{6154.23}       &
\colhead{6160.75}       &
\colhead{6696.03}       &
\colhead{6698.67}       &
\colhead{6155.13}       &
\colhead{6237.32}       &
\colhead{6161.29}       &
\colhead{6162.18}       &
\colhead{6166.44}       &
\colhead{6169.04}       &
\colhead{6169.56}       &
\colhead{6210.67}	&
\colhead{6305.66}	\\
\colhead{Ion}  &
\colhead{[O I]}  &
\colhead{Na I}  &
\colhead{Na I}  &
\colhead{Al I}  &
\colhead{Al I}  &
\colhead{Si I}  &
\colhead{Si I}  &
\colhead{Ca I}  &
\colhead{Ca I}  &
\colhead{Ca I}  &
\colhead{Ca I}  &
\colhead{Ca I}  &
\colhead{Sc I}	&
\colhead{Sc I}	\\
\colhead{E.P. (eV)}  &
\colhead{0.00}  &
\colhead{2.10}  &
\colhead{2.10}  &
\colhead{3.14}  &
\colhead{3.14}  &
\colhead{5.62}  &
\colhead{5.61}  &
\colhead{2.52}  &
\colhead{1.90}  &
\colhead{2.52}  &
\colhead{2.52}  &
\colhead{2.52}  &
\colhead{0.00}	&
\colhead{0.02}	\\
\colhead{Log gf}	&
\colhead{$-$9.75}       &
\colhead{$-$1.57}       &
\colhead{$-$1.27}       &
\colhead{$-$1.57}       &
\colhead{$-$1.89}       &
\colhead{$-$0.78}       &
\colhead{$-$1.07}       &
\colhead{$-$1.28}       &
\colhead{$-$0.07}       &
\colhead{$-$1.11}       &
\colhead{$-$0.69}       &
\colhead{$-$0.42}       &
\colhead{$-$1.53}	&
\colhead{$-$1.30}	\\
\colhead{Log $\epsilon$$_{\rm \odot}$}	&
\colhead{8.93}  &
\colhead{6.33}  &
\colhead{6.33}  &
\colhead{6.47}  &
\colhead{6.47}  &
\colhead{7.55}  &
\colhead{7.55}  &
\colhead{6.36}  &
\colhead{6.36}  &
\colhead{6.36}  &
\colhead{6.36}  &
\colhead{6.36}  &
\colhead{3.10}	&
\colhead{3.10}	\\
\hline
\colhead{Star (LEID)}	&
\multicolumn{14}{c}{Equivalent Widths (m\AA)}
}
\startdata
9	&	syn\tablenotemark{a}	&	5	&	14	&	\nodata	&	\nodata	&	37	&	36	&	59	&	202	&	63	&	88	&	110	&	\nodata	&	\nodata	\\
5009	&	syn	&	\nodata	&	3	&	\nodata	&	\nodata	&	22	&	9	&	35	&	150	&	33	&	48	&	78	&	3	&	\nodata	\\
6017	&	syn	&	52	&	80	&	\nodata	&	\nodata	&	51	&	\nodata	&	109	&	248	&	101	&	129	&	144	&	\nodata	&	\nodata	\\
8014	&	syn	&	\nodata	&	14	&	\nodata	&	\nodata	&	20	&	18	&	28	&	148	&	35	&	56	&	68	&	\nodata	&	\nodata	\\
9013	&	syn	&	\nodata	&	7	&	syn	&	\nodata	&	26	&	20	&	28	&	\nodata	&	33	&	57	&	68	&	\nodata	&	\nodata	\\
10009	&	syn	&	24	&	34	&	\nodata	&	\nodata	&	38	&	40	&	74	&	197	&	70	&	93	&	116	&	16	&	\nodata	\\
10012	&	syn	&	31	&	51	&	\nodata	&	\nodata	&	30	&	33	&	100	&	254	&	107	&	140	&	144	&	62	&	121	\\
11019	&	syn	&	9	&	16	&	\nodata	&	\nodata	&	34	&	28	&	69	&	189	&	69	&	92	&	110	&	10	&	\nodata	\\
\enddata

\tablenotetext{a}{Lines displaying the ``syn" designation indicate those that
were determined via spectrum synthesis instead of equivalent width 
measurements.}

\end{deluxetable}

\clearpage
\pagestyle{empty}
\setlength{\voffset}{1in}

\tablenum{4b}
\tablecolumns{16}
\tablewidth{0pt}

\begin{deluxetable}{cccccccccccccccc}
\rotate
\tabletypesize{\scriptsize}
\tablecaption{O--Eu Atomic Parameters, Equivalent Widths, and Solar Abundances}
\tablehead{
\colhead{Wavelength (\AA)}	&
\colhead{6245.62}       &
\colhead{6309.88}       &
\colhead{6146.23}       &
\colhead{6258.11}       &
\colhead{6261.10}       &
\colhead{6303.76}       &
\colhead{6312.24}       &
\colhead{6336.11}       &
\colhead{6175.37}       &
\colhead{6176.81}       &
\colhead{6177.25}       &
\colhead{6186.71}       &
\colhead{6327.60}       &
\colhead{6262.29}	&
\colhead{6645.06}	\\
\colhead{Ion}  &
\colhead{Sc II} &
\colhead{Sc II} &
\colhead{Ti I} &
\colhead{Ti I} &
\colhead{Ti I} &
\colhead{Ti I} &
\colhead{Ti I} &
\colhead{Ti I} &
\colhead{Ni I} &
\colhead{Ni I} &
\colhead{Ni I} &
\colhead{Ni I} &
\colhead{Ni I} &
\colhead{La II}	&
\colhead{Eu II}	\\
\colhead{E.P. (eV)}  &
\colhead{1.51}  &
\colhead{1.50}  &
\colhead{1.87}  &
\colhead{1.44}  &
\colhead{1.43}  &
\colhead{1.44}  &
\colhead{1.46}  &
\colhead{1.44}  &
\colhead{4.09}  &
\colhead{4.09}  &
\colhead{1.83}  &
\colhead{4.11}  &
\colhead{1.68}  &
\colhead{0.40}	&
\colhead{1.38}	\\
\colhead{Log gf}	&
\colhead{$-$1.07} &
\colhead{$-$1.58} &
\colhead{$-$1.51} &
\colhead{$-$0.38} &
\colhead{$-$0.49} &
\colhead{$-$1.56} &
\colhead{$-$1.55} &
\colhead{$-$1.69} &
\colhead{$-$0.55} &
\colhead{$-$0.42} &
\colhead{$-$3.53} &
\colhead{$-$0.96} &
\colhead{$-$3.14} &
\colhead{$-$1.22}	&
\colhead{$+$0.12}	\\
\colhead{Log $\epsilon$$_{\rm \odot}$}	&
\colhead{3.10}  &
\colhead{3.10}  &
\colhead{4.99}  &
\colhead{4.99}  &
\colhead{4.99}  &
\colhead{4.99}  &
\colhead{4.99}  &
\colhead{4.99}  &
\colhead{6.25}  &
\colhead{6.25}  &
\colhead{6.25}  &
\colhead{6.25}  &
\colhead{6.25}  &
\colhead{1.22}	&
\colhead{0.51}	\\
\hline
\colhead{Star (LEID)}	&
\multicolumn{15}{c}{Equivalent Widths (m\AA)}
}
\startdata
9	&	\nodata	&	18	&	\nodata	&	83	&	72	&	\nodata	&	16	&	\nodata	&	27	&	35	&	21	&	9	&	71	&	syn	&	\nodata	\\
5009	&	45	&	18	&	\nodata	&	40	&	35	&	\nodata	&	\nodata	&	4	&	\nodata	&	19	&	5	&	\nodata	&	26	&	syn	&	\nodata	\\
6017	&	\nodata	&	\nodata	&	\nodata	&	151	&	135	&	61	&	49	&	50	&	35	&	52	&	38	&	10	&	99	&	syn	&	\nodata	\\
8014	&	38	&	\nodata	&	\nodata	&	38	&	29	&	\nodata	&	7	&	\nodata	&	14	&	13	&	\nodata	&	\nodata	&	25	&	syn	&	\nodata	\\
9013	&	36	&	\nodata	&	\nodata	&	36	&	25	&	\nodata	&	\nodata	&	\nodata	&	12	&	16	&	\nodata	&	\nodata	&	29	&	syn	&	9	\\
10009	&	64	&	16	&	6	&	96	&	86	&	25	&	12	&	12	&	\nodata	&	36	&	13	&	\nodata	&	65	&	syn	&	\nodata	\\
10012	&	\nodata	&	36	&	\nodata	&	\nodata	&	\nodata	&	72	&	65	&	60	&	27	&	\nodata	&	\nodata	&	16	&	115	&	syn	&	\nodata	\\
11019	&	54	&	\nodata	&	\nodata	&	88	&	81	&	29	&	21	&	13	&	23	&	36	&	16	&	9	&	60	&	syn	&	\nodata	\\
\enddata

\end{deluxetable}

\clearpage
\pagestyle{empty}

\tablenum{5}
\tablecolumns{18}
\tablewidth{0pt}

\begin{deluxetable}{cccccccccccccccccc}
\rotate
\tabletypesize{\scriptsize}
\tablecaption{Chemical Abundance Ratios}
\tablehead{
\colhead{Star}	&
\colhead{Alt. ID}         &
\colhead{[FeI/H]}	&
\colhead{[FeII/H]}       &
\colhead{[Fe/H]}       &
\colhead{[O/Fe]}       &
\colhead{[Na/Fe]}       &
\colhead{[Al/Fe]}       &
\colhead{[Si/Fe]}       &
\colhead{[Ca/Fe]}       &
\colhead{[ScI/Fe]}       &
\colhead{[ScII/Fe]}       &
\colhead{[Sc/Fe]}       &
\colhead{[Ti/Fe]}       &
\colhead{[Ni/Fe]}	&
\colhead{[LaII/Fe]}	&
\colhead{[EuII/Fe]}	&
\colhead{Code\tablenotemark{a}}	\\
\colhead{LEID}	&
\colhead{ROA}	&
\colhead{}      &
\colhead{}      &
\colhead{Avg.}      &
\colhead{}      &
\colhead{}      &
\colhead{}      &
\colhead{}      &
\colhead{}      &
\colhead{}      &
\colhead{}      &
\colhead{Avg.}      &
\colhead{}      &
\colhead{}      &
\colhead{}      &
\colhead{}      &
\colhead{}
}
\startdata
9	&	370	&	$-$1.34	&	$-$1.35	&	$-$1.35	&	$+$0.57	&	$-$0.38	&	\nodata	&	$+$0.23	&	$+$0.19	&	\nodata	&	$-$0.33	&	$-$0.33	&	$+$0.11	&	$-$0.12	&	$-$0.01	&	\nodata	&	\nodata	\\
5009	&	548	&	$-$1.87	&	$-$1.84	&	$-$1.86	&	$+$0.51	&	$-$0.42	&	\nodata	&	$+$0.20	&	$+$0.30	&	$-$0.06	&	$+$0.12	&	$+$0.03	&	$+$0.14	&	$-$0.08	&	$+$0.16	&	\nodata	&	\nodata	\\
6017	&	240	&	$-$1.32	&	$-$1.05	&	$-$1.19	&	$-$0.77	&	$+$0.37	&	\nodata	&	$+$0.31	&	$+$0.23	&	\nodata	&	\nodata	&	\nodata	&	$+$0.11	&	$-$0.24	&	$+$0.54	&	\nodata	&	1	\\
8014	&	6734	&	$-$1.65	&	$-$1.80	&	$-$1.73	&	$+$0.61	&	$+$0.23	&	\nodata	&	$+$0.30	&	$+$0.32	&	\nodata	&	$+$0.15	&	$+$0.15	&	$+$0.23	&	$-$0.02	&	$+$0.18	&	\nodata	&	\nodata	\\
9013	&	6771	&	$-$1.74	&	$-$1.60	&	$-$1.67	&	$+$0.56	&	$-$0.10	&	$+$0.52	&	$+$0.34	&	$+$0.18	&	\nodata	&	$-$0.06	&	$-$0.06	&	$+$0.04	&	$+$0.01	&	$+$0.42	&	$+$0.30	&	\nodata	\\
10009	&	\nodata	&	$-$1.48	&	$-$1.38	&	$-$1.43	&	$-$0.32	&	$+$0.26	&	\nodata	&	$+$0.40	&	$+$0.37	&	$-$0.09	&	$-$0.09	&	$-$0.09	&	$+$0.16	&	$-$0.14	&	$+$0.45	&	\nodata	&	\nodata	\\
10012	&	43	&	$-$1.53	&	$-$1.35	&	$-$1.44	&	$+$0.24	&	$+$0.11	&	\nodata	&	$+$0.33	&	$+$0.19	&	$-$0.29	&	$-$0.11	&	$-$0.20	&	$+$0.10	&	$-$0.10	&	$+$0.09	&	\nodata	&	\nodata	\\
11019	&	537	&	$-$1.50	&	$-$1.57	&	$-$1.54	&	$+$0.49	&	$-$0.09	&	\nodata	&	$+$0.35	&	$+$0.44	&	$-$0.19	&	$+$0.13	&	$-$0.03	&	$+$0.36	&	$-$0.04	&	$+$0.46	&	\nodata	&	\nodata	\\
\enddata

\tablenotetext{a}{1 = [O/Fe] upper limit; 2 = [Na/Fe] upper limit; 
3 = [Al/Fe] upper limit; 4 = [La/Fe] upper limit.}

\end{deluxetable}

\clearpage
\pagestyle{empty}

\tablenum{6a}
\tablecolumns{11}
\tablewidth{0pt}

\begin{deluxetable}{ccccccccccc}
\tabletypesize{\scriptsize}
\tablecaption{Chemical Abundance Measurement Uncertainties}
\tablehead{
\colhead{Star}	&
\colhead{N--FeI}	&
\colhead{$\sigma$/$\surd$(N)}	&
\colhead{N--FeII}        &
\colhead{$\sigma$/$\surd$(N)}   &
\colhead{N--Na}        &
\colhead{$\sigma$/$\surd$(N)}   &
\colhead{N--Al}        &
\colhead{$\sigma$/$\surd$(N)}   &
\colhead{N--Si}        &
\colhead{$\sigma$/$\surd$(N)} 
}
\startdata
9	&	21	&	0.02	&	2	&	0.02	&	2	&	0.07	&	\nodata	&	\nodata	&	2	&	0.12	\\
5009	&	22	&	0.02	&	2	&	0.06	&	1	&	\nodata	&	\nodata	&	\nodata	&	2	&	0.12	\\
6017	&	19	&	0.02	&	2	&	0.11	&	2	&	0.04	&	\nodata	&	\nodata	&	1	&	\nodata	\\
8014	&	22	&	0.03	&	2	&	0.14	&	1	&	\nodata	&	\nodata	&	\nodata	&	2	&	0.09	\\
9013	&	22	&	0.03	&	2	&	0.13	&	1	&	\nodata	&	1	&	\nodata	&	2	&	0.06	\\
10009	&	22	&	0.02	&	2	&	0.02	&	2	&	0.05	&	\nodata	&	\nodata	&	2	&	0.16	\\
10012	&	24	&	0.02	&	1	&	\nodata	&	2	&	0.00	&	\nodata	&	\nodata	&	2	&	0.17	\\
11019	&	24	&	0.02	&	2	&	0.06	&	2	&	0.01	&	\nodata	&	\nodata	&	2	&	0.06	\\
\enddata

\end{deluxetable}

\clearpage
\pagestyle{empty}

\tablenum{6b}
\tablecolumns{11}
\tablewidth{0pt}

\begin{deluxetable}{ccccccccccc}
\tabletypesize{\scriptsize}
\tablecaption{Chemical Abundance Measurement Uncertainties}
\tablehead{
\colhead{Star}	&
\colhead{N--Ca}        &
\colhead{$\sigma$/$\surd$(N)}   &
\colhead{N--ScI}        &
\colhead{$\sigma$/$\surd$(N)}   &
\colhead{N--ScII}        &
\colhead{$\sigma$/$\surd$(N)}   &
\colhead{N--Ti}        &
\colhead{$\sigma$/$\surd$(N)}   &
\colhead{N--Ni}        &
\colhead{$\sigma$/$\surd$(N)} 
}
\startdata
9	&	5	&	0.04	&	\nodata	&	\nodata	&	1	&	\nodata	&	3	&	0.04	&	5	&	0.06	\\
5009	&	5	&	0.07	&	1	&	\nodata	&	2	&	0.04	&	3	&	0.05	&	3	&	0.10	\\
6017	&	5	&	0.05	&	\nodata	&	\nodata	&	\nodata	&	\nodata	&	5	&	0.04	&	5	&	0.09	\\
8014	&	5	&	0.05	&	\nodata	&	\nodata	&	1	&	\nodata	&	3	&	0.09	&	3	&	0.06	\\
9013	&	4	&	0.03	&	\nodata	&	\nodata	&	1	&	\nodata	&	2	&	0.05	&	3	&	0.05	\\
10009	&	5	&	0.05	&	1	&	\nodata	&	2	&	0.25	&	6	&	0.05	&	3	&	0.14	\\
10012	&	5	&	0.04	&	2	&	0.19	&	1	&	\nodata	&	3	&	0.02	&	3	&	0.12	\\
11019	&	5	&	0.03	&	1	&	\nodata	&	1	&	\nodata	&	5	&	0.04	&	5	&	0.06	\\
\enddata

\end{deluxetable}

\clearpage

\tablenum{7}
\tablecolumns{5}
\tablewidth{0pt}

\begin{deluxetable}{ccccc}
\tablecaption{Abundance Sensitivity to Model Atmosphere Parameters}
\tablehead{
\colhead{Ion}         &
\colhead{T$_{\rm eff}$$\pm$100}   &
\colhead{log g$\pm$0.30}       &
\colhead{[M/H]$\pm$0.30}     &
\colhead{v$_{\rm t}$$\pm$0.30}	\\
\colhead{}	&
\colhead{(K)}      &
\colhead{(cgs)}      &
\colhead{(dex)}      &
\colhead{(km s$^{\rm -1}$)} 
}
\startdata
\hline
\multicolumn{5}{c}{T$_{\rm eff}$=4200, log g=0.75, [Fe/H]=$-$2.00, v$_{\rm t}$=2.00}	\\
\hline
Fe I	&	$\pm$0.13	&	$\mp$0.01	&	$\mp$0.03	&	$\mp$0.10	\\
Fe II	&	$\mp$0.10	&	$\pm$0.13	&	$\pm$0.08	&	$\mp$0.03	\\
O I	&	$\pm$0.03	&	$\pm$0.11	&	$\pm$0.08	&	$\mp$0.03	\\
Na I	&	$\pm$0.09	&	$\mp$0.04	&	$\mp$0.05	&	$\pm$0.00	\\
Al I	&	$\pm$0.07	&	$\mp$0.02	&	$\mp$0.03	&	$\mp$0.02	\\
Si I	&	$\mp$0.01	&	$\pm$0.02	&	$\pm$0.02	&	$\mp$0.01	\\
Ca I	&	$\pm$0.14	&	$\mp$0.05	&	$\mp$0.06	&	$\mp$0.10	\\
\enddata
\end{deluxetable}

\clearpage

\tablenum{8}
\tablecolumns{2}
\tablewidth{0pt}

\begin{deluxetable}{ll}
\tablecaption{Literature References for Figures \ref{f15}--\ref{f18}}
\tablehead{
\colhead{Object}         &
\colhead{Reference}
}
\startdata
Thin/Thick Disk	&	Bensby et al. (2003)	\\
Thin/Thick Disk	&	Bensby et al. (2005)	\\
Thin/Thick Disk &	Edvardsson et al. (1993)	\\
Thin/Thick Disk &	Fulbright (2000)	\\
Thin/Thick Disk &	Fulbright et al. (2007)	\\
Thin/Thick Disk &	Reddy et al. (2003)	\\
Thin/Thick Disk &	Reddy et al. (2006)	\\
Thin/Thick Disk &	Simmerer et al. (2004)	\\
\enddata

\end{deluxetable}

\end{document}